\documentclass[11pt]{article}
\usepackage[final]{acl}

\usepackage{times}
\usepackage{latexsym}
\usepackage{amsmath,amssymb,amsfonts}
\usepackage{graphicx}
\usepackage{booktabs}
\usepackage{longtable}
\usepackage{caption}
\usepackage{stfloats}
\usepackage{multirow} 
\usepackage{comment}
\usepackage{mfirstuc,textcase,xspace,graphicx}
\usepackage{longtable}
\usepackage{wrapfig}
\usepackage{comment}  
\usepackage{algorithm}
\usepackage{algorithmic}
\usepackage{float}
\usepackage{xcolor,colortbl} 
\usepackage{amsmath,amsfonts} 
\usepackage{graphicx}
\usepackage{textcomp}
\usepackage{array}
\usepackage{pifont}
\usepackage{amsthm}
\usepackage{bm}
\usepackage{setspace}%
\usepackage{multirow}%
\usepackage{xltabular}
\usepackage{siunitx}
\usepackage{footnote}%
\usepackage{blindtext}
\usepackage{ctable}
\usepackage{pifont}
\usepackage{booktabs}
\usepackage{makecell}
\usepackage{tabularx}
\usepackage{enumitem}
\graphicspath{ {image/} }
\usepackage{titlesec}
\usepackage{soul}
\usepackage{url}
\usepackage{pifont}
\usepackage[
  protrusion=true,
  expansion=true,
  activate={true,nocompatibility},
  final,
  tracking=true,
  kerning=true,
  stretch=10,
  shrink=10
]{microtype}
\newcommand{\cmark}{\ding{51}}
\newcommand{\xmark}{\ding{55}}
\DeclareUnicodeCharacter{2009}{,}
\DeclareUnicodeCharacter{2248}{\ensuremath{\approx}}

\newcolumntype{P}[1]{>{\raggedright\arraybackslash}p{#1}}
\title{UDSS-BWE: {U}ncertainty- and {D}ecision-{S}cience Inspired {S}win {B}and{W}idth {E}xtension\thanks{Accepted to the main conference of EMNLP 2026, Budapest, Hungary.}}

\author{Tarikul Islam Tamiti, Sajid Fardin Dipto, David Vergano, \\
\textbf{Luke Baja-Ricketts, Anomadarshi Barua} \\
Department of Cyber Security Engineering, 
George Mason University, USA.
}
\date{}
\hypersetup{
  pdftitle={UDSS-BWE: Uncertainty- and Decision-Science Inspired Swin BandWidth Extension},
  pdfauthor={Tarikul Islam Tamiti, Sajid Fardin Dipto, David Vergano, Luke Baja-Ricketts, Anomadarshi Barua}
}

\begin{document}
\maketitle

\begin{abstract}

Bandwidth extension (BWE) is fundamentally localized: the most perceptual distortions are not average-case distortions, but rare high-frequency (HF) transients that standard, risk-neutral objectives tend to smooth away. To close this gap, we seek solutions in the risk-sensitive and uncertainty-aware decision
science rules and present \textbf{UDSS-BWE}, which introduces five decision-science and uncertainty-aware discriminators: CVaRD (does tail pooling to amplify HF artifacts), CCD (a primal-dual augmented Lagrangian to prevent HF overboost), MCUD (a learnable utility over spectral flatness/ centroid/ rolloff), EDD (captures epistemic uncertainty), and DROD (captures entropic KL-DRO aggregation). UDSS-BWE is also designed as a complex-valued adversarial BWE framework that uses Swin-based generators, a lightweight dual-stream shifted-window backbone, to capture local and long-range structure efficiently, while learnable \emph{lattice coupling} provides controlled cross-stream exchange. UDSS-BWE is optimized extensively and achieves better perceptual-quality 
with 3.89x fewer parameters (72M vs. 18.5M) over two English and French datasets under clean and noisy conditions. To the best of our knowledge, this work shows how multi disciplinary decision-science-inspired and uncertainty theories can be successfully used to design efficient discriminators for producing more nuanced audios, establishing a new baseline in the BWE task. 


\end{abstract}

\vspace{-0.0em}
\section{Introduction and Related Work}
\label{sec:intro}
\vspace{-0.0em}

Speech bandwidth extension (BWE) reconstructs missing frequency contents from bandwidth-limited speech 
 and improves recognition accuracy on
low-bandwidth inputs for \text{Text-to-Speech (TTS)} and \text{Automatic Speech Recognition (ASR)} tasks.

\textbf{Research gap:} Human speech is an inherently \textit{epistemically uncertain and stochastic} process. While traditional average spectral recovery is important, the perceptually most difficult part of BWE is the restoration of sparse high-frequency (HF) contents, such as fricatives, sibilants, stop bursts, and sharp onsets, that strongly affect naturalness, intelligibility, and downstream usability. These cues are only weakly implied by narrowband observations, making HF reconstruction intrinsically ambiguous. {This is supported by our phoneme analysis in Appendix}~\ref{app:hf_phoneme_benchmarks_complexity}: overall voiced speech contains relatively weak upper-band energy (see Table~\ref{tab:hfe_overall_supported}), whereas fricatives such as /s/ and /sh/ exhibit much stronger and highly phoneme-dependent HF structure (see Table~\ref{tab:hfe_fricatives_supported}). Therefore, successful BWE must determine not only \emph{which} HF content to restore, but also \emph{when} it should occur and \emph{how} strongly it should be expressed.


This issue is not addressed, as generic adversarial critics mostly aggregate errors in an average sense \cite{10246854}. Therefore, supervision is dominated by easy low-frequency or vowel-dominated regions, while rare but perceptually important HF failures receive limited emphasis. In contrast, sharper adversarial feedback can encourage implausible HF amplification or hiss-like artifacts. These issues are particularly important in \textit{complex-valued} BWE, where realistic BWE depends on jointly modeling magnitude and phase \cite{gerkmann2012phase,yin2020phasen,lu2025explicit,11464384,tamiti-etal-2026-cis}. \textit{Although recent models, such as EBEN \cite{hauret2023eben}, AERO \cite{mandel2023aero}, and AP-BWE \cite{lu2024towards} have advanced the field, open challenges remain in HF-sensitive supervision, training stability, and efficiency.}

\textbf{Contribution:} To close these gaps, we introduce \textbf{UDSS-BWE} (\textbf{{U}}ncertainty and \textbf{{D}}ecision - \textbf{{S}}cience guided \textbf{{S}}win \textbf{{BWE}}). 
UDSS-BWE introduces a \textit{parameter-efficient dual-stream Swin-1D lattice generator} that uses shifted-window multihead self-attention (MSA) to model local-to-nonlocal temporal structure and employs \textit{lattice coupling} \cite{Luo2020LatticeNet} to control cross-stream interaction, refining magnitude \& phase in parallel. 

On the supervision side, to address \textit{which, when and how} the HF artifacts can be restored under \textit{epistemically uncertainty with confidence}, to emphasize perceptually critical \textit{tail errors}, to \textit{guard-rail} against HF over-boost, we seek solutions in risk-sensitive and uncertainty-aware decision-science rules. And, for the first time, we introduce five decision-science and uncertainty-inspired critics: \textbf{CVaRD} (Conditional Value at Risk Discriminator), which emphasizes tail errors through CVaR-style pooling \cite{rockafellar2002conditional}; \textbf{CCD} (Chance-Constraint HF Discriminator), which discourages persistent HF over-boost through a chance-constrained barrier \cite{erdougan2006ambiguous}; \textbf{MCUD} (Multi-Criteria Utility Discriminator), which learns an interpretable utility over spectral flatness, centroid, and rolloff \cite{keeney1993decisions}; \textbf{EDD} (Evidential Dirichlet Discriminator), which exposes epistemic uncertainty through evidential Dirichlet outputs \cite{kendall2017uncertainties}; and \textbf{DROD} (Entropic KL-DRO Discriminator), which robustly emphasizes hard segments under shift through entropic KL-DRO aggregation \cite{ahmadi2012entropic}. 


UDSS-BWE achieves the best perceptual score (NISQA-MOS) on VCTK English dataset, while remaining competitive across other objective metrics (LSD, STOI, PESQ, SI-SNR, SI-SDR, WER, refer to Section \ref{subsec:EvaluationMetrics} for full forms) and generalizes to MLS French dataset. We further show the robustness of UDSS-BWE in noisy conditions and subjective evaluation. 
Most importantly, we show that these gains are obtained with substantially 3.89x fewer parameters compared to AP-BWE (18.5M vs. 72M) with 2.66x MACs and FLOPs reduction. 
Therefore, for the first time, this paper shows that decision-science and uncertainty theories can be used as effective inductive biases for designing parameter-efficient BWE frameworks.




\section{UDSS-BWE Architecture}
\label{sec:overview}
\vspace{-0.0em}

\begin{figure}[t]
  \centering
  \includegraphics[width=0.485\textwidth,height=0.16\textheight]{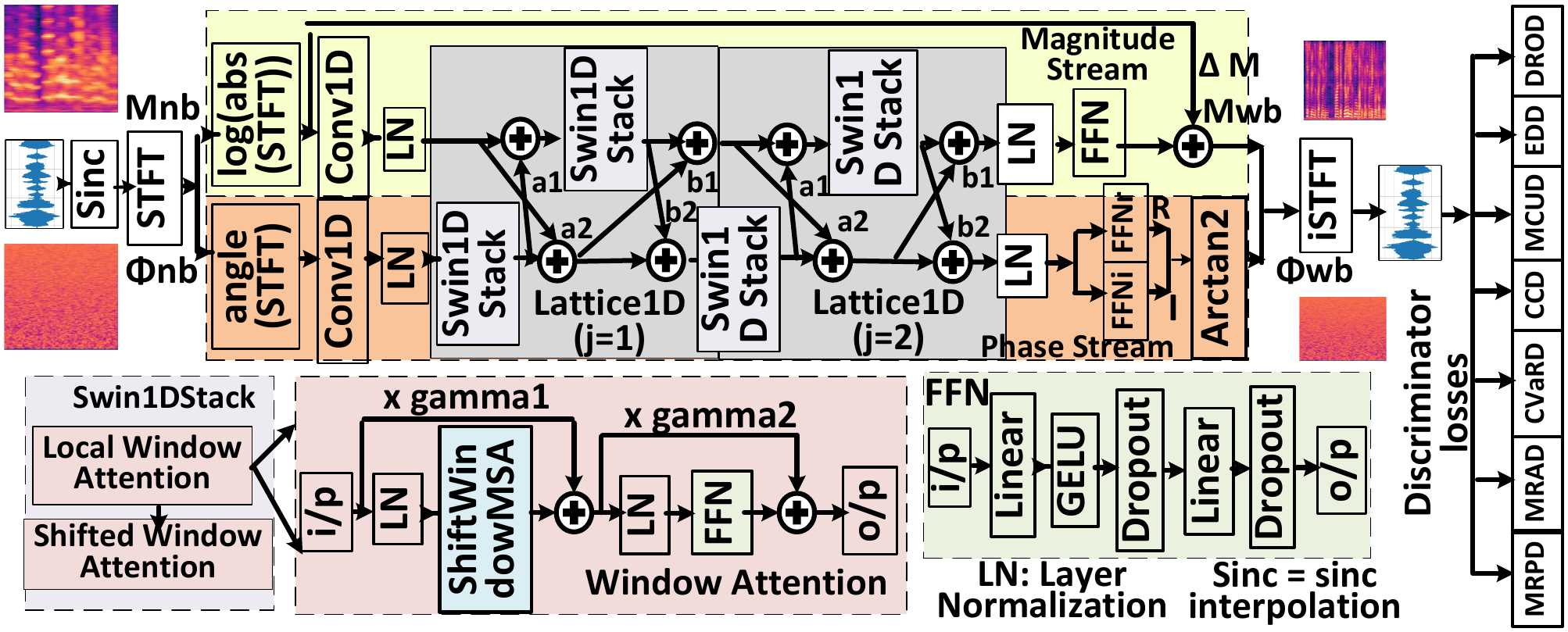}
  \vspace{-02.0em}
   \caption{{\color{black}The overall architecture of UDSS-BWE.}}
  \label{fig:overall_architecture}
  \vspace{-1.1em}
\end{figure}

\subsection{Generator Design (see Fig.   \ref{fig:overall_architecture} \& Alg.~\ref{alg:gen})}
\label{subsec:gen}
\vspace{-0.0em}


As phase and magnitude are important \cite{yin2020phasen}, the generator is designed as a dual-stream architecture that processes magnitude and phase in parallel while enabling controlled cross-stream interaction after the Short-Term Fourier Transformation (STFT) of the narrowband audio $\mathbf{S}_{\mathrm{nb}}$. Let $\mathbf{M}_{\mathrm{nb}}$ is the \textit{log-magnitude} of narrow-band spectrogram (i.e., $\mathbf{M}_{\mathrm{nb}}=\log|\mathbf{S}_{\mathrm{nb}}|$), and $\boldsymbol{\Phi}_{\mathrm{nb}}$ is the wrapped narrow-band phase after STFT.
The generator's forward pass has the following two stages:

\textbf{a) {Swin 1-D Block with Lattice Interaction:}} To explicitly exchange information between magnitude and phase streams without collapsing them into a single representation, we employ \textit{two successive} one-dimensional (1D) lattice blocks (\textbf{Lattice1D}). Each lattice block \cite{Luo2020LatticeNet} does (i) \textit{amplitude-phase stream refinement} and (ii) \textit{cross-stream injection} via four learnable coupling coefficients $(a_1, a_2, b_1, b_2)$. Concretely, within the $j$-th lattice block (i.e, $j\epsilon\{1,2\}$, we use a \textit{single shared} Swin-1D shifted-window self-attention $\mathrm{Swin1DStack}_j(\cdot)$ (i.e., the \textit{same parameters} are reused for both refinement calls inside that lattice block, see Alg.~\ref{alg:gen} in Appendix \ref{appen:alg:gen} for details).


Unlike standard global attention, Swin-1D performs self-attention within window size $W$, and alternates shift offsets $\{0,\lfloor W/2\rfloor\}$ across successive blocks, enabling cross-window propagation while maintaining windowed complexity $O(TW)$ instead of $O(T^2)$.
We set the shift size to $\lfloor W/2\rfloor$,  meaning an even $W$ equals the canonical half-window shift $W/2$.
{This design 
replaces heavier sequence refiners with an efficient local-to-nonlocal Swin operator.} The proposed lattice-coupled Swin refinement mitigates  wideband failure by learning \textit{where} and \textit{how much} phase cues should modulate magnitude reconstruction (vice versa) by efficient information exchange via Lattice coupling (controllable inter-stream mixing).

\textbf{b) Residual Prediction and Synthesis:} After the final lattice stage, a residual narrowband $\Delta\mathbf{M}$ 
is added to the magnitude stream to reconstruct wideband magnitude output $\mathbf{M}_{\mathrm{wb}}$. This residual connection isolates missing high-frequency content rather than re-modeling the full spectrum.

In contrast, direct phase regression is unstable due to the noisy nature of phase \cite{yin2020phasen}. Therefore, the phase stream is mapped to a complex-valued representation by predicting ``pseudo-real $(R)$'' and ``pseudo-imaginary $(I)$'' components using linear networks (FFN) and LN. Later, the wrapped wideband phase $\Phi_{\mathrm{wb}}$ is recovered via $\operatorname{atan2}(\cdot,\cdot)$ as $\Phi_{\mathrm{wb}} = \operatorname{atan2}(I, R)$.
Finally, the complex wideband spectrogram is formed from the $\mathbf{M}_{\mathrm{wb}}$ and $\Phi_{\mathrm{wb}}$ via inverse STFT.

\subsection{Decision-Science Discriminators}
\label{subsec:dsdisc}
\vspace{-0.0em}


\textbf{Why:} Wideband reconstruction errors are not uniformly distributed over time. Most perceptually harmful artifacts (e.g., high-frequency chirps, transient bursts) appear as \textit{rare tail events}. \textit{Standard waveform critics typically aggregate temporal evidence using mean pooling, which implicitly optimizes an average-case objective.} As a result, supervision for short, low-duty-cycle failures is diluted, allowing the generator to hide artifacts in a few frames while still matching global statistics.

To solve these, we seek solutions in risk-sensitive and constraint-aware decision-science rules, and, for the first time, we introduce three decision-science-inspired critics -- CVaRD, CCD, and MCUD -- below to encode interpretable HF artifacts. 
All critics are kept compact via depthwise-separable 1D convolutions (SepConv1D). 

\begin{figure}[t]
  \centering
  \includegraphics[width=0.48\textwidth,height=0.2\textheight]{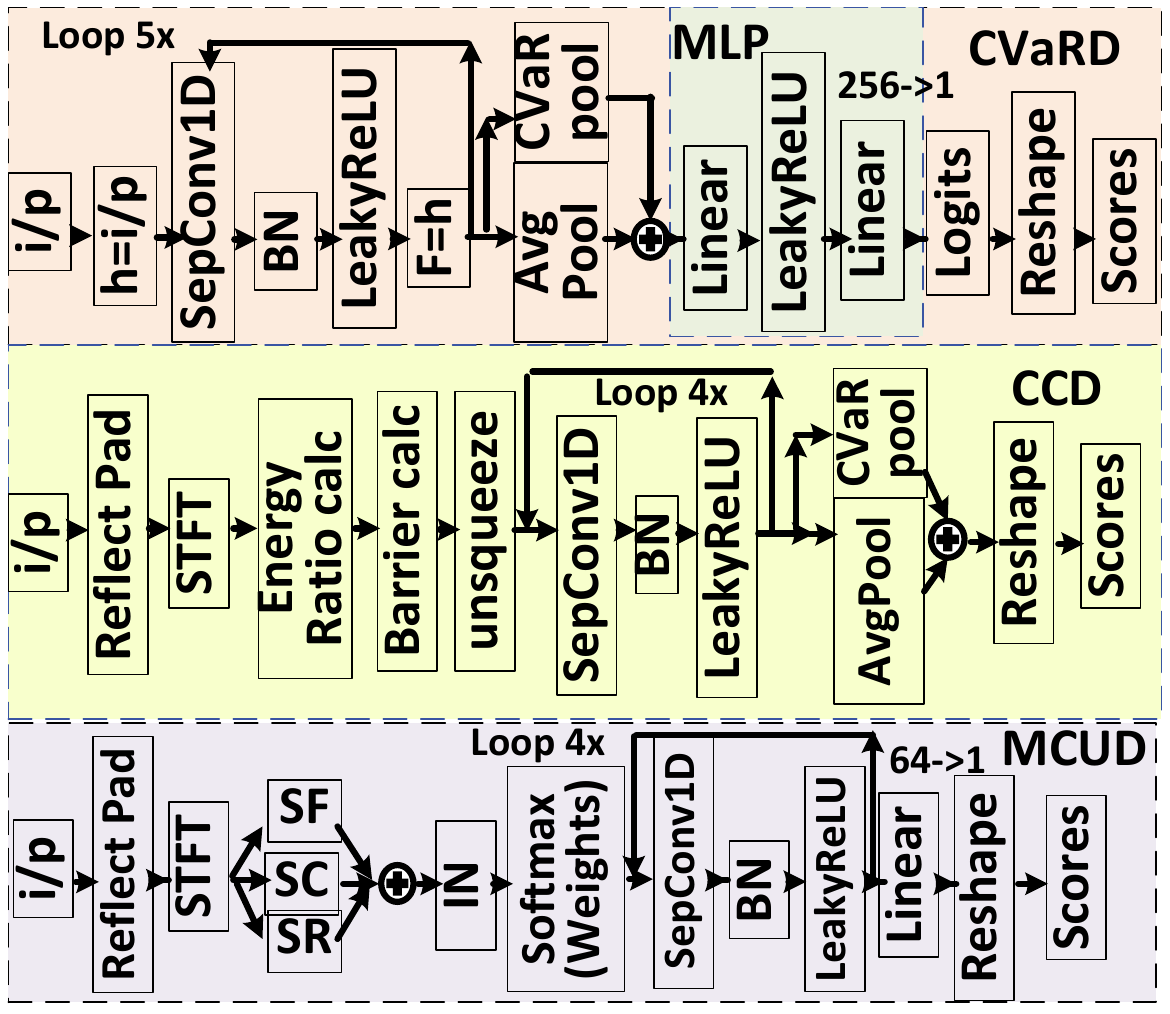}
  \vspace{-1.7em}
   \caption{Architecture of CVaRD, CCD, and MCUD.}
  \label{fig:ds_discrim}
  \vspace{-1.20em}
\end{figure}

\subsubsection{Conditional Value at Risk Dis. (CVaRD)}
\label{subsec:disc_cvar}
\vspace{-0.0em}


\textbf{Why:}
CVaR can capture localized HF tail fraction like it does for local responses  \cite{rockafellar2002conditional}. Therefore, we design CVaR-inspired adversarial supervision to emphasize short high-risk temporal regions instead of averaging them out, useful for brief \& missing HF details.

\textbf{Design (see Alg.~\ref{alg:cvar} \& Fig.~\ref{fig:ds_discrim}}): 
The generator's output $y$ passes through $L$ SepConv1D blocks with batch normalization (BN) and LeakyReLU activations and stores intermediate features $\{F^{(\ell)}\}_{\ell=1}^{L}$ for feature matching, where $\ell$ indexes the blocks (\textbf{Row \textcircled{\scriptsize 2}}--\textbf{\textcircled{\scriptsize 5}}). \textbf{Row \textcircled{\scriptsize 6}} sets $k=\max(1,\lfloor \alpha T' \rfloor)$ using tail fraction $\alpha\in(0,1]$. Here, $T'$ is the temporal length of the final feature map $h$, and $k$ is the number of time steps selected from the tail. The term $\alpha$ controls how much of the highest-magnitude activations are treated as rare high-risk events. Conceptually, this reflects the idea that perceptually harmful artifacts are rare events, so the critic focuses on the worst temporal regions. \textbf{Row \textcircled{\scriptsize 7}} computes the CVaR-style summary $h_{\mathrm{cvar}}$ by averaging the Top-$k$ values of $|h|$ along time, so the discriminator focuses on the strongest temporal responses. We compute the global summary $h_{\mathrm{gap}}$ through average pooling. \textbf{Row \textcircled{\scriptsize 8}} fuses local risk and global context as $z$=$h_{\mathrm{cvar}}+h_{\mathrm{gap}}$, and maps $z$ to scalar critic score $s$ using a multi layer perceptron (MLP). \textbf{Row \textcircled{\scriptsize 10}}--\textbf{\textcircled{\scriptsize 11}} apply the ForwardSingle() to calculate $y_{\mathrm{real}}$ and $y_{\mathrm{fake}}$. The decision-science principle is used in 
\textbf{Row \textcircled{\scriptsize 7}}, where CVaR-inspired tail selection makes the critic risk-sensitive in time. Overall, CVaRD yields sharper gradients on short transients and reduces the case where average spectra look reasonable but brief HF events are missing.

{
\setlength{\textfloatsep}{0pt plus 0pt minus 0pt}   
\begin{algorithm}[t]
\caption{Pseudo code - CVaRD (Table \ref{tab:disc_risk_constraint_hparams_app}, \ref{tab:riskcvar_breakdown})}
\label{alg:cvar}
{\scriptsize
\begin{algorithmic}[1]
\STATE \textbf{Require:} $y_{\mathrm{real}}, y_{\mathrm{fake}} \in \mathbb{R}^{B \times 1 \times T}$,
tail fraction $\alpha \in (0,1]$
\STATE \textbf{ForwardSingle}$(y)$:
\STATE \hspace{1.2em} $h \gets y$\\
\hspace{1.2em} \textbf{FOR {$\ell = 1$ to $L$}}
    \STATE \hspace{1.8em} $h \gets
    \mathrm{LeakyReLU}\!\big(
    \mathrm{BN}^{(\ell)}(
    \mathrm{SepConv1D}^{(\ell)}(h))\big)$
    \STATE \hspace{1.8em} Store $F^{(\ell)} \gets h$\\
\hspace{1.2em} \textbf{ENDFOR}
\STATE \hspace{1.2em} $T' \gets \mathrm{length}(h)$; $k \gets \max(1,\lfloor \alpha T' \rfloor)$
\STATE \hspace{1.2em} $h_{\mathrm{cvar}} \gets
\frac{1}{k}\sum \mathrm{TopK}(|h|, k)$; $h_{\mathrm{gap}} \gets \mathrm{AvgPool}_{t}(h)$
\STATE \hspace{1.2em} $z \gets h_{\mathrm{cvar}} + h_{\mathrm{gap}}$; $s \gets \mathrm{MLP}(z)$
\STATE \hspace{1.2em} \textbf{return} $(s,\{F^{(\ell)}\}_{\ell=1}^{L})$
\STATE $(r,\{F_r^{(\ell)}\}) \gets \textbf{ForwardSingle}(y_{\mathrm{real}})$; 
\STATE $(f,\{F_f^{(\ell)}\}) \gets \textbf{ForwardSingle}(y_{\mathrm{fake}})$; 
\STATE \textbf{return} $([r],[f],\{F_r^{(\ell)}\}_{\ell=1}^{L},\{F_f^{(\ell)}\}_{\ell=1}^{L})$
\end{algorithmic}
}
\end{algorithm}
}

\subsubsection{Chance-Constraint HF Discri. (CCD)}
\label{subsec:disc_cc}
\vspace{-0.0em}

\textbf{Why:}
CCD applies adversarial pressure against \emph{persistent} HF boost inspired by a chance constraint perspective \cite{erdougan2006ambiguous}. Instead of rewarding any increase in HF energy that may correlate with perceptual sharpness, CCD treats excessive HF ratio as a rare-event violation and penalizes it only when it occurs too often over time, discouraging systematic HF amplification while still allowing brief, consistent HF  bursts.


\textbf{Design (see Alg.~\ref{alg:ccd} \& Fig.~\ref{fig:ds_discrim}}):  
\textbf{Row \textcircled{\scriptsize 3}} does  reflect padding to input $y$. The power spectrogram $P$ is calculated from the STFT of $y$. \textbf{Row \textcircled{\scriptsize 4}} computes the HF start bin $f_0$ from the frequency bins $F$ and HF fraction $\rho$. Here, $\rho$ specifies the fraction of the highest frequencies included in the HF band. By restricting attention to the top $\rho$ fraction of frequencies, the CCD critic focuses on HF spectral violations. \textbf{Row \textcircled{\scriptsize 5}} computes the total energy $E_{\mathrm{tot}}(\tau)$ and the HF energy $E_{\mathrm{hf}}(\tau)$ for each $\tau$ indexes from the STFT frames. \textbf{Row \textcircled{\scriptsize 6}} forms the per-frame HF ratio $q(\tau)$=$E_{\mathrm{hf}}/E_{\mathrm{tot}}$ and maps it through the smooth softplus barrier
$b(\tau)$=$\log\!\big(1+\exp(\kappa(q(\tau)$-$\tau_0))\big)$,
yielding a differentiable chance-constraint violation trace. Here, $\kappa$ is the barrier slope, which helps the softplus barrier
$b(\tau)$ to convert HF-threshold exceedance into a smooth temporal violation signal. \textbf{Row \textcircled{\scriptsize 7}} passes this trace through a SepConv1D head and produces the final feature map $h$. 
We compute global average $g_{\mathrm{avg}}$ and max averages $g_{\mathrm{max}}$ over time. \textbf{Row \textcircled{\scriptsize 8}} combines these averages as $z$ and maps to scalar critic score $s$. \textbf{Row \textcircled{\scriptsize 10}-\textcircled{\scriptsize 11}} apply the \textbf{ForwardSingle()} to calculate $y_{\mathrm{real}}$ and $y_{\mathrm{fake}}$. 
Overall, CCD, through Row \textcircled{\scriptsize 6}, applies stronger pressure against persistent HF over-boost and to reduce cases where perceived sharpness arises mainly from systematic HF amplification rather than genuinely consistent HF detail.

{
\setlength{\textfloatsep}{0pt plus 0pt minus 0pt}   
\begin{algorithm}[t]
\caption{Pseudo code of CCD (Table \ref{tab:disc_risk_constraint_hparams_app}, \ref{tab:chance_utility_breakdown})}
\label{alg:ccd}
{\scriptsize
\begin{algorithmic}[1]
\STATE \textbf{Require:} $y_{\mathrm{real}}, y_{\mathrm{fake}}$, STFT $(n_{\mathrm{fft}},h,w)$, HF fraction $\rho$, barrier $\kappa,\tau_0$ 
\STATE \textbf{ForwardSingle}$(y)$:
\STATE \hspace{1.2em} \textbf{if} $T<w$ \textbf{then} reflect-pad $y$;
 \hspace{0em} $X \gets \mathrm{STFT}(y)$; $P \gets |X|^2$
\STATE \hspace{1.2em} $F \gets \#\mathrm{freqbins}(P)$; $f_0 \gets \lfloor(1-\rho)F\rfloor$
\STATE \hspace{1.2em} $E_{\mathrm{tot}}(\tau) \gets \sum_f P+\varepsilon$; $E_{\mathrm{hf}}(\tau) \gets \sum_{f=f_0}^{F} P$
\STATE \hspace{1.2em} $q(\tau) \gets E_{\mathrm{hf}}/E_{\mathrm{tot}}$; $b(\tau) \gets \log\!\big(1+\exp(\kappa(q(\tau)$-$\tau_0))\big)$
\STATE \hspace{1.2em} $h \gets \mathrm{SepConv1D}(b(\tau))$;
 \hspace{1.2em} $g_{\mathrm{avg}} \gets \mathrm{AvgPool}_t(h)$
 \STATE \hspace{1.2em} $g_{\mathrm{max}} \gets \mathrm{MaxPool}_t(h)$;
 \hspace{0em} $z \gets \mathrm{concat}(g_{\mathrm{avg}},g_{\mathrm{max}})$; $s \gets \sum z$
\STATE \hspace{1.2em} \textbf{return} $(s,\{h^{(0)},h\})$
\STATE $(r,F_r) \gets \textbf{ForwardSingle}(y_{\mathrm{real}})$ 
\STATE $(f,F_f) \gets \textbf{ForwardSingle}(y_{\mathrm{fake}})$
\STATE \textbf{return} $([r],[f],F_r,F_f)$
\end{algorithmic}
}
\end{algorithm}
}

In generator optimization, the same barrier trace $b(\tau)$ is reused as a differentiable constraint monitor (primal dual augmented Lagrangian). Let $\bar{b}$ denote the batch mean of the barrier trace on generated audio and let $\bar{b}_0$ be a target exceedance level. The residual is $g$=$\bar{b}$-$\bar{b}_0$, and the generator receives the penalty $\mathcal{L}_{\mathrm{CC}}$=$\lambda_{\mathrm{cc}}g+\frac{\rho_{\mathrm{cc}}}{2}g^2$, scaled by the adversarial warmup factor so constraint enforcement activates progressively with GAN training (see Table \ref{tab:lossfunction_uds_swinbwe}). The dual variable is updated at each step as $\lambda_{\mathrm{cc}}\leftarrow\mathrm{clip}\!\left(\lambda_{\mathrm{cc}}+\eta_{\mathrm{cc}}g,\;0,\;\lambda_{\max}\right)$. This synchronization discourages persistent HF over-boost while preserving legitimate transient HF events.

\subsubsection{Multi-Criteria Utility Discri. (MCUD)}
\label{subsec:disc_mcu}
\vspace{-0.0em}

\textbf{Why:}
MCUD applies adversarial pressure against unrealistic HF structure inspired by a multi-criteria utility perspective \cite{keeney1993decisions}. Instead of collapsing HF realism into a single opaque scalar that can entangle distinct artifacts, it explicitly evaluates several interpretable spectral criteria and learns a convex utility over them. This is useful for distinguishing cases where generated HF content may appear sharper overall but remains implausible in specific ways, such as being overly noisy, overly bright, or excessively spread.

{
\setlength{\textfloatsep}{0pt}   
\begin{algorithm}[t]
\caption{Pseudo code - MCUD (Table \ref{tab:disc_utility_evidential_hparams_app}, \ref{tab:chance_utility_breakdown})}
\label{alg:mcud}
{\scriptsize
\begin{algorithmic}[1]
\STATE \textbf{Require:} $y_{\mathrm{real}}, y_{\mathrm{fake}} \in \mathbb{R}^{B\times1\times T}$, STFT $(n_{\mathrm{fft}},h,w)$, rate $f_s$
\STATE \textbf{ForwardSingle}$(y)$:
\STATE \hspace{1.2em} \textbf{if} $T<w$ \textbf{then} reflect-pad $y$
\STATE \hspace{1.2em} $X \gets \mathrm{STFT}(y)$; $A \gets |X|+\varepsilon$
\STATE \hspace{1.2em} Compute $\mathrm{SF}(\tau)$, normalized $\mathrm{SC}(\tau)$, and $\mathrm{SR}_{85}(\tau)$ from $A$
\STATE \hspace{1.2em} $C \gets [\mathrm{SF},\mathrm{SC},\mathrm{SR}_{85}]$; $C_n \gets \mathrm{InstanceNormTime}(C)$
\STATE \hspace{1.2em} $w \gets \mathrm{softmax}(\theta)$; $u(\tau) \gets \sum_{k=1}^{3} w_k\,C_n(k,\tau)$
\STATE \hspace{1.2em} $s \gets \mathrm{SepConv1D}(u)$ 
\STATE \hspace{1.2em} \textbf{return} $(s,\{C_n,u\})$
\STATE $(r,F_r) \gets \textbf{ForwardSingle}(y_{\mathrm{real}})$
\STATE $(f,F_f) \gets \textbf{ForwardSingle}(y_{\mathrm{fake}})$
\STATE \textbf{return} $([r],[f],F_r,F_f)$
\end{algorithmic}
}
\end{algorithm}
}

\textbf{Design (see Alg.~\ref{alg:mcud} \& Fig.~\ref{fig:ds_discrim}}): 
We do reflect padding on input $y$ and compute the magnitude spectrogram $A$ from the STFT of $y$ (\textbf{Row \textcircled{\scriptsize 2}}--\textbf{\textcircled{\scriptsize 4}}). \textbf{Row \textcircled{\scriptsize 5}} computes three per-frame spectral criteria from $A$: spectral flatness $\mathrm{SF}(\tau)$, normalized spectral centroid $\mathrm{SC}(\tau)$, and spectral rolloff at $85\%$ cumulative magnitude $\mathrm{SR}_{85}(\tau)$. \textbf{Row \textcircled{\scriptsize 6}} stacks these criteria into $C$ and applies instance normalization over time, yielding $C_n$, which reduces scale mismatch among criteria and stabilizes their joint use during training. \textbf{Row \textcircled{\scriptsize 7}} computes simplex-constrained weights $w$ and forms the convex utility trace $u(\tau)$=$\sum_{k=1}^{3} w_k\,C_n(k,\tau)$, which measures the overall spectral quality of frame $\tau$ according to three criteria: SF (tonal vs noise structure), SC (brightness), and SR (energy distribution). \textbf{Row \textcircled{\scriptsize 8}} passes this utility trace through a SepConv1D head and produces the final critic score $s$. \textbf{Row \textcircled{\scriptsize 10}}--\textbf{\textcircled{\scriptsize 11}} applies the same routine to $y_{\mathrm{real}}$ and $y_{\mathrm{fake}}$. The decision-science-inspired principle is used 
especially in \textbf{Row \textcircled{\scriptsize 5}}, where MCUD learns a convex utility over explicit criteria rather than relying on a single entangled HF statistic, 
reducing cases where improvement in one HF characteristic hides degradation in another.
\begin{figure}[t]
\vspace{-0.0em}
  \centering
  \includegraphics[width=0.37\textwidth,height=0.14\textheight]{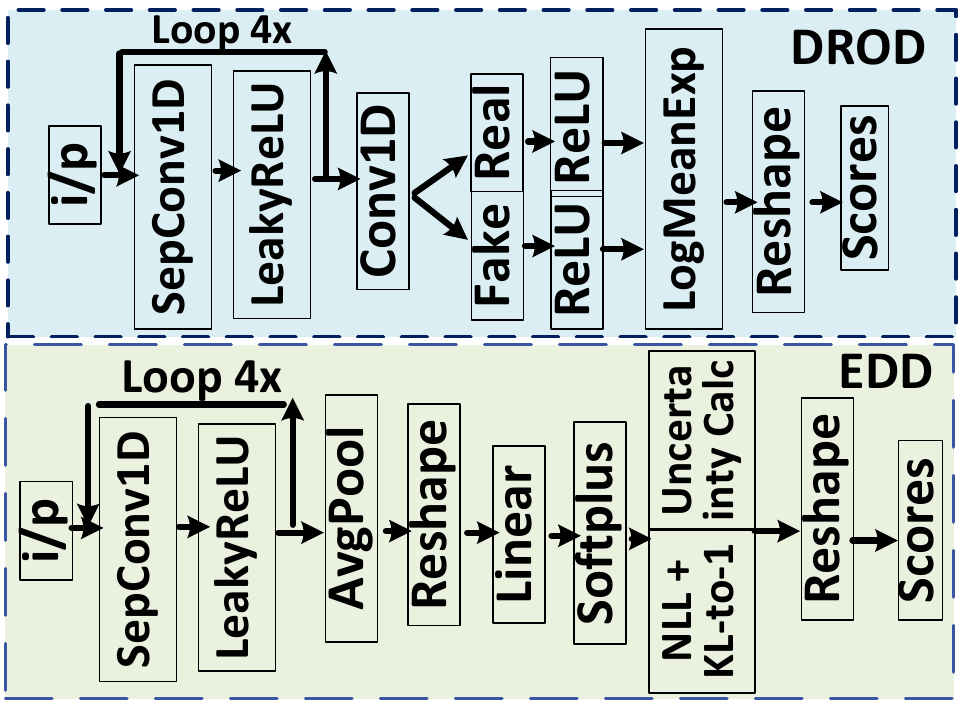}
  \vspace{-0.400em}
   \caption{Architecture of EDD and DROD.}
  \label{fig:un_discr}
  \vspace{-0.95em}
\end{figure}

\subsection{Uncertainty-Aware Discriminators}
\label{subsec:uadisc}
\vspace{-0.0em}

\textbf{Why:} 
In speech, HF bands depend on lower fundamental frequencies and harmonics. These mappings are difficult to model due to the inherent uncertain relationships and \textit{epistemic ambiguity}. Some HF regions contain multiple plausible wide-band realizations (e.g., speaker-dependent fricatives, noisy bursts, and rare out-of-distribution phonetic events). A standard critic can only output a real/fake score and does not indicate whether that decision is made with strong evidence under uncertainty. To solve this problem, we introduce two lightweight uncertainty-aware discriminators, built on SepConv as a backbone: (i) \textit{EDD} explicitly models evidence and epistemic uncertainty, and (ii) \textit{DROD} emphasizes hard temporal segments through a smooth entropic-risk objective.

{
\setlength{\textfloatsep}{0pt}   
\begin{algorithm}[t]
\caption{Pseudo code of EDD (Table \ref{tab:disc_utility_evidential_hparams_app}, \ref{tab:evidential_dro_breakdown})}
\label{alg:edd}
{\scriptsize
\begin{algorithmic}[1]
\STATE \textbf{Require:} $y_{\mathrm{real}}, y_{\mathrm{fake}} \in \mathbb{R}^{B\times1\times T}$, $\varepsilon>0$
\STATE \textbf{ForwardSingle}$(y)$:
\STATE \hspace{0.5em} $(h,\{F^{(\ell)}\}_{\ell=1}^{4}) \gets \mathrm{SepConv1D}(y)$
\STATE \hspace{0.5em} $\ell \gets \mathrm{Linear}(\mathrm{Flatten}(\mathrm{GAP}_t(h))) \in \mathbb{R}^{B\times2}$; $e \gets \mathrm{softplus}(\ell)$ \STATE \hspace{0.5em} $\alpha \gets e+\mathbf{1}$;
 $S \gets \sum_{k=1}^{2}\alpha_k$;  $u \gets 2/(S+\varepsilon)$
\STATE \hspace{0.5em} \textbf{return} $(\alpha,u,\{F^{(\ell)}\}_{\ell=1}^{4}\cup\{\ell\})$
\STATE $(\alpha_r,u_r,F_r) \gets \textbf{ForwardSingle}(y_{\mathrm{real}})$
\STATE $(\alpha_f,u_f,F_f) \gets \textbf{ForwardSingle}(y_{\mathrm{fake}})$
\STATE \textbf{return} $([\alpha_r],[\alpha_f],[u_r],[u_f],F_r,F_f)$
\end{algorithmic}
}
\end{algorithm}
}

\vspace{-0.0em}
\subsubsection{Evidential Dirichlet Discri. (EDD)}
\label{subsec:disc_edd}
\vspace{-0.0em}

\textbf{Why:}
EDD applies adversarial pressure against unreliable HF decisions inspired by an evidential-uncertainty perspective \cite{kendall2017uncertainties}. Instead of collapsing the discriminator output into a single scalar score, it predicts Dirichlet \cite{sensoy2018evidential} evidence over $\{\mathrm{real},\mathrm{fake}\}$ and distinguishes \emph{lack of evidence} from \emph{strong evidence for the wrong class}. This is useful because some HF regions are intrinsically ambiguous, so the critic should reveal not only \textit{which} class it prefers, but also \textit{how confidently} that decision is supported.

\textbf{Design (see Alg.~\ref{alg:edd} \& Fig.~\ref{fig:un_discr}}): The function \textbf{ForwardSingle()} takes input $y$ and passes it through 
four SepConv1D blocks with 
LeakyReLU activations, 
producing final temporal representation $h$ with intermediate feature maps $\{F^{(\ell)}\}_{\ell=1}^{4}$ for feature matching (\textbf{Row \textcircled{\scriptsize 2}}--\textbf{\textcircled{\scriptsize 3}}). \textbf{Row \textcircled{\scriptsize 4}-\textcircled{\scriptsize 5}} apply global average pooling (GAP) over time, flattening, and a linear head to obtain two logits $\ell\in\mathbb{R}^{B\times2}$. We convert these logits into nonnegative evidence $e=\mathrm{softplus}(\ell)$ and then into Dirichlet concentration parameters $\alpha=e+\mathbf{1}$, where larger total concentration indicates stronger evidential support. We compute the total concentration $S=\sum_{k=1}^{2}\alpha_k$ and the predictive uncertainty
$
u=\frac{2}{S+\varepsilon},
$
so uncertainty decreases as total evidence increases. \textbf{Row \textcircled{\scriptsize 6}} returns $\alpha$, $u$, and the backbone feature maps together with the final logit map for feature matching. \textbf{Row \textcircled{\scriptsize 7}}--\textbf{\textcircled{\scriptsize 8}} apply the same \textbf{ForwardSingle()} to $y_{\mathrm{real}}$ and $y_{\mathrm{fake}}$. The uncertainty principle is used mainly in \textbf{Row \textcircled{\scriptsize 4}}--\textbf{\textcircled{\scriptsize 5}}, where the discriminator outputs evidential mass rather than a single unconstrained score and explicitly exposes uncertainty through the inverse total concentration.

During discriminator optimization (see Table \ref{tab:lossfunction_uds_swinbwe}), real and fake samples are supervised with the evidential objective
$\mathcal{L}_{\mathrm{EDD}}(\alpha,\mathbf{y})$
=
$\sum_k y_k\!\left(\psi\!\left(\sum_j \alpha_j\right)-\psi(\alpha_k)\right)$
+
$\lambda_{\mathrm{kl}} 
\mathrm{KL}\!\left(\mathrm{Dir}(\alpha)\,\|\,\mathrm{Dir}(\mathbf{1})\right),
$
so EDD learns both class preference and evidence calibration. In generator optimization, the same evidential loss is applied to generated samples with the \emph{real} target, while the mean uncertainty on generated audio is penalized as an additional term. These adversarial terms are then scaled by the adversarial warmup factor so evidential supervision activates progressively with GAN training. 
Overall, EDD separates \emph{incorrect} from \emph{uncertain} HF decisions, reduces overconfident learning from weak cues, and pressures the generator 
to accumulate strong evidential support for natural HF reconstruction.

{
\setlength{\textfloatsep}{0pt}   
\begin{algorithm}[t]
\caption{Pseudo code of DROD (Table \ref{tab:disc_dro_weights_hparams_app}, \ref{tab:evidential_dro_breakdown})}
\label{alg:drod}
{\scriptsize
\begin{algorithmic}[1]
\STATE \textbf{Require:} $y_{\mathrm{real}}, y_{\mathrm{fake}} \in \mathbb{R}^{B\times1\times T}$, risk parameter $\lambda \ge 0$, numerical constant $\varepsilon>0$
\STATE \textbf{ForwardSingle}$(y)$:
\STATE \hspace{0.5em} $(h,\{F^{(\ell)}\}_{\ell=1}^{4}) \gets \mathrm{SepConv1D}(y)$; $s(\tau) \gets \mathrm{Conv1D}_{1\times1}(h)$ 
\STATE \hspace{0em} \textbf{return} $(s,\{F^{(\ell)}\}_{\ell=1}^{4}\cup\{s\})$
\STATE $(s_r,F_r) \gets \textbf{ForwardSingle}(y_{\mathrm{real}})$
\STATE $(s_f,F_f) \gets \textbf{ForwardSingle}(y_{\mathrm{fake}})$
\STATE $z_D(\tau) \gets \mathrm{ReLU}(1-s_r(\tau))+\mathrm{ReLU}(1+s_f(\tau))$
\STATE $z_G(\tau) \gets -s_f(\tau)$
\STATE $\rho_\lambda(z) \gets
\begin{cases}
\mathrm{mean}_\tau(z), & \lambda\le 0\\
m+\frac{1}{\lambda}\log(\mathrm{mean}_\tau(\exp(\lambda(z-m)))+\varepsilon), & \lambda>0,\;
\end{cases}$
\STATE $\mathcal{L}_D^{\mathrm{DRO}} \gets \mathrm{mean}_B(\rho_\lambda(z_D))$; $\mathcal{L}_G^{\mathrm{DRO}} \gets \mathrm{mean}_B(\rho_\lambda(z_G))$
\STATE \textbf{return} $([s_r],[s_f],F_r,F_f)$
\end{algorithmic}
}
\end{algorithm}
}

\subsubsection{Entropic KL-DRO Discri. (DROD)}
\label{subsec:disc_drod}
\vspace{-0.0em}

\textbf{Why:}
Instead of averaging temporal errors uniformly, DROD preserves per-segment critic scores and aggregates their violations inspired by an entropic-risk function \cite{ahmadi2012entropic}. It is useful as short-lived HF artifacts can be severe while remaining diluted under ordinary mean pooling. By emphasizing hard temporal regions through KL-DRO style aggregation, it becomes more sensitive to sparse yet harmful local failures.


\textbf{Design (see Alg.~\ref{alg:drod} \& Fig.~\ref{fig:un_discr}}): The function \textbf{ForwardSingle()} takes input $y$ and passes it through 
four SepConv1D blocks with 
LeakyReLU activations with slope $0.1$. This produces a final temporal representation $h$ together with intermediate feature maps $\{F^{(\ell)}\}_{\ell=1}^{4}$ for feature matching. We apply a $1\times1$ Conv1D to obtain per-segment ($\tau$) critic scores, $s(\tau)$, thereby retaining localized temporal evidence instead of collapsing it immediately into a scalar (\textbf{Row \textcircled{\scriptsize 2}}--\textbf{\textcircled{\scriptsize 3}}). \textbf{Row \textcircled{\scriptsize 4}} returns these segment scores together with the intermediate feature maps and the score trace itself for feature matching and diagnostics. \textbf{Row \textcircled{\scriptsize 5}}--\textbf{\textcircled{\scriptsize 6}} apply the same \textbf{ForwardSingle()} to $y_{\mathrm{real}}$ and $y_{\mathrm{fake}}$. \textbf{Row \textcircled{\scriptsize 7}}--\textbf{\textcircled{\scriptsize 8}} computes the per-segment discriminator hinge loss $z_D(\tau)$ and generator loss $z_G(\tau)$ at each time step. \textbf{Row \textcircled{\scriptsize 9}} applies the entropic-risk aggregation across time and calculates entropic-risk function
$\rho_\lambda(z)$. \textit{As $\lambda$ increases, the objective smoothly shifts from average-case behavior toward tail-sensitive adversarial supervision.} 
\textbf{Row \textcircled{\scriptsize 10}} forms the batch-mean discriminator and generator losses $\mathcal{L}_{D}^{\mathrm{DRO}}$ and $\mathcal{L}_{G}^{\mathrm{DRO}}$ from $\rho_\lambda(z)$ and \{$z_D$, $z_G$\}. Thus, during discriminator optimization, DROD emphasizes temporal regions with large hinge violations, while during generator optimization it pushes the generator to repair its hardest local failures rather than merely improving average realism. Overall, DROD gives more weight to difficult temporal segments without unstable hard-max pooling. 

\vspace{-0.0em}
\subsection{Loss Functions and Training Objectives}
\label{subsec:losses}
\vspace{-0.0em}


Given paired narrowband/wideband waveforms $(y,x)$, the generator predicts $\hat{x}=G(y)$. The total generator loss is
$
\mathcal{L}_{G}={}
\mathcal{L}_{\mathrm{mag}}
+\mathcal{L}_{\mathrm{pha}}
+\mathcal{L}_{\mathrm{com}}
+\mathcal{L}_{\mathrm{stft}} 
+\gamma_{\mathrm{adv}}
\Big(
\mathcal{L}_{\mathrm{FM}}
+\mathcal{L}_{\mathrm{adv}}
+w_{g}^{\mathrm{EDD}}\mathcal{L}_{\mathrm{EDD}}^{G}
+w_{g}^{\mathrm{DRO}}\mathcal{L}_{\mathrm{DRO}}^{G}
\Big) 
+\mathcal{L}_{\mathrm{CC}}
\;+\,\mathcal{L}_{\mathrm{UW}},
$
where $\mathcal{L}_{\mathrm{mag}}$, $\mathcal{L}_{\mathrm{pha}}$, $\mathcal{L}_{\mathrm{com}}$, and $\mathcal{L}_{\mathrm{stft}}$ are reconstruction losses (see Appendix \ref{appen:loss_recon}); $\mathcal{L}_{\mathrm{FM}}$ and $\mathcal{L}_{\mathrm{adv}}$ are the feature-matching and hinge adversarial terms; $\mathcal{L}_{\mathrm{EDD}}^{G}$ and $\mathcal{L}_{\mathrm{DRO}}^{G}$ are the generator-side EDD and DROD losses (see Sections \ref{subsec:disc_edd} \& \ref{subsec:disc_drod}); $\mathcal{L}_{\mathrm{CC}}$ is the CCD augmented-Lagrangian penalty (see Sections \ref{subsec:disc_cc}); and $\mathcal{L}_{\mathrm{UW}}$ is an optional uncertainty-weighted waveform $L_1$ term. Here, $\gamma_{\mathrm{adv}}\in[0,1]$ is the adversarial warmup factor, and $w_{g}^{\mathrm{EDD}}, w_{g}^{\mathrm{DRO}}$ weight the EDD and DROD generator terms (see Table~\ref{tab:lossfunction_uds_swinbwe} and Appendix \ref{appen:loss_recon} for details on losses).

The total discriminator loss is $\mathcal{L}_{D} $=$ \sum_{d\in\mathcal{D}_{\mathrm{hinge}}}\mathcal{L}_{D}^{(d)}$ + $w_{d}^{\mathrm{EDD}}\mathcal{L}_{\mathrm{EDD}}^{D}$ + $w_{d}^{\mathrm{DRO}}\mathcal{L}_{\mathrm{DRO}}^{D}$, where $\mathcal{D}_{\mathrm{hinge}}$ = \{{MRAD}, {MRPD}, {CVaRD}, {CCD}, {MCUD}\} and $w_{d}^{\mathrm{EDD}}, w_{d}^{\mathrm{DRO}}$ weight the discriminator-side EDD \& DROD losses (see Sections \ref{subsec:disc_edd} \& \ref{subsec:disc_drod}). The MRAD and MRPD losses are adopted from \cite{lu2024towards}.


\vspace{-0.2em}
\begin{table*}[ht]
\centering
\scriptsize
\setlength{\tabcolsep}{0.85pt}
\renewcommand{\arraystretch}{0.99}

\begin{tabular}{l l l
  ccc  ccc  ccc  ccc  ccc  ccc ccc}
\toprule
\multirow{2}{*}{Method}
& \multirow{2}{*}{Size}
& \multirow{2}{*}{Data}
  & \multicolumn{3}{c}{NISQA-MOS$\uparrow$}
  & \multicolumn{3}{c}{STOI$\uparrow$}
  & \multicolumn{3}{c}{PESQ$\uparrow$}
  & \multicolumn{3}{c}{SI-SDR$\uparrow$}
  & \multicolumn{3}{c}{SI-SNR$\uparrow$}
  & \multicolumn{3}{c}{LSD$\downarrow$} 
  & \multicolumn{3}{c}{WER \%$\downarrow$}\\
  \cmidrule(lr){4-6}\cmidrule(lr){7-9}\cmidrule(lr){10-12}
\cmidrule(lr){13-15}\cmidrule(lr){16-18}\cmidrule(lr){19-21}\cmidrule(lr){22-24}
  & &
  & 4–16 & 8–16 & 16–48
  & 4–16 & 8–16 & 16–48
  & 4–16 & 8–16 & 16–48
  & 4–16 & 8–16 & 16–48
  & 4–16 & 8–16 & 16–48
  & 4–16 & 8–16 & 16–48 
  & 4–16 & 8–16 & 16–48\\
\midrule
\multirow{2}{*} {Unprocessed}   & \multirow{2}{*} {-}     & VCTK & 2.79 & 3.67 & 4.43 & 0.55 & 0.61 & 0.61 & 1.15  & 1.51  & 1.41  & -11.03 & -8.07 & -6.07 & -10.53 & -7.62 & -5.63 & 3.27 & 2.27 & 2.85  & 90.0 & 6.1 & 2.1\\

&   & MLS  & 2.29 & 3.12 & - & 0.51 & 0.57 & - & 1.05  & 1.34  & -  & -14.15 & -18.89 & - & -13.67 & -17.71 & - & 3.48 & 2.48 & - & 91.1 & 6.8 & -\\

\hline

\multirow{2}{*} {EBEN, 2023}   & \multirow{2}{*} {29.7M} & VCTK & 2.59 & 2.69 & 2.53     & 0.89 & 0.98 & 0.98      & 2.64  & 3.69  & 3.71      & 11.94  & 19.94 & 20.82      & 11.94  & 19.94 & 20.83      & 1.03  & 0.78  & 0.92   & 19.1 & 12.4 & 8.5 \\
   &   & MLS & 2.14 & 2.28 & -     & 0.87 & 0.96 & -      & 2.47  & 3.55  & -      & 11.81  & 18.24 & -      & 11.74  & 18.38 & -      &  1.17 &  0.88 & -   & 19.4 & 12.7 & - \\

\hline

\multirow{2}{*} {AERO, 2023}  & \multirow{2}{*} {36.4M} & VCTK & 2.79 & 2.75 & 2.88 & 0.83 & 0.94 & 0.99  & 2.62  & 3.65  & 3.69  & \textbf{13.60}  & \textbf{20.70} & \textbf{21.56} & \textbf{13.60}  & \textbf{20.70} & \textbf{21.56} & 1.09  & 0.97  & 0.75 & 20.3  & 13.4 & 9.8\\
&   & MLS  & 2.38 & 2.37 & - & 0.81 & 0.92 & -  & 2.49  & 3.45  & -  &  13.47 & \textbf{19.41} & - & \textbf{13.37}  & \textbf{19.19} & - & 1.21  & 1.04  & - & 20.7  & 13.7 & -\\

\hline

\multirow{2}{*} {AP-BWE, 2024} & \multirow{2}{*} {72M}  & VCTK & 3.86 & 3.97 & 4.49 & 0.94 & 0.99 & 0.99  & 2.55  & \textbf{3.69}  & 3.72  & 13.42  & 18.26 & 20.86 & 13.35  & 18.07 & 20.74 & 0.96  & 0.74  & 0.75 & 13.7 & 4.2  & 1.7\\ 
&   & MLS & 3.18& 3.42 & - & 0.92&0.99  & -  & 2.51  & 3.62  & -  & \textbf{13.57}  & 17.39 & - & 13.17  & 16.33 & - & 1.11  &  0.84 & - & 13.9 & 4.3 & -\\  

\hline



\multirow{2}{*} {UDSS-BWE}         & \multirow{2}{*} {18.5M} & VCTK & \textbf{4.35} & \textbf{4.37} & \textbf{4.52} & \textbf{0.94} & \textbf{0.99} & \textbf{0.99}  & \textbf{2.56}  & {3.68}  & \textbf{4.46}  & {13.19}  & {18.15} & {19.87} & {13.13}  & {17.92} & {19.32} & \textbf{0.96}  & \textbf{0.74}  & \textbf{0.75} & \textbf{13.6} & \textbf{4.2} & \textbf{1.61}\\

(our) &   & MLS & \textbf{3.3} & \textbf{3.65} & - & \textbf{0.92} & \textbf{0.99} & -  & \textbf{2.51}  & \textbf{3.63} & -  & {13.17}  & {18.11} & - & {12.91}  & {16.98} & - & \textbf{1.11}  & \textbf{0.83}  & - & \textbf{13.8} & \textbf{4.2} & \textbf{-}\\

\bottomrule
\end{tabular}
\vspace{-0.71em}
\caption{Comparison for clean English-VCTK and MLS French dataset (MLS does not have 48 kHz speech).}
\label{tab:Comparative_analysis}
\vspace{-01.30em}
\end{table*}

\vspace{-0.2em}
\section{Comprehensive Analysis}
\label{sec:Comprehensive_Result_Analysis}
\vspace{-0.2em}
\subsection{Metrics, Hyperparameters, and Datasets}
\label{subsec:EvaluationMetrics}
\vspace{-0.0em}

We use  Log-Spectral Distance (LSD),  Short-Time Objective Intelligibility (STOI), Perceptual Evaluation of Speech Quality (PESQ), Scale-Invariant SDR (SI-SDR), Scale-Invariant SNR (SI-SNR), Non-Intrusive Speech Quality Assessment (NISQA-MOS), and Word Error Rate (WER) for comprehensive evaluation (see Appendix \ref{subsec:EvaluationMetricsAppendix}). 


We train for 50 epochs using AdamW. The generator is set with a learning rate of $2\times10^{-4}$, and the discriminator learning rate is $3.5\times10^{-4}$. 
Both optimizers use $\beta_1=0.8$, $\beta_2=0.99$, and a weight-decay coefficient 0.01. An exponential learning-rate scheduler with a decay factor 0.999/epoch is used. The effective batch size is 16. 
A random seed is 1234. Training is performed on an NVIDIA RTX 4090 GPU and an AMD Ryzen 7950X3D CPU. 



We use the English VCTK (v0.92) ~\cite{yamagishi2019cstr} (110 speakers) and MLS French corpus \cite{Pratap2020MLSAL} (114 speakers), to test in a multilingual and cross-speaker setup. There are in total $\sim$1045 hours of speech in the two datasets with sampling rates of 16 and 48 kHz. We load 16/48 kHz files, convert to mono channel, remove silent parts, downsample to simulate band-limited audio, sinc interpolate, and align length-wise. 
We refer to Appendix \ref{sec:data_pipeline} and \ref{subsec:hyperparameters} for details.

\vspace{-0.0em}
\subsection{Ablation Study} 
\label{subsec:Ablation_Study}
\vspace{-0.0em}

Table~\ref{table:Ablation_Study} reports the discriminator-only ablations. 

\noindent\textbf{a) Row\textcircled{\scriptsize 1}--\textcircled{\scriptsize 5}:}
Each discriminator alone induces different results.
\textbf{Row\textcircled{\scriptsize 1} (CVaRD)} gives the best perceptual quality among single critics (N-MOS $=3.07$). 
\textbf{Row\textcircled{\scriptsize 2} (CCD)} achieves the best PESQ/SNR but not the best N-MOS, indicating that HF-energy control alone does not ensure naturalness.
\textbf{Row\textcircled{\scriptsize 3} (MCUD)} preserves PESQ ($2.62$) but sharply worsens LSD ($1.99$), suggesting utility-based HF shaping can drift without global alignment.
Among the uncertainty-aware critics, \textbf{Row\textcircled{\scriptsize 4} (EDD)} is the weakest, hurting intelligibility and quality, while \textbf{Row\textcircled{\scriptsize 5} (DROD)} is objectively strong but yields  moderate perceptual gains.

\begin{table}[t]
\centering
{\scriptsize

\setlength{\tabcolsep}{1.2pt}
\renewcommand{\arraystretch}{0.85}
\begin{tabular}{l c c c c c c c |c| c |c| c |c}
\toprule
SL & A & P & CVaR & CC & MCU & ED & DRO & L$\downarrow$ & S$\uparrow$ & P$\uparrow$ & SN$\uparrow$ & N$\uparrow$ \\
\midrule
\multicolumn{13}{c}{Single discriminators (no MRAD/MRPD)} \\[-1pt]
\midrule
1  & \xmark & \xmark & \cmark & \xmark & \xmark & \xmark & \xmark & 1.11 & 0.95 & 2.53 & 14.31 & 3.07 \\
2  & \xmark & \xmark & \xmark & \cmark & \xmark & \xmark & \xmark & 1.14 & 0.95 & 2.65 & 14.46 & 3.05 \\
3  & \xmark & \xmark & \xmark & \xmark & \cmark & \xmark & \xmark & 1.99 & 0.95 & 2.62 & 14.16 & 3.02 \\
4  & \xmark & \xmark & \xmark & \xmark & \xmark & \cmark & \xmark & 1.20 & 0.92 & 1.97 & 10.22 & 2.98 \\
5  & \xmark & \xmark & \xmark & \xmark & \xmark & \xmark & \cmark & 1.13 & 0.95 & 2.54 & 14.42 & 3.03 \\
\midrule
\multicolumn{13}{c}{With global feedback (MRAD+MRPD)} \\[-1pt]
\midrule
\rowcolor{red!30}
6 & \cmark & \cmark & \xmark & \xmark & \xmark & \xmark & \xmark & 1.02 & 0.94 & 2.32 & 13.31 & 4.22 \\
7  & \cmark & \cmark & \cmark & \xmark & \xmark & \xmark & \xmark & 0.98 & 0.94 & 2.35 & 13.23 & 4.21 \\
8  & \cmark & \cmark & \xmark & \xmark & \xmark & \cmark & \xmark & 0.98 & 0.94 & 2.28 & 12.64 & 4.31 \\
9  & \cmark & \cmark & \xmark & \xmark & \xmark & \xmark & \cmark & 0.98 & 0.94 & 2.30 & 12.69 & 4.23 \\
10  & \cmark & \cmark & \xmark & \xmark & \xmark & \cmark & \cmark & 0.99 & 0.94 & 2.22 & 12.38 & 4.29 \\
11 & \cmark & \cmark & \cmark & \cmark & \cmark & \xmark & \xmark & 0.99 & 0.94 & 2.29 & 13.10 & 4.32 \\
\midrule
\multicolumn{13}{c}{AP-BWE's Generator vs. our Swin-based Generator} \\[-1pt]
\midrule
\rowcolor{gray!30}
12 & \cmark & \cmark & \cmark & \cmark & \cmark & \cmark & \cmark & 1.44 & 0.88 & 1.38 & 2.56 & 2.01 \\
\rowcolor{green!25}
13 & \cmark & \cmark & \cmark & \cmark & \cmark & \cmark & \cmark & 1.00 & 0.94 & 2.23 & 12.75 & 4.35 \\
\midrule
\multicolumn{13}{c}{Parameter count (per discriminator)} \\[-1pt]
\midrule
14 & 0.6M & 0.6M & 147.6k & 7.6k & 7.4k & 15.8k & 15.7k &  &  &  &  &  \\
\bottomrule
\end{tabular}
}
\vspace{-0.6em}
\caption{Ablation on discriminators for 4$\rightarrow$16~kHz BWE.
Acronyms: A=MRAD, P=MRPD, L=LSD, S=STOI, P=PESQ, SN=SI-SNR, N= NISQA-MOS.}
\label{table:Ablation_Study}
\vspace{-0.20em}
\end{table}

\noindent\textbf{b) Row\textcircled{\scriptsize 6}--\textcircled{\scriptsize 11}:}
Adding MRAD \& MRPD markedly improves results: N-MOS rises from $\sim$ 3.0 to  4.21--4.32, LSD falls to 0.98--0.99, and STOI stabilizes at 0.94. \textbf{Row\textcircled{\scriptsize 6}--\textcircled{\scriptsize 7}} show that, once global structure is recovered, tail-risk guidance by CVaRD can focus on sparse HF defects.

\textbf{Row\textcircled{\scriptsize 7}} gives the best N-MOS among the uncertainty variants, EDD and DROD, despite lower SNR, suggesting a preference for perceptual plausibility over strict waveform matching.
\textbf{Row\textcircled{\scriptsize 8}} remains competitive, whereas \textbf{Row\textcircled{\scriptsize 9}} offers little added benefit, implying overlap between EDD and DROD.
\textbf{Row\textcircled{\scriptsize 10}} has the best perceptual score (N-MOS $=4.32$) with competitive fidelity, supporting \textit{global structure first (MRAD/MRPD), then local HF in-painting (CVaRD/CCD/MCUD)}.

\noindent\textbf{c) Row \textcircled{\scriptsize 12} vs. \textcircled{\scriptsize 13}:}
Under the full discriminator suite, the AP-BWE's generator (our baseline) in \textbf{Row\textcircled{\scriptsize 11}} collapses, whereas our Swin-based generator in \textbf{Row\textcircled{\scriptsize 12}} remains high quality, indicating our Swin-based generator is better matched to multi-critic (decision-science + uncertainty) feedback.

\noindent\textbf{d) Row\textcircled{\scriptsize 14} (parameter cost):}
MRAD and MRPD are parameter-heavy (0.6 M each; total 1.2 M). \ul{Whereas, our proposed discriminators are lightweight, totaling only $\sim$194.1k parameters} (see Table~\ref{tab:riskcvar_breakdown}, \ref{tab:chance_utility_breakdown}, \ref{tab:evidential_dro_breakdown}, \ref{tab:mrad_breakdown}, \ref{tab:mrpd_breakdown}), \ul{indicating they add only minimum overhead with improved performance.}

\begin{table}[t]
\centering
\scriptsize
\setlength{\tabcolsep}{1.2pt}
\renewcommand{\arraystretch}{0.95}
\begin{tabular}{ll ccc ccc ccc}
\toprule
\multirow{2}{*}{Method} & \multirow{2}{*}{Data}
& \multicolumn{3}{c}{NISQA-MOS$\uparrow$}
& \multicolumn{3}{c}{SI-SNR$\uparrow$}
& \multicolumn{3}{c}{LSD$\downarrow$} \\
\cmidrule(lr){3-5}\cmidrule(lr){6-8}\cmidrule(lr){9-11}
& & 4--16 & 8--16 & 16--48
  & 4--16 & 8--16 & 16--48
  & 4--16 & 8--16 & 16--48 \\
\midrule
\multirow{1}{*}{EBEN}
& VCTK & 1.01 & 1.08 & 1.15 & 4.23 & 5.31 & 6.01 & 1.41 & 1.11 & 1.00 \\
\multirow{1}{*}{AERO}
& VCTK & 1.52 & 1.12 & 1.19 & \textbf{4.39} & 5.38 & 6.15 & 1.43 & 1.13 & 1.01 \\
\multirow{1}{*}{AP-BWE}
& VCTK & 2.74 & 3.15 & 3.71 & 4.36 & 5.31 & 6.11 & 1.37 & 1.07 & 0.89 \\
\multirow{1}{*}{UDSS-BWE}
& VCTK & \textbf{3.91} & \textbf{3.87} & \textbf{3.94}
       & {4.01} & \textbf{5.65} & \textbf{7.01}
       & \textbf{1.22} & \textbf{1.02} & \textbf{0.89} \\
\bottomrule
\end{tabular}
\vspace{-01.2em}
\caption{Study on the noisy VCTK dataset}
\label{tab:noiseComparative_analysis}
\vspace{-02.0em}
\end{table}

\subsection{Comparative Analysis with Baselines} 
\label{subsec:ComparativeAnalysiswithBaselines}
\vspace{-0.2em}

Table~\ref{tab:Comparative_analysis} compares UDSS-BWE against three baselines---EBEN \cite{hauret2023eben}, AERO \cite{mandel2023aero}, and AP-BWE \cite{lu2024towards}---across three extension ranges on the English-VCTK and MLS French datasets. EBEN is a pseudo-QMF-based model, AERO is a complex-valued model, and AP-BWE is a dual-stream amplitude-phase prediction model.

Compared to unprocessed speech on VCTK for 4$\rightarrow$16~kHz, UDSS-BWE achieves a 3.41$\times$ reduction in LSD, a 1.72$\times$ increase in STOI, a 2.23$\times$ increase in PESQ, and a 1.56$\times$ increase in NISQA-MOS. Moreover, UDSS-BWE provides the highest NISQA-MOS, outperforming all baselines. STOI is also higher for UDSS-BWE than EBEN/AERO and on par with AP-BWE. In terms of PESQ, UDSS-BWE performs far better than all the other baselines for 16-48 kHz range. In terms of LSD, UDSS-BWE outperforms EBEN and AERO, but has lower SI-SDR/SI-SNR than AERO. Since LSD reflects over-smoothing and NISQA-MOS, PESQ, and STOI capture perceptual quality, these results support that UDSS-BWE improves perceptual reconstruction more effectively \textit{while using 3.89$\times$ fewer parameters compared to the overall strongest baseline AP-BWE (72M vs.\ 18.5M).} The same trend also holds for WER and the MLS French dataset, indicating that the gains generalize to multilingual settings.


\begin{figure}[t]
  \centering
 \includegraphics[width=0.48\textwidth,height=0.13\textheight]{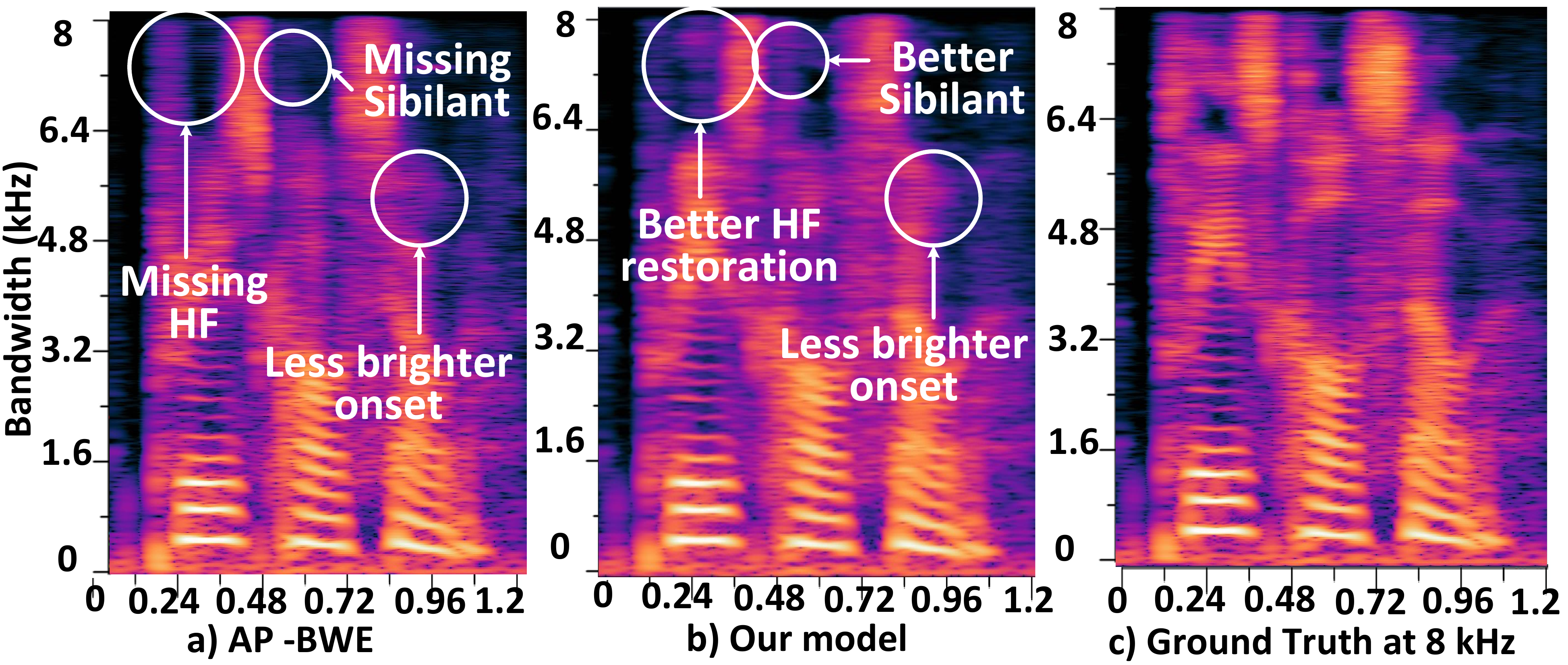} 
 \vspace{-1.9em}
  \caption{4-16 kHz extended speech by UDSS-BWE.}
  \label{fig:figcomparison}
  \vspace{-0.5em}
\end{figure}

\noindent\textbf{Qualitative observation:}
Fig.~\ref{fig:figcomparison} further shows that UDSS-BWE reconstructs subtle HF contents---such as missing sibilants and brighter onsets---more precisely than AP-BWE, and is almost similar to ground-truth. Hence, the combined effect of specialized decision-science-based discriminators are more effective than general purpose parameter heavy discriminators. {\color{black}We also refer to Table~\ref{tab:SoTA_different_fr_range} for a comparison across different frequency ranges.}

\vspace{-0.0em}
\subsection{Evaluation under Noisy Conditions}
\label{subsec:noisy_conditions}
\vspace{-0.0em}

We use 8 noise sources from the AURORA dataset \cite{pearce00_icslp} and five SNR levels
($-10$, $-5$, $0$, $5$, $10$ dB). 
The selected noises cover both stationary and non-stationary conditions, including airport, babble, car, exhibition, station, street, speech-shaped noise (SSN), and white Gaussian noise (AWGN). 
Table~\ref{tab:noiseComparative_analysis} reports the \textit{average performance} on all eight noise sources and all five SNR levels. 
\textbf{\textit{UDSS-BWE consistently outperforms the strongest baseline, AP-BWE, on the noisy benchmark.}} 
See Appendix~\ref{sec:appendix_noisy_analysis} for details.

\subsection{Robustness to Channel Degradation}
\label{subsec:telephony_codec_summary}
\vspace{-0.1em}

We evaluate UDSS-BWE under five different channel degradation settings:
R0 uses sinc resampling only; R1 adds PCM; R2 adds G.711; R3 uses a broader codec mixture; and R4 further adds telephone filtering and packet loss. Table~\ref{tab:telephony_summary_main} reports the results across R0--R4. 
Detailed analysis are provided in Appendix~\ref{subsec:telephony_codec_results}.

\begin{table}[t]
\centering
\scriptsize
\setlength{\tabcolsep}{2.6pt}
\renewcommand{\arraystretch}{0.92}
\begin{tabular}{lccc}
\toprule
Condition & LSD$\downarrow$ & SI-SNR$\uparrow$ & N-MOS$\uparrow$ \\
\midrule
Ideal/PCM-24 & 0.95 (R0) & 12.98 (R0) & 4.42 (R3) \\
PCM-8        & 1.00 (R1) & 12.37 (R0) & 3.76 (R1) \\
G.711 $\mu$-law & 0.95 (R0) & 12.93 (R0) & 4.40 (R3) \\
G.711 A-law  & 0.95 (R0) & 12.95 (R0) & 4.38 (R3) \\
Loss 3/5\%   & 1.10/1.10 (R2) & 7.79/6.93 (R4) & 3.53/3.30 (R1) \\
\bottomrule
\end{tabular}
\vspace{-0.8em}
\caption{Best channel-robustness results across R0--R4.} 
\label{tab:telephony_summary_main}
\vspace{-0.0em}
\end{table}

\vspace{-0.1em}
\subsection{Computational Complexity}
\label{subsec:ncomp_comp}
\vspace{-0.2em}

Table \ref{tab:computation} summarizes computational complexity in terms of MACs, FLOPs, RTF, and inference time, while Table \ref{tab:feature_extractor_complexity_main} reports feature-wise complexity for the proposed discriminators (see Section \ref{sec:limitation}). Due to optimized generator and discriminator (see Sections \ref{subsec:Ablation_Study}, \ref{subsec:Parameter Breakdown for Discriminators}, and \ref{subsec:Parameter Breakdown for Generators}), \textbf{\textit{UDSS-BWE uses 3.89$\times$ fewer parameters and 2.66$\times$ fewer MACs and FLOPs than AP-BWE}}, reducing memory usage while still achieving better perceptual quality, as reflected by the NISQA-MOS results in Table~\ref{tab:Comparative_analysis}. UDSS-BWE also shows  comparable RTF and inference time, supporting its suitability for real-time applications (see Appendix \ref{append:analysisoncomputationcomplexity} for details).

\begin{table}[t]
\centering
\scriptsize
\setlength{\tabcolsep}{1.5pt}
\begin{tabular}{l|c|c|c|c|c|c}
\toprule
Model & Freq. & Par.(M) & MAC(M) & FLOP(M) & RTF(GPU) & Inf.(ms) \\
\hline
AP-BWE & 16-48 kHz & 72.07& 14236.65 & 28473.31 &  0.0025x & 16.60 \\
UDSS-BWE & 16-48 kHz & \textbf{18.5} & \textbf{5334.72} & \textbf{10669.45} & \textbf{0.0028x} & \textbf{16.94}\\
\bottomrule
\end{tabular}
\vspace{-0.94em}
\caption{Overall computational complexity.} 
\label{tab:computation}
\vspace{-0.5em}
\end{table}

\subsection{Subjective Mean Opinion Score (MOS)}
\label{subsec:Subjective Test}
\vspace{-0.0em}

We use 5-point (1=bad to 5=excellent) MOS ratings by a selected panel of 10 persons. The unprocessed audio has MOS = 2.805, meaning a severe loss of recognizability. \textit{Our UDSS-BWE (MOS = 4.25) outperforms AP-BWE (MOS = 4.10).}  
Results provide strong evidence that UDSS-BWE consistently generates perceptually higher quality audio, favored by a wide range of listeners.  We refer to Appendix \ref{subsec:subj_eval_details} for details on MOS analysis.


\begin{table}[t]
\centering
\scriptsize
\setlength{\tabcolsep}{2.5pt}
\begin{tabular}{clcccccc}
\toprule
Row & Band &
\multicolumn{2}{c}{$D_{\mathrm{SF},i} \downarrow$} &
\multicolumn{2}{c}{$D_{\sigma L,i} \downarrow$} &
\multicolumn{2}{c}{$D_{\mathrm{miss},i} \downarrow$} \\
\cmidrule(lr){3-4}\cmidrule(lr){5-6}\cmidrule(lr){7-8}
& & AP & UDSS
& AP & UDSS
& AP & UDSS \\
\midrule
\textcircled{\scriptsize 1} & 0--4
& \textbf{$1.04{\times}10^{-4}$} & $1.23{\times}10^{-4}$
& 0.0521 & \textbf{0.0514}
& \textbf{0.227} & 0.302 \\
\textcircled{\scriptsize 2} & 4--8 
& \textbf{$1.57{\times}10^{-4}$} & $1.93{\times}10^{-4}$
& 0.375 & \textbf{0.171}
& 0.704 & \textbf{0.561} \\
\textcircled{\scriptsize 3} & 8--12 
& 0.00114 & \textbf{0.000906}
& 0.831 & \textbf{0.774}
& \textbf{3.979} & 4.233 \\
\textcircled{\scriptsize 4} & 12--16 
& 0.00124 & \textbf{0.000866}
& \textbf{0.689} & 1.018
& 3.214 & \textbf{2.961} \\
\textcircled{\scriptsize 5} & 16--20 
& 0.00113 & \textbf{0.000795}
& \textbf{0.775} & 0.793
& 3.616 & \textbf{2.560} \\
\textcircled{\scriptsize 6} & 20--24
& 0.00133 & \textbf{0.000817}
& 0.942 & \textbf{0.711}
& 3.342 & \textbf{2.979} \\
\bottomrule
\end{tabular}
\vspace{-1em}
\caption{Sub-band analysis by ITU-inspired metrics.}
\vspace{-0.2em}
\label{tab:ap_udss_v18_selected_itu_metrics_}
\end{table}

\subsection{HF Analysis by ITU-Inspired Metrics}
\vspace{-0.0em}
We use three ITU (International Telecommunication Union)-derived diagnostic metrics to analyze HF reconstruction accuracy : \textbf{1.} the spectral-flux error $\mathrm{D}_{SF,i}$ derived from ITU-R BS.1387 \cite{bs1387_2023}; \textbf{2.} the K-weighted loudness error $D_{\sigma L,i}$ derived from ITU-BS1770 \cite{itu_bs1770_2023}; and \textbf{3.} the missing disturbance error $D_{miss,i}$ derived from P.862/PESQ \cite{itu_p862_2001}. They capture complementary HF reconstruction cues: frame-to-frame spectral continuity, short-term loudness stability, and missing perceptual components.  
Table \ref{tab:ap_udss_v18_selected_itu_metrics_} shows that UDSS-BWE achieves lower error in 9 of 12 comparisons within the generated HF region above 8 kHz. These results indicate that UDSS-BWE recovers more coherent HF structure than AP-BWE, especially in the HF bands where BWE is most challenging. 
\textbf{See Appendix \ref{subsec:itu_inspired_udss_ap_evidence} 
for details.}

\begin{table}[t]
\centering
\resizebox{\linewidth}{!}{
\setlength{\tabcolsep}{4pt}
\renewcommand{\arraystretch}{1}
\begin{tabular}{llccccc}
\hline
Agg. & Method & $N$ & Mean $\uparrow$ & SD & SEM & 95\% CI $\uparrow$\\
\hline
Utt. & AP-BWE   & 1000 & 4.458 & 0.472 & 0.015 & [4.429, 4.488] \\
Utt. & UDSS-BWE & 1000 & 4.497 & 0.486 & 0.015 & [4.466, 4.527] \\
Spk. & AP-BWE   & 7    & 4.448 & 0.196 & 0.074 & [4.267, 4.630] \\
Spk. & UDSS-BWE & 7    & 4.487 & 0.196 & 0.074 & [4.305, 4.668] \\
\hline
\end{tabular}
}
\vspace{-0.5em}
\caption{MOS descriptive statistics.} 
\vspace{-0.0em}
\label{tab:nisqa_mos_stats_udss_1000_}
\end{table}

\begin{table}[ht!]
\centering
\resizebox{\linewidth}{!}{
\setlength{\tabcolsep}{4pt}
\renewcommand{\arraystretch}{1}
\begin{tabular}{lccccccc}
\hline
Agg. & $N$ & $\overline{d}$ & Median$(d)$ & $W^+$ & $W^-$ & $T$ & $p$ \\
\hline
Utt. & 1000 & 0.038 & 0.038 & 363749 & 136751 & 136751 & $1.94{\times}10^{-35}$ \\
Spk. & 7    & 0.038 & 0.039 & 27     & 1      & 1      & 0.0313 \\
\hline
\end{tabular}
}
\vspace{-0.5em}
\caption{Paired Wilcoxon signed-rank test.}
\label{tab:wilcoxon_nisqa_udss_1000_}
\vspace{-0.85em}
\end{table}

\subsection{Statistical Analysis of MOS Scores}
Tables \ref{tab:nisqa_mos_stats_udss_1000_} and \ref{tab:wilcoxon_nisqa_udss_1000_} show statistical analysis using mean, standard deviation (SD), standard error of the
mean (SEM), 95\% confidence interval (CI), positive ($W^+$) and negative ($W^-$) rank sums, and paired Wilcoxon signed-rank test over 1000 utterances. UDSS-BWE shows statistically significant Wilcoxon paired difference $p$ over AP-BWE in both utterance (Utt.) and speaker (Spk.) levels. A detailed analysis is provided in Appendix \ref{subsec:nisqa_stats_udss_ap}.


\vspace{-0.0em}
\section{Conclusion}
\vspace{-0.0em}

We introduce \textbf{UDSS-BWE}, an uncertainty- and decision-science-guided framework for speech BWE that combines a parameter-efficient dual-stream generator with specialized discriminator sets to focus on perceptually critical high-frequency failures. Extensive experiments show that UDSS-BWE consistently improves wideband speech reconstruction across objective, perceptual, and downstream evaluation metrics. We have demonstrated that decision science and uncertainty theories can be used as effective inductive biases for designing novel and parameter efficient BWE frameworks for ASR and TTS tasks. \textbf{\textit{We hope this perspective motivates further interdisciplinary research on uncertainty-aware, risk-sensitive, and resource-efficient architectures in NLP.}}

\vspace{-0.0em}
\section{Limitations}
\label{sec:limitation}

A limitation of UDSS-BWE is its slight increase in  training  time relative to AP-BWE (17 minutes/epoch vs. 20 minutes/epoch). The slightly increase in training cost mainly comes from the additional discriminator-side supervision used during training. In particular, UDSS-BWE augments the standard adversarial setup with decision-science and uncertainty-aware critics, and the main overhead is introduced by the feature extraction stages inside these branches. 

\begin{table}[ht]
\centering
\scriptsize
\small
\setlength{\tabcolsep}{1pt}
\begin{tabular}{lccc}
\toprule
Feature Extractor (FE) & MACs (M) & FLOPs (M) & Latency (ms) \\
\midrule
CVaR\_FE              & 49.696 & 99.392 & 0.659 \\
CC\_FE    & 0.141  & 0.282  & 0.823 \\
MCU\_FE  & 0.139  & 0.278  & 0.889 \\
ED\_FE   & 14.376 & 28.752 & 0.441 \\
DRO\_FE           & 14.248 & 28.496 & 0.348 \\
\bottomrule
\end{tabular}
\vspace{-0.0em}
\caption{Discriminator-wise computational complexity.}
\label{tab:feature_extractor_complexity_main}
\vspace{-0.50em}
\end{table}

As shown in Table \ref{tab:feature_extractor_complexity_main}, the most expensive decision-side components are the Chance-Constraint feature extractor (CC\_FE) and the Multi-Criteria Utility feature extractor (MCU\_FE), since both require explicit time--frequency analysis and additional nonlinear processing before discrimination. The CVaR feature extractor (CVaR\_FE) also contributes noticeable overhead due to its tail-focused aggregation. By contrast, the uncertainty critics, namely the Evidential Dirichlet feature extractor (ED\_FE) and the Entropic KL-DRO feature extractor (DRO\_FE), are comparatively lighter, but they still add extra forward and backward computations during training. Therefore, UDSS-BWE shifts computational burden to training to obtain tail-aware, constraint-aware, and uncertainty-calibrated supervision.


\vspace{-0.0em}
\section{Potential Risks / Ethical Considerations}

Although UDSS-BWE is developed for speech restoration research, it could be misused in harmful ways, including covert enhancement of recorded speech, speaker impersonation, or downstream deepfake-style audio manipulation. Such misuse may introduce privacy, consent, and security concerns. For this reason, any practical use of the model should be accompanied by clear disclosure, informed consent, and appropriate safeguards to reduce the risk of abuse.

\vspace{-0.0em}
\section*{Acknowledgment}

We sincerely thank the VCTK and MLS data collection teams for developing and releasing the VCTK and MLS corpora. 
In addition, we are grateful to the participants in the subjective evaluation for their time and valuable contributions to the listening tests. Finally, we acknowledge the use of Elicit for identifying relevant literature, and ChatGPT for debugging code and assisting with grammar correction. 

This work is supported by the Office of Naval Research (ONR) with an award number 000142612163.

\vspace{-1.5em}
\bibliography{custom}

\appendix

\section{Appendix}
\label{sec:appendix}

\subsection{Algorithms for the generator}
\label{appen:alg:gen}

Please refer to Alg. \ref{alg:gen} for more details on the generator architecture.

\begin{algorithm}[ht!]
\caption{Generator: Swin-1D with Lattice Coupling (shared Swin stack per lattice block)}
\label{alg:gen}
\begin{algorithmic}[1]
\scriptsize
\STATE \textbf{Require:} log-mag $M^{\mathrm{nb}}$, phase $\Phi^{\mathrm{nb}}$; Swin window $W$; two lattice blocks
\STATE \textbf{Ensure:} log-mag $M^{\mathrm{wb}}$, phase $\Phi^{\mathrm{wb}}$

\STATE $x \gets \mathrm{LN}(\mathrm{Conv1D}(M^{\mathrm{nb}}))$ \COMMENT{pre-project mag stream}
\STATE $y \gets \mathrm{LN}(\mathrm{Conv1D}(\Phi^{\mathrm{nb}}))$ \COMMENT{pre-project phase stream}

\STATE \COMMENT{two lattice blocks; within each block, the same $\mathrm{Swin1D}_j(\cdot)$ is reused twice (shared weights)}
\FOR{$j=1$ \TO $2$}
    \STATE $u \gets \mathrm{Swin1D}_j(y; W,\{0,\lfloor W/2\rfloor\})$ \COMMENT{refine phase (windowed + shifted)}
    \STATE $m_1 \gets x + a_1 u$
    \STATE $n_1 \gets a_2 x + u$ \COMMENT{AP coupling}
    \STATE $v \gets \mathrm{Swin1D}_j(m_1; W,\{0,\lfloor W/2\rfloor\})$ \COMMENT{refine mag (same Swin weights as above)}
    \STATE $x \gets b_1 n_1 + v$
    \STATE $y \gets n_1 + b_2 v$ \COMMENT{couple back}
\ENDFOR

\STATE $M^{\mathrm{wb}} \gets M^{\mathrm{nb}} + \mathrm{Linear}(\mathrm{LN}(x))$ \COMMENT{log-mag residual prediction}
\STATE $\Phi^{\mathrm{wb}} \gets \operatorname{atan2}(\mathrm{Lin}_i(\mathrm{LN}(y)), \mathrm{Lin}_r(\mathrm{LN}(y)))$
\STATE \textbf{return} $M^{\mathrm{wb}}, \Phi^{\mathrm{wb}}$
\end{algorithmic}
\end{algorithm}

\subsection{Algorithms for the training}
\label{appen:uds_train} Please refer to Alg. \ref{alg:uds_train} for more details on the training.

\vspace{-0.35em}
\subsection{Detailed Reconstruction Losses}
\label{appen:loss_recon}
\vspace{-0.2em}

This subsection continues Section~\ref{subsec:losses} by defining the spectral reconstruction terms used in the generator objective. Let $C$ and $\hat{C}$ denote the target and generated complex STFTs \cite{11604748,11461864}, with magnitudes $M$ and $\hat{M}$ and generated phase $\hat{\phi}$. We use $\mathrm{MSE}(\cdot,\cdot)$ to denote mean squared error.

The magnitude reconstruction loss is
\begin{equation}
\label{eq:loss_mag}
\mathcal{L}_{\mathrm{mag}}
=
\lambda_{\mathrm{mag}}\,
\mathrm{MSE}(\hat{M},M),
\end{equation}
where $\lambda_{\mathrm{mag}}$ controls the contribution of the magnitude
term.

The phase reconstruction loss is
\begin{equation}
\label{eq:loss_phase}
\mathcal{L}_{\mathrm{pha}}
=
\lambda_{\mathrm{pha}}
\left(
\mathcal{L}_{\mathrm{IP}}
+
\mathcal{L}_{\mathrm{GD}}
+
\mathcal{L}_{\mathrm{IAF}}
\right),
\end{equation}
where $\mathcal{L}_{\mathrm{IP}}$, $\mathcal{L}_{\mathrm{GD}}$, and
$\mathcal{L}_{\mathrm{IAF}}$ denote the instantaneous-phase, group-delay,
and instantaneous-amplitude-frequency losses, respectively, and
$\lambda_{\mathrm{pha}}$ is the phase-loss weight.

The complex STFT reconstruction loss is
\begin{equation}
\label{eq:loss_complex}
\mathcal{L}_{\mathrm{com}}
=
\lambda_{\mathrm{com}}\,
\mathrm{MSE}(\hat{C},C),
\end{equation}
where $\lambda_{\mathrm{com}}$ weights the complex-domain consistency term.

To further encourage STFT-domain self-consistency, we reconstruct the STFT
from the generated magnitude and phase as
\begin{equation}
\label{eq:reconstructed_stft}
\tilde{C}
=
\mathrm{STFT}
\bigl(
\mathrm{ISTFT}(\hat{M},\hat{\phi})
\bigr),
\end{equation}
and define
\begin{equation}
\label{eq:loss_stft}
\mathcal{L}_{\mathrm{stft}}
=
\lambda_{\mathrm{stft}}\,
\mathrm{MSE}(\hat{C},\tilde{C}),
\end{equation}
where $\lambda_{\mathrm{stft}}$ controls the strength of this
self-consistency regularization.
Detailed definitions of other losses are provided in Table~\ref{tab:lossfunction_uds_swinbwe}.

\begin{table*}[ht]
  \centering
  \scriptsize
  \setlength{\tabcolsep}{2pt}
  \renewcommand{\arraystretch}{0.5}
  \begin{tabular}{m{2.5cm} | m{5.6cm} | m{7.55cm}}
    \toprule
    \textbf{Loss function}
    & \textbf{Equation \,(MSE = Mean Squared Error)}
    & \textbf{Terms} \\
    \midrule

    Feature Matching Loss
    & $\mathcal{L}_{\mathrm{FM}}
      =\displaystyle\sum_{d\in\mathcal{D}_{\mathrm{FM}}}\lambda_d
      \sum_{\ell}\mathbb{E}\!\left[\left\lVert f^{\mathrm{real}}_{d,\ell}-f^{\mathrm{fake}}_{d,\ell}\right\rVert_1\right]$
    & $\mathcal{D}_{\mathrm{FM}}=\{\mathrm{MRAD},\mathrm{MRPD},\mathrm{CVaRD},\mathrm{CCD},\mathrm{MCUD}, \mathrm{EDD},\mathrm{DROD}\}$ \; $f_{d,\ell}$: layer-$\ell$ feature map from $d$; $\lambda_d$: per-d FM weight. \\
    \midrule

    Generator Hinge Loss\\(base adversarial set)
    & $\mathcal{L}_{\mathrm{adv}}
      $=$\displaystyle\sum_{d\in\mathcal{D}_{\mathrm{hinge}}}\lambda_d\,
      \mathbb{E}_{\hat{y}\sim p_G}\!\left[\max\!\big(0,1-D_d(\hat{y})\big)\right]$
    & $\mathcal{D}_{\mathrm{hinge}}=\{\mathrm{MRAD},\mathrm{MRPD},\mathrm{CVaRD}, \mathrm{CCD},\mathrm{MCUD}\}$; $D_d(\cdot)$: scalar hinge critic score of discriminator $d$. \\
    \midrule

    Discriminator Hinge Loss\\(base adversarial set)
    & $\mathcal{L}_{D}^{(d)}
      $=$\mathbb{E}_{y\sim p_{\mathrm{data}}}\!\left[\max\!\big(0,1-D_d(y)\big)\right]
      +\mathbb{E}_{\hat{y}\sim p_G}\!\left[\max\!\big(0,1+D_d(\hat{y})\big)\right]$
    & For each $d\in\mathcal{D}_{\mathrm{hinge}}$: real-hinge enforces $D_d(y)\ge 1$ and fake-hinge enforces $D_d(\hat{y})\le -1$. \\
    \midrule

    EDD Discriminator Loss
    & $\mathcal{L}_{\mathrm{EDD}}^{D}
      =\mathcal{L}_{\mathrm{EDD}}(\alpha_r,\mathbf{y}_{\mathrm{real}})
      +\mathcal{L}_{\mathrm{EDD}}(\alpha_f,\mathbf{y}_{\mathrm{fake}})$
    & $\alpha\in\mathbb{R}_{+}^{2}$: Dirichlet concentration over $\{\mathrm{real},\mathrm{fake}\}$;
      $\mathbf{y}_{\mathrm{real}}=[1,0]$, $\mathbf{y}_{\mathrm{fake}}=[0,1]$; 
      $\mathcal{L}_{\mathrm{EDD}}(\alpha,\mathbf{y})
      =\sum_k y_k\!\left(\psi\!\left(\sum_j\alpha_j\right)-\psi(\alpha_k)\right)
      +\lambda_{\mathrm{kl}}\,\mathrm{KL}\!\big(\mathrm{Dir}(\alpha)\Vert \mathrm{Dir}(\mathbf{1})\big)$. \\
    \midrule

    EDD Generator Loss\\+ Uncertainty Penalty
    & $\mathcal{L}_{\mathrm{EDD}}^{G}
      =\mathcal{L}_{\mathrm{EDD}}(\alpha_f,\mathbf{y}_{\mathrm{real}})
      +\lambda_u\,\mathbb{E}\!\left[u(\alpha_f)\right]$
    & Generator pushes fake samples toward the real evidential target; 
      $u(\alpha)=\dfrac{K}{\sum_{k=1}^{K}\alpha_k+\varepsilon}$ with $K=2$;
      $\lambda_u$: uncertainty-penalty weight. \\
    \midrule

    DROD Discriminator Loss
    & $\mathcal{L}_{\mathrm{DRO}}^{D}
      =\mathrm{mean}_{B}\!\left[\rho_{\lambda}(z_D)\right],\qquad
      z_D(\tau)=\mathrm{ReLU}\!\big(1-s_r(\tau)\big)+\mathrm{ReLU}\!\big(1+s_f(\tau)\big)$
    & $s_r(\tau),s_f(\tau)$: real/fake per-segment score traces from DROD; 
      $\rho_{\lambda}(z)=
      \begin{cases}
      \mathrm{mean}_{\tau}(z), & \lambda\le 0\\
      m+\dfrac{1}{\lambda}\log\!\big(\mathrm{mean}_{\tau}(\exp(\lambda(z-m)))+\varepsilon\big), & \lambda>0
      \end{cases}$,
      with $m=\max_{\tau}z$; $\lambda$: risk parameter. \\
    \midrule

    DROD Generator Loss
    & $\mathcal{L}_{\mathrm{DRO}}^{G}
      =\mathrm{mean}_{B}\!\left[\rho_{\lambda}(z_G)\right],\qquad
      z_G(\tau)=-s_f(\tau)$
    & Entropic-risk aggregation emphasizes hard temporal segments and pushes the generator to increase fake segment scores under DROD. \\
    \midrule

    CCD Augmented-Lagrangian Loss (Generator)
    & $\mathcal{L}_{\mathrm{CC}}
      =\gamma_{\mathrm{adv}}
      \left(\lambda_{\mathrm{cc}}\,g+\dfrac{\rho_{\mathrm{cc}}}{2}\,g^{2}\right),\qquad
      g=\bar{b}-\bar{b}_{0}$
    & $\bar{b}$: batch mean of the CCD barrier trace on generated audio; $\bar{b}_{0}$: target exceedance level; $\lambda_{\mathrm{cc}}$: dual variable; $\rho_{\mathrm{cc}}$: quadratic penalty; $\gamma_{\mathrm{adv}}$: adversarial warmup factor. The dual update is
      $\lambda_{\mathrm{cc}}\leftarrow \mathrm{clip}\!\big(\lambda_{\mathrm{cc}}+\eta_{\mathrm{cc}}g,\,0,\,\lambda_{\max}\big)$. \\
    \midrule

    Uncertainty-Weighted Waveform \(L_1\)
    & $\mathcal{L}_{\mathrm{UW}}
      =\lambda_{\mathrm{uw}}\cdot
      \mathbb{E}\!\left[w(u)\,\lVert y-\hat{y}\rVert_1\right],\qquad
      w(u)=\mathrm{clip}\!\left(\dfrac{u}{\mathbb{E}[u]},\,w_{\min},\,w_{\max}\right)$
    & $u$: epistemic uncertainty from EDD; $w(u)$ emphasizes harder or more uncertain samples; $\lambda_{\mathrm{uw}}$: weighting coefficient. \\

    \midrule

  \end{tabular}
  \vspace{-1.20em}
  \caption{Details of UDSS-BWE's generator and discriminator loss functions and training objectives.}
  \label{tab:lossfunction_uds_swinbwe}
  \vspace{-1.50em}
\end{table*}

\vspace{-0.25em}
\begin{algorithm}[t]
\caption{Training UDSS-BWE: Hybrid Reconstruction + Decision/Uncertainty Critics + HF Chance Constraint}
\label{alg:uds_train}
\begin{algorithmic}[1]
\scriptsize
\STATE \textbf{Require:} paired data $(y,x)$; generator $G$; logit critics $\mathcal{D}_0=\{\mathrm{CVaRD,CCD,MCUD,MRAD,MRPD}\}$; uncertainty critics $\{\mathrm{EDD,DROD}\}$;
warm-up $N_{\mathrm{warm}}$; CCD parameters $(b^\star,\rho_{\mathrm{cc}},\eta_{\mathrm{cc}},\lambda_{\max})$;
weights $(w^{G}_{\mathrm{evid}},w^{D}_{\mathrm{evid}},w^{G}_{\mathrm{dro}},w^{D}_{\mathrm{dro}})$.
\STATE \textbf{Ensure:} trained generator $G$ (inference uses generator only).
\STATE Initialize CCD dual variable $\lambda_{\mathrm{cc}}\gets 0$.

\FOR{$\mathrm{step}=1$ \textbf{to} $T$}
  \STATE $w_{\mathrm{adv}} \gets \min(1,\mathrm{step}/N_{\mathrm{warm}})$

  \STATE \textbf{/* Discriminator update */} \\
  \STATE $\hat{x}\gets G(y)$; \ $\hat{x}\gets \mathrm{detach}(\hat{x})$
  \STATE Compute hinge discriminator loss for each $d\in\mathcal{D}_0$:
  \STATE \hspace{1.0em} $\mathcal{L}^{d}_{D}\gets \mathbb{E}[\mathrm{ReLU}(1-D_d(x))]+\mathbb{E}[\mathrm{ReLU}(1+D_d(\hat{x}))]$
  \STATE Compute evidential discriminator loss:
  \STATE \hspace{1.0em} $\mathcal{L}_{\mathrm{EDD},D}\gets \mathcal{L}_{\mathrm{evid}}(\alpha(x),y_{\mathrm{real}};\lambda_{\mathrm{KL}})+\mathcal{L}_{\mathrm{evid}}(\alpha(\hat{x}),y_{\mathrm{fake}};\lambda_{\mathrm{KL}})$
  \STATE Compute entropic-risk discriminator loss for DROD (Algorithm~\ref{alg:drod}):
  \STATE \hspace{1.0em} $\mathcal{L}^{\mathrm{DRO}}_{D}\gets \mathrm{DROD}\_\mathrm{LossD}(x,\hat{x})$
  \STATE $\mathcal{L}_{D}\gets \sum_{d\in\mathcal{D}_0}\mathcal{L}^{d}_{D}+w^{D}_{\mathrm{evid}}\mathcal{L}_{\mathrm{EDD},D}+w^{D}_{\mathrm{dro}}\mathcal{L}^{\mathrm{DRO}}_{D}$
  \STATE Update $\{\mathcal{D}_0,\mathrm{EDD,DROD}\}$ with AdamW; clip gradients (thr.=10).

  \STATE \textbf{/* Generator update */} \\
  \STATE $\hat{x}\gets G(y)$
  \STATE Compute reconstruction loss $\mathcal{L}_{\mathrm{rec}}$ (mag/phase/complex/self-consistency; fixed weights).
  \STATE Feature matching over selected critics:
  \STATE \hspace{1.0em} $\mathcal{L}_{\mathrm{fm}}\gets \sum_{d\in\mathcal{D}_0}\sum_{\ell}\|f_{d,\ell}(x)-f_{d,\ell}(\hat{x})\|_1 \;+\; w^{G}_{\mathrm{evid}}\sum_{\ell}\|f_{\mathrm{EDD},\ell}(x)-f_{\mathrm{EDD},\ell}(\hat{x})\|_1$
  \STATE Hinge generator adversarial loss for $d\in\mathcal{D}_0$:
  \STATE \hspace{1.0em} $\mathcal{L}^{d}_{G}\gets \mathbb{E}[\mathrm{ReLU}(1-D_d(\hat{x}))]$
  \STATE EDD generator term (evidential + uncertainty penalty):
  \STATE \hspace{1.0em} $\mathcal{L}_{\mathrm{EDD},G}\gets w^{G}_{\mathrm{evid}}\Big(\mathcal{L}_{\mathrm{evid}}(\alpha(\hat{x}),y_{\mathrm{real}};0)+\lambda_u\,\mathcal{U}(\alpha(\hat{x}))\Big)$
  \STATE DROD generator term (entropic-risk over per-segment scores; Algorithm~\ref{alg:drod}):
  \STATE \hspace{1.0em} $\mathcal{L}^{\mathrm{DRO}}_{G}\gets w^{G}_{\mathrm{dro}}\cdot \mathrm{DROD}\_\mathrm{LossG}(\hat{x})$
  \STATE $\mathcal{L}_{\mathrm{adv}}\gets \sum_{d\in\mathcal{D}_0}\mathcal{L}^{d}_{G}+\mathcal{L}_{\mathrm{EDD},G}+\mathcal{L}^{\mathrm{DRO}}_{G}$

  \STATE CCD chance-constraint (augmented Lagrangian):
  \STATE \hspace{1.0em} $\Delta b\gets \bar{b}(\hat{x})-b^\star$
  \STATE \hspace{1.0em} $\mathcal{L}_{\mathrm{cc}}\gets \lambda_{\mathrm{cc}}\Delta b+\frac{\rho_{\mathrm{cc}}}{2}(\Delta b)^2$
  \STATE $\mathcal{L}_{G}\gets \mathcal{L}_{\mathrm{rec}}+w_{\mathrm{adv}}\Big(\mathcal{L}_{\mathrm{adv}}+\mathcal{L}_{\mathrm{fm}}+\mathcal{L}_{\mathrm{cc}}\Big)$
  \STATE Update $G$ with AdamW; clip gradients (thr.=10).
  \STATE Dual update (projected ascent): $\lambda_{\mathrm{cc}}\gets \Pi_{[0,\lambda_{\max}]}\big(\lambda_{\mathrm{cc}}+\eta_{\mathrm{cc}}\Delta b\big)$
\ENDFOR
\end{algorithmic}
\end{algorithm}

\subsection{Phoneme-wise HF energy benchmarks, discriminator pressure, and computational overhead}
\label{app:hf_phoneme_benchmarks_complexity}

\subsubsection{HF-band definition and phoneme-wise priors}

A convenient empirical landmark in the upper voice spectrum is the spectral dip around 4--5~kHz, identified by Ternstr\"om as the piriform-fossa (PF) notch. In that convention, octave~8 spans 5--10~kHz and octave~9 spans 10--20~kHz. Above about 4--5~kHz, resonance density increases; in sustained vowels, octave~8 still shows discernible resonances and antiresonances, whereas octave~9 tends to smear into broader resonance clusters \cite{ternstrom2008hi}.

For sustained vowels and voiced running speech, energy in octaves~8 and~9 is typically weak relative to total SPL. Ternstr\"om reports that the relative level in octaves~8 and~9 is typically about $-30$ to $-45$~dB, with octave~9 on average 5--8~dB weaker than octave~8. He also notes that, for sustained vowels, octave~6--7 levels are typically 20--30~dB higher than octave~8. Taken together, these observations indicate that steady voiced segments usually carry only weak high-frequency energy above the PF notch. This pattern is also consistent with the on-axis upper-octave levels shown in Table~\ref{tab:hfe_overall_supported}, where the 8-kHz and 16-kHz bands remain substantially below the overall SPL across speech and singing conditions \cite{ternstrom2008hi}.
By contrast, \cite{monson2012horizontal} report strong phoneme dependence for voiceless fricatives. In their horizontal-plane directivity measurements, /s/ and /sh/ are more directional than /f/ and /th/ in the 4-, 8-, and 16-kHz octave bands. They further note that most of the energy in /s/ lies in the 8-kHz octave, whereas /sh/ is influenced more by the 4-kHz octave, while /f/ and /th/ are more broadband. These phoneme-specific differences are reflected in Table~\ref{tab:hfe_fricatives_supported}, where /s/ and /sh/ show markedly different upper-band levels from /f/ and /th/. Together, these findings support the use of phoneme-aware HF priors: constraints appropriate for steady voiced segments should not be applied indiscriminately to voiceless fricatives.

\begin{table}[t]
\centering
\small
\caption{On-axis overall and upper-octave levels from the appendix of \cite{monson2012horizontal} (dB SPL at 1~m).}
\label{tab:hfe_overall_supported}
\begin{tabular}{lccc}
\toprule
Condition & Overall & 8-kHz & 16-kHz \\
\midrule
Soft speech    & 54.8 & 41.8 & 32.2 \\
Normal speech  & 62.0 & 46.4 & 38.0 \\
Loud speech    & 73.8 & 54.7 & 47.6 \\
Normal singing & 73.9 & 50.2 & 42.3 \\
\bottomrule
\end{tabular}
\end{table}

\begin{table}[t] 
\centering
\small
\caption{On-axis octave-band levels for voiceless fricatives from the appendix of \cite{monson2012horizontal} (dB SPL at 1~m).}
\label{tab:hfe_fricatives_supported}
\begin{tabular}{lcc}
\toprule
Fricative & 8-kHz & 16-kHz \\
\midrule
/s/  & 57.0 & 48.2 \\
/sh/ & 54.9 & 37.7 \\
/f/  & 39.2 & 38.7 \\
/th/ & 36.7 & 39.5 \\
\bottomrule
\end{tabular}
\end{table}

\subsubsection{Decision-aware discriminator pressure on HF structure}

Let $X(f,t)$ denote the complex STFT and $P(f,t)=|X(f,t)|^2$ the framewise power spectrum. We define the per-frame HF ratio as
$
r_{\mathrm{HF}}(t)=\frac{\sum_{f\in\mathcal{F}_{\mathrm{HF}}} P(f,t)}{\sum_{f\in\mathcal{F}} P(f,t)+\epsilon},
$
where $\mathcal{F}_{\mathrm{HF}}$ is the set of highest-frequency bins. For 16~kHz audio, this corresponds approximately to the 5--8~kHz band. The discriminators in UDSS-BWE are designed to act on this structure in a complementary and synchronized way.

\paragraph{CVaRD: tail emphasis for rare HF artifacts.}
CVaRD uses Conditional Value at Risk-style pooling to emphasize the largest temporal responses rather than their mean. This raises adversarial pressure on rare but perceptually dominant events such as brief HF clicks, hiss bursts, and boundary artifacts.

\paragraph{CCD: chance-constrained HF control.}
CCD penalizes exceedances of an HF ceiling $\tau$ through a smooth barrier. By discouraging only excessive HF excursions, CCD suppresses spurious HF over-boost in vowels while preserving legitimate fricative bursts.

\paragraph{MCUD: multi-criteria HF utility.}
MCUD evaluates HF realism through interpretable spectral descriptors, including roll-off, flatness, and centroid. Rather than reacting to energy alone, it raises adversarial pressure when the generator produces HF spectra that are inconsistent with real phoneme structure, such as over-flat vowels, over-damped /s/, or incorrect /s/ versus /S/ peak placement.

\paragraph{EDD: evidential uncertainty.}
EDD models evidence for the \{real, fake\} decision through Dirichlet parameters and exposes uncertainty explicitly. This is especially useful in ambiguous or OOD segments, where overly confident discriminator gradients can destabilize learning.

\paragraph{DROD: entropic tail-risk aggregation.}
DROD aggregates per-segment discriminator violations through entropic risk, which smoothly interpolates between average-case and max-like tail emphasis. It is therefore well matched to artifact distributions that are heavy-tailed across time.

Overall, these mechanisms encourage the discriminator to pull down spurious HF energy in steady voiced segments, pull up and reshape HF noise in fricatives and affricates, and preserve relative HF distinctions across phoneme classes.

\subsubsection{Details of Discriminator Design}

To quantify the training-time overhead introduced by these decision-science and uncertainty-aware feature computations, we profile the discriminator-side modules used in the UDSS-BWE training pipeline: (i) CVaR tail pooling in \textbf{CVaRD}, (ii) HF chance-constraint barrier in \textbf{CCD}, (iii) multi-criteria spectral utility in \textbf{MCUD}, (iv) evidential Dirichlet uncertainty in \textbf{EDD}, and (v) entropic-risk aggregation in \textbf{DROD}. Training is performed on VCTK train partition. All discriminator-side features are computed \emph{on-the-fly during training} on the GPU.
\textbf{Feature implementation:}
\textbf{CVaRD}: CVaR pooling via \texttt{torch.topk} over the final temporal feature map.
\textbf{CCD}: STFT-based HF-energy ratio and smooth barrier via \texttt{log1p(exp(.))}.
\textbf{MCUD}: STFT-based flatness, centroid, rolloff, and utility aggregation.
\textbf{EDD}: evidential parameters $\alpha$ and uncertainty $u=\tfrac{K}{\sum_k \alpha_k}$ with $K=2$ classes.
\textbf{DROD}: score-map aggregation via numerically stable \texttt{log-mean-exp}.

In the inference-time, only the \emph{generator} has been used, whereas the present analysis concerns training-time discriminator-side supervision. Standard MAC/FLOP tools capture parameterized backbones such as EDD and DROD, but they do not fully attribute the cost of parameter-free operations such as STFT-derived criteria and top-$k$ tail selection. For CVaRD, CCD, and MCUD, we therefore report analytic approximations in addition to measured timings.

\subsubsection{Analytic cost model}

Let the waveform length be $T$, batch size $B$, and STFT parameters $(n_{\mathrm{fft}},h,w)$, where $n_{\mathrm{fft}}$ is the FFT size, $h$ the hop, and $w$ the window length. Let
$L(T)\approx\left\lceil \frac{T}{h}\right\rceil$, $F=\frac{n_{\mathrm{fft}}}{2}+1$ denote the approximate number of frames and frequency bins, respectively.
\vspace{-0.5em}
\paragraph{CVaRD:} CVaRD replaces purely average-case temporal aggregation with CVaR-style tail pooling, so that brief but severe artifacts receive disproportionate attention. This is particularly important for transient HF defects that occupy only a small fraction of frames yet dominate perceived quality. CVaRD applies tail pooling over the final temporal feature map $x\in\mathbb{R}^{B\times C\times T'}$, with $k=\max(1,\lfloor \alpha T'\rfloor)$: 
\[
\mathrm{CVaR}_{\alpha}(x)=\frac{1}{k}\sum_{j\in\mathrm{Top}\text{-}k} |x_j|.
\]
Its dominant cost is top-$k$ selection,
$\mathrm{Ops}_{\mathrm{CVaR}}(T')\approx\mathcal{O}\!\bigl(B\,C\,T'\log k\bigr)$
followed by negligible absolute-value and averaging operations.
\vspace{-0.5em}
\paragraph{CCD:} CCD addresses the failure mode in which adversarial training rewards spurious HF amplification. By penalizing exceedances of an HF ceiling only when they become excessive, it suppresses systematic over-boost in vowels and sonorants while preserving short, natural HF bursts in fricatives. CCD computes the STFT, the HF ratio,
\[
r_{\mathrm{HF}}(\ell)=\frac{\sum_{f=f_0}^{F-1}|X(f,\ell)|^2}{\sum_{f=0}^{F-1}|X(f,\ell)|^2},
\]
and the smooth barrier
\[
b(\ell)=\log\!\bigl(1+\exp(k(r_{\mathrm{HF}}(\ell)-\tau))\bigr),
\]
where $f_0=\lfloor(1-\rho)F\rfloor$ is the HF start bin. The dominant cost is the STFT:
\[
\mathrm{MACs}_{\mathrm{STFT}}(T)\approx \beta\,B\,L(T)\,n_{\mathrm{fft}}\log_2(n_{\mathrm{fft}}),
\]
with post-STFT reductions scaling as
\[
\mathrm{MACs}_{\mathrm{CCD,post}}(T)=\mathcal{O}\!\bigl(B\,L(T)\,F\bigr).
\]
\vspace{-0.5em}
\paragraph{MCUD:} MCUD formalizes HF realism as a multi-objective problem. Brightness, noisiness, spread, and phoneme-dependent HF placement cannot be reduced to a single scalar energy target. By aggregating spectral flatness, centroid, and rolloff through simplex-constrained weights, MCUD encourages the generator to match the \emph{structure} of HF detail rather than only its gross magnitude. MCUD also begins with an STFT and then computes three frame-level spectral descriptors: spectral flatness (SF), spectral centroid (SC), and spectral rolloff (SR). Let
\[
M(f,\ell)=|X(f,\ell)|+\varepsilon,
\]
where \(X(f,\ell)\) denotes the STFT coefficient at frequency bin \(f\in\{1,\dots,F\}\) and frame \(\ell\), and \(\varepsilon>0\) is a small constant for numerical stability. The three descriptors are defined as
\[
\mathrm{SF}(\ell)=
\frac{
\exp\!\left(\frac{1}{F}\sum_{f=1}^{F}\log M(f,\ell)\right)
}{
\frac{1}{F}\sum_{f=1}^{F} M(f,\ell)
},
\]
\[
\mathrm{SC}(\ell)=
\frac{\sum_{f=1}^{F}\omega_f\,M(f,\ell)}
{\sum_{f=1}^{F}M(f,\ell)},
\]
\[
\mathrm{SR}(\ell)=\frac{f_r(\ell)}{F-1},
\]
\[
f_r(\ell)=
\min\!\left\{
f:\sum_{j=1}^{f}M(j,\ell)\ge \rho \sum_{j=1}^{F}M(j,\ell)
\right\},
\]
where \(\omega_f\) is the center frequency of bin \(f\), and \(\rho\in(0,1)\) is the rolloff ratio (e.g., \(\rho=0.85\)). These features are concatenated as
\[
\mathbf{c}(\ell)=\bigl[\mathrm{SF}(\ell),\,\mathrm{SC}(\ell),\,\mathrm{SR}(\ell)\bigr],
\]
and then combined using learned simplex weights:
\[
u(\ell)=\sum_{i=1}^{3}\pi_i\,\tilde{c}_i(\ell), 
\qquad 
\boldsymbol{\pi}=\mathrm{softmax}(\mathbf{w}).
\]
Its post-STFT complexity is again dominated by reductions across \(F\) bins,
\[
\mathrm{MACs}_{\mathrm{MCUD,post}}(T)=\mathcal{O}\!\bigl(B\,L(T)\,F\bigr),
\]
with additional \texttt{log/exp} operations for flatness and \texttt{cumsum/compare} operations for rolloff.
\vspace{-0.52em}
\paragraph{EDD:} EDD distinguishes confident from ambiguous discrimination by outputting Dirichlet evidence over $\{\text{real},\text{fake}\}$. This is important for atypical phonetic contexts, noisy segments, and OOD examples, where overly confident gradients may destabilize training or drive pathological HF behavior. Uncertainty-aware supervision therefore improves calibration and moderates discriminator pressure when evidence is weak. EDD relies mainly on a lightweight separable-convolution backbone, followed by evidential quantities
\[
\alpha=\mathrm{softplus}(\mathrm{logits})+1,
\qquad
u=\frac{2}{\alpha_1+\alpha_2}.
\]
These feature-side computations are negligible relative to the convolutional backbone.
\vspace{-0.5em}
\paragraph{DROD} DROD treats discriminator training as a robust risk-aggregation problem. By replacing simple averaging with entropic risk, it emphasizes high-loss segments while remaining smoother than a hard maximum. This makes it well suited to artifact distributions that are heavy-tailed across time. DROD uses a shallow separable-convolution backbone and computes an entropic-risk aggregation
\[
\rho_{\lambda}(z)=\frac{1}{\lambda}\log \mathbb{E}\!\left[\exp(\lambda z)\right],
\]
implemented in practice through a stable \texttt{logsumexp}. Because this operates over a short score map, its extra overhead is small relative to the backbone.

\subsubsection{Measured per-segment cost and overall interpretation}

Per-segment latency is reported for discriminator segments of length $8{,}000$ samples, corresponding to 0.5~s at 16~kHz. Across the five modules, the dominant \emph{parameter-free} training overhead comes from the STFT-based feature computations in \textbf{CCD} and \textbf{MCUD}. By contrast, \textbf{CVaRD} adds only lightweight tail pooling over an already-computed temporal map, while \textbf{EDD} and \textbf{DROD} add modest evidential and robust aggregation on top of shallow separable-convolution backbones. Accordingly, the main training-time cost of phoneme-aware HF supervision is not the decision logic itself, but the repeated spectral analysis needed to compute HF-sensitive descriptors.

\subsection{Decision- and uncertainty-aware properties of speech generation}
\label{subsec:Properties_of_Speech_Generation}

\subsubsection{Why speech generation requires tail-aware supervision}

Speech production is a coupled aeroelastic--acoustic process with strong localized nonstationarity: voiced regions are dominated by quasi-periodic high-frequency harmonic structure shaped by time-varying vocal-tract resonances, whereas unvoiced consonants and transitions contain turbulence, bursts, and rapid spectral reconfiguration \cite{titze2008nonlinear}. For bandwidth extension, this means that perceptually important errors are not uniformly distributed across time. Instead, they often occur as brief, high-impact events such as fricative smearing, metallic chirps, transient HF spikes, and discontinuities at phonetic boundaries.

A discriminator based only on mean pooling or global spectral similarity therefore tends to optimize an average-case criterion. Such a criterion can under-penalize low-duty-cycle artifacts and may encourage over-smoothed outputs that match coarse spectral statistics while missing the micro-transients that dominate perceived realism. This mismatch motivates discriminator designs that explicitly emphasize tail events, soft constraints, interpretable HF structure, uncertainty, and robustness.

\subsubsection{Connection to phoneme-wise HF benchmarks} The phoneme-wise HF benchmarks in Appendix~\ref{app:hf_phoneme_benchmarks_complexity} provide an interpretable calibration target for these mechanisms. The vowel prior implies that steady voiced frames should only rarely produce large HF ratios; accordingly, CCD acts as a barrier against artificial broadband filling-in. The fricative distributions in Table~\ref{tab:hfe_fricatives_supported} show, however, that strong HF energy can be entirely natural and highly phoneme dependent; this is precisely why CVaRD, MCUD, and DROD must emphasize \emph{where} and \emph{how} HF content appears, not merely whether it is present. EDD complements this by moderating supervision in segments where phonetic or acoustic ambiguity makes the real/fake boundary inherently uncertain.

\subsubsection{Overall implication for UDSS-BWE} Decision- and uncertainty-aware discriminators in UDSS-BWE are designed to reflect three empirical facts about speech reconstruction. First, perceptual failures are often rare but dominant, so supervision should be tail-sensitive. Second, HF realism is multi-objective and phoneme dependent, so supervision should encode more than aggregate energy. Third, ambiguous and OOD segments should not elicit overconfident gradients, so supervision should be uncertainty aware.

Under this view, adversarial training is no longer framed as matching only average spectral statistics. Instead, it is organized around \emph{tail risk} (CVaRD, DROD), \emph{soft HF constraint satisfaction} (CCD), \emph{interpretable HF utility trade-offs} (MCUD), \emph{calibrated epistemic uncertainty} (EDD), and \emph{entropic risk-aware (DROD)}. This synchronized design guides the generator toward preserving micro-transients, realistic phoneme-wise HF behavior, and stable learning dynamics without producing spurious amplification or brittle discriminator confidence.

\subsection{Data Preparation, Spectral Representation, and Inference}
\label{sec:data_pipeline}

Our experiments use the English VCTK corpus (version 0.92) \cite{yamagishi2019cstr} as the main benchmark, with training and evaluation following the corpus-provided speaker-wise split. In this setup, recordings are indexed from the official training and test lists by extracting the utterance identifiers needed to build paired file lists for the custom PyTorch dataset. This yields a cross-speaker protocol with 102 speakers used for training and 8 speakers reserved for testing, for a total of 88{,}329 utterances across both partitions. The same overall preprocessing strategy is also applied to MLS French \cite{Pratap2020MLSAL}, where we generate a 35 hour version to train the model to match the VCTK dataset since the reduced subset was found to be inadequate for strong BWE performance.

At data loading time, each utterance is read from disk with optional in-memory waveform reuse to reduce repeated I/O overhead. Leading and trailing silence are removed when corpus metadata are available, and multi-channel recordings are converted to mono by averaging channels. The clean target signal is then resampled to the desired high-rate bandwidth used for reconstruction. To simulate a narrowband observation, this high-rate waveform is first downsampled to a lower sampling rate and then upsampled back to the high-rate domain using sinc-based resampling. The resulting band-limited waveform is used as the model input, while the high-rate waveform serves as the supervision target.

For optimization, training examples are formed by cropping fixed-length waveform segments (8000 samples) from both the narrowband input and the corresponding wideband target. In the 16\,kHz setting, the chosen segment duration is 0.5\,s. Utterances shorter than the required segment length are zero-padded. During evaluation, random cropping is disabled so that the loader returns deterministic full utterances, or fixed evaluation segments depending on the evaluation configuration. The dataset is wrapped by a standard PyTorch data loader, with optional multi-process loading to overlap CPU-side preprocessing with GPU execution. In distributed training, the dataset is partitioned across workers with a distributed sampler and reshuffled at the start of every epoch to preserve correct stochasticity.

After waveform preparation, each input segment is transformed into a time--frequency representation using a short-time Fourier transform with a Hann window. For the 16\,kHz experiments in this work, we use a 1024-point transform, a hop size of 80 samples, and a window length of 320 samples. From the complex spectrum, we construct two complementary streams: a log-magnitude representation and an explicit phase representation. The generator predicts the corresponding wideband spectral magnitude and phase, which are then recombined and converted back to waveform samples through an inverse short-time Fourier transform using the same analysis parameters. To maintain numerical stability under mixed-precision training, the spectral transforms themselves are executed in full precision even when the rest of the network operates with automatic mixed precision.

Inference follows the same signal path as training-time preprocessing, but without cropping. For each input waveform, the system first loads the audio, converts it to mono if necessary, and resamples it to the target high-rate domain. A narrowband conditioning signal is then synthesized by low-rate downsampling followed by upsampling back to the target rate, after which its length is aligned with the target waveform. The model extracts log-magnitude and phase features from this conditioning signal, predicts the corresponding wideband spectral components, and reconstructs the enhanced waveform through inverse spectral synthesis. Output files are written as 16-bit PCM waveforms at the target sampling rate, while total inference time is accumulated over all processed files. Evaluation is performed with gradient computation disabled and the generator set to inference mode.

\subsection{Evaluation Metrics}
\label{subsec:EvaluationMetricsAppendix}
\vspace{-0.3em}
We assess quality of reconstructed speech using seven metrics: Log‐Spectral Distance (LSD) \cite{115972} to quantify fine‐grained spectral deviations; Short‐Time Objective Intelligibility (STOI) \cite{taal2011algorithm} to evaluate speech intelligibility; Perceptual Evaluation of Speech Quality (PESQ) \cite{rix2001perceptual} to predict quality in line with human judgments; Scale‐Invariant SDR (SI‐SDR) \cite{le2019sdr} as a general distortion metric invariant to amplitude scaling; Scale‐Invariant SNR (SI‐SNR) \cite{luo2018tasnet} to specifically gauge noise‐related distortion; Non‐Intrusive Speech Quality Assessment (NISQA‐MOS) \cite{mittag2021nisqa} for reference‐free estimation of perceptual speech quality; and Word Error Rate (WER) for ASR evaluation with the open-source deepspeech-0.9.3 model (see Appendix~\ref{sec:WERCalculation}). 

\subsection{WER Calculation for ASR Tasks}
\label{sec:WERCalculation}

UDSS-BWE is also evaluated on a downstream automatic speech recognition task. Following earlier studies on silent-speech voicing and the DeepSpeech framework \cite{gaddy2021improved,hannun2014deep}, we transcribe both the band-limited inputs and the UDSS-BWE-enhanced outputs using the open-source \texttt{deepspeech-0.9.3-models} checkpoint, and then measure word error rate, character error rate, and word accuracy using the same evaluation protocol as \cite{gaddy2021improved}. The results show that the proposed model restores high-frequency information that is critical for recognition, effectively converting degraded narrowband speech into signals that are much more ASR-compatible. For instance, under the 4--16~kHz setting in Table~\ref{tab:ASR_eval}, character error rate decreases from $68.59\%$ for the unprocessed input to $10.19\%$, word error rate drops from $90.01\%$ to $13.6\%$, and word accuracy rises from $10.12\%$ to $82.74\%$. Similar gains are observed across all evaluated bandwidth settings, indicating that UDSS-BWE successfully preserves and reconstructs phonetic detail that remains highly beneficial for downstream speech recognition.

\begin{table}[t]
\centering
\scriptsize
\setlength{\tabcolsep}{3pt}
\renewcommand{\arraystretch}{0.85}

\begin{tabular}{lcccccc}
\toprule
Model &  Range & WER  & CER  & Word Accuracy\\
\midrule
Unprocessed & 4 kHz & 90.01\% & 68.59\% & 10.12\% \\
UDSS-BWE & 4-16 kHz & 13.6 \% & 10.19\% & 82.74\% \\
Unprocessed & 8 kHz & 6.14\% & 3.15\% & 93.96\%\\
UDSS-BWE & 8-16 kHz & 4.18\% & 2.69\% & 94.87\% \\
Unprocessed & 16 kHz & 2.1\% & 1.21\% & 97.47\%\\
UDSS-BWE & 16-48 kHz & 1.61\% & 0.79\% & 98.4\%\\
\bottomrule
\end{tabular}
\caption{WER on the VCTK dataset.}
\label{tab:ASR_eval}
\end{table}
\vspace{-0.5em}

\subsection{Noisy Dataset Preparation and Detailed Noise Analysis}
\label{sec:appendix_noisy_analysis}
We use eight AURORA noise types and five SNR levels to simulate realistic environments. Each noise recording is concatenated in a loop until it reaches five minutes, without cross-fading. For each clean utterance, one noise type and
one SNR are randomly selected from the 40 possible combinations; a noise segment of identical duration is then cropped, scaled to satisfy the selected SNR, and mixed with the clean signal. A CSV log stores the noise type, SNR, and crop indices for reproducibility.

Tables~\ref{tab:appendix_snr_wise} and~\ref{tab:appendix_noise_wise} provide the detailed VCTK
decompositions corresponding to the compact main-text summaries in
Table~\ref{tab:noiseComparative_analysis}. 
Table~\ref{tab:appendix_snr_wise} reports detailed SNR-wise results for AP-BWE and UDSS-BWE,
whereas Table~\ref{tab:appendix_noise_wise} reports the detailed breakdown by noise type.

\begin{table*}[t]
\centering
\scriptsize
\setlength{\tabcolsep}{1.2pt}
\renewcommand{\arraystretch}{0.85}
\resizebox{\textwidth}{!}{%
\begin{tabular}{ll ccc ccc ccc ccc ccc ccc ccc}
\toprule
\multirow{2}{*}{Model} & \multirow{2}{*}{SNR}
& \multicolumn{3}{c}{Samples}
& \multicolumn{3}{c}{NISQA-MOS}
& \multicolumn{3}{c}{STOI}
& \multicolumn{3}{c}{PESQ}
& \multicolumn{3}{c}{SI-SDR}
& \multicolumn{3}{c}{SI-SNR}
& \multicolumn{3}{c}{LSD} \\
\cmidrule(lr){3-5}\cmidrule(lr){6-8}\cmidrule(lr){9-11}\cmidrule(lr){12-14}
\cmidrule(lr){15-17}\cmidrule(lr){18-20}\cmidrule(lr){21-23}
& & 4--16 & 8--16 & 16--48
  & 4--16 & 8--16 & 16--48
  & 4--16 & 8--16 & 16--48
  & 4--16 & 8--16 & 16--48
  & 4--16 & 8--16 & 16--48
  & 4--16 & 8--16 & 16--48
  & 4--16 & 8--16 & 16--48 \\
\midrule
\multirow{5}{*}{AP-BWE}
& -10 & 558 & 558 & 593 & 2.02 & 2.55 & 2.98 & 0.60 & 0.72 & 0.73 & 1.08 & 1.19 & 1.15 & -0.52 & 0.05 & 1.24 & -0.60 & -0.04 & 1.17 & 1.49 & 1.18 & 0.91 \\
& -5  & 585 & 585 & 608 & 2.52 & 2.86 & 3.54 & 0.69 & 0.80 & 0.80 & 1.15 & 1.38 & 1.26 & 2.96 & 3.62 & 4.54 & 2.86 & 3.48 & 4.44 & 1.45 & 1.11 & 0.88 \\
& 0   & 606 & 606 & 601 & 2.83 & 3.21 & 3.80 & 0.76 & 0.86 & 0.85 & 1.24 & 1.63 & 1.41 & 5.13 & 6.03 & 6.87 & 5.06 & 5.93 & 6.80 & 1.39 & 1.06 & 0.86 \\
& 5   & 580 & 580 & 576 & 3.05 & 3.42 & 4.07 & 0.81 & 0.90 & 0.89 & 1.35 & 1.93 & 1.61 & 6.57 & 7.81 & 8.68 & 6.53 & 7.74 & 8.64 & 1.34 & 1.02 & 0.84 \\
& 10  & 608 & 608 & 559 & 3.22 & 3.70 & 4.22 & 0.84 & 0.93 & 0.92 & 1.49 & 2.26 & 1.80 & 7.63 & 9.05 & 9.77 & 7.62 & 9.02 & 9.77 & 1.29 & 0.97 & 0.82 \\
\midrule
\multirow{5}{*}{UDSS-BWE}
& -10 & 558 & 533 & 563 & 2.959 & 3.076 & 3.303 & 0.595 & 0.712 & 0.731 & 1.081 & 1.183 & 1.202 & -0.844 & 0.592 & 1.482 & -0.953 & 0.542 & 1.381 & 1.599 & 1.187 & 0.931 \\
& -5  & 585 & 583 & 581 & 3.682 & 3.547 & 3.706 & 0.695 & 0.791 & 0.800 & 1.136 & 1.336 & 1.404 & 2.705 & 3.999 & 4.822 & 2.576 & 3.909 & 4.695 & 1.527 & 1.150 & 0.905 \\
& 0   & 606 & 579 & 578 & 4.112 & 3.956 & 4.030 & 0.760 & 0.854 & 0.856 & 1.203 & 1.590 & 1.697 & 4.858 & 6.323 & 7.576 & 4.791 & 6.250 & 7.497 & 1.451 & 1.082 & 0.882 \\
& 5   & 580 & 599 & 613 & 4.301 & 4.233 & 4.219 & 0.805 & 0.898 & 0.901 & 1.295 & 1.890 & 2.061 & 6.383 & 7.986 & 9.801 & 6.328 & 7.916 & 9.743 & 1.384 & 1.028 & 0.864 \\
& 10  & 608 & 643 & 602 & 4.415 & 4.383 & 4.399 & 0.843 & 0.928 & 0.931 & 1.418 & 2.177 & 2.403 & 7.402 & 8.866 & 11.312 & 7.389 & 8.832 & 11.270 & 1.332 & 1.003 & 0.847 \\
\bottomrule
\end{tabular}
}
\caption{Detailed SNR-wise analysis on the noisy VCTK dataset for AP-BWE and UDSS-BWE.} 
\label{tab:appendix_snr_wise}
\end{table*}

\begin{table*}[t]
\centering
\scriptsize
\setlength{\tabcolsep}{1.2pt}
\renewcommand{\arraystretch}{0.85}
\resizebox{\textwidth}{!}{%
\begin{tabular}{ll ccc ccc ccc ccc ccc ccc ccc}
\toprule
\multirow{2}{*}{Model} & \multirow{2}{*}{Noise Type}
& \multicolumn{3}{c}{Samples}
& \multicolumn{3}{c}{NISQA-MOS}
& \multicolumn{3}{c}{STOI}
& \multicolumn{3}{c}{PESQ}
& \multicolumn{3}{c}{SI-SDR}
& \multicolumn{3}{c}{SI-SNR}
& \multicolumn{3}{c}{LSD} \\
\cmidrule(lr){3-5}\cmidrule(lr){6-8}\cmidrule(lr){9-11}\cmidrule(lr){12-14}
\cmidrule(lr){15-17}\cmidrule(lr){18-20}\cmidrule(lr){21-23}
& & 4--16 & 8--16 & 16--48
  & 4--16 & 8--16 & 16--48
  & 4--16 & 8--16 & 16--48
  & 4--16 & 8--16 & 16--48
  & 4--16 & 8--16 & 16--48
  & 4--16 & 8--16 & 16--48
  & 4--16 & 8--16 & 16--48 \\
\midrule
\multirow{8}{*}{AP-BWE}
& AWGN       & 348 & 348 & 378 & 2.68 & 2.73 & 3.76 & 0.75 & 0.81 & 0.80 & 1.25 & 1.52 & 1.35 & 4.99 & 6.33 & 6.54 & 4.90 & 6.22 & 6.52 & 1.40 & 2.73 & 0.88 \\
& Airport    & 372 & 372 & 411 & 2.73 & 3.27 & 3.89 & 0.74 & 0.86 & 0.85 & 1.28 & 1.75 & 1.49 & 4.44 & 5.34 & 6.28 & 4.41 & 5.28 & 6.22 & 1.39 & 3.27 & 0.83 \\
& Babble     & 384 & 384 & 348 & 2.67 & 3.26 & 3.82 & 0.74 & 0.85 & 0.84 & 1.28 & 1.70 & 1.45 & 4.28 & 5.01 & 5.93 & 4.23 & 4.94 & 5.88 & 1.38 & 3.26 & 0.83 \\
& Car        & 354 & 354 & 370 & 2.87 & 3.23 & 3.66 & 0.75 & 0.85 & 0.84 & 1.26 & 1.71 & 1.45 & 4.55 & 5.42 & 6.26 & 4.49 & 5.34 & 6.21 & 1.38 & 3.23 & 0.85 \\
& Exhibition & 333 & 333 & 366 & 2.91 & 3.24 & 3.94 & 0.77 & 0.86 & 0.85 & 1.30 & 1.79 & 1.47 & 5.41 & 6.15 & 6.91 & 5.37 & 6.07 & 6.87 & 1.34 & 3.24 & 0.84 \\
& SSN        & 386 & 386 & 359 & 2.60 & 3.14 & 3.20 & 0.73 & 0.83 & 0.82 & 1.23 & 1.64 & 1.39 & 3.15 & 4.31 & 4.81 & 3.05 & 4.18 & 4.65 & 1.47 & 3.14 & 0.97 \\
& Station    & 380 & 380 & 349 & 2.64 & 3.19 & 3.78 & 0.74 & 0.85 & 0.85 & 1.25 & 1.68 & 1.49 & 4.08 & 5.09 & 6.39 & 4.01 & 4.99 & 6.33 & 1.39 & 3.19 & 0.83 \\
& Street     & 380 & 380 & 356 & 2.81 & 3.18 & 3.63 & 0.75 & 0.84 & 0.83 & 1.26 & 1.72 & 1.41 & 4.67 & 5.66 & 6.11 & 4.63 & 5.59 & 6.09 & 1.37 & 3.17 & 0.84 \\
\midrule
\multirow{8}{*}{UDSS-BWE}
& AWGN       & 348 & 397 & 368 & 3.877 & 3.174 & 3.578 & 0.744 & 0.801 & 0.821 & 1.208 & 1.418 & 1.700 & 4.742 & 6.253 & 7.731 & 4.646 & 6.130 & 7.669 & 1.481 & 1.351 & 0.908 \\
& Airport    & 372 & 354 & 346 & 3.886 & 4.028 & 4.043 & 0.742 & 0.856 & 0.856 & 1.247 & 1.728 & 1.791 & 4.181 & 5.597 & 6.888 & 4.136 & 5.527 & 6.817 & 1.462 & 1.014 & 0.855 \\
& Babble     & 384 & 364 & 361 & 3.856 & 3.962 & 4.106 & 0.731 & 0.845 & 0.850 & 1.232 & 1.684 & 1.795 & 3.788 & 5.238 & 6.869 & 3.729 & 5.170 & 6.799 & 1.449 & 1.024 & 0.857 \\
& Car        & 354 & 359 & 379 & 4.043 & 4.032 & 3.895 & 0.741 & 0.840 & 0.845 & 1.212 & 1.666 & 1.774 & 4.293 & 5.501 & 6.891 & 4.225 & 5.478 & 6.812 & 1.458 & 1.038 & 0.878 \\
& Exhibition & 333 & 379 & 407 & 4.254 & 4.255 & 4.243 & 0.773 & 0.857 & 0.860 & 1.253 & 1.739 & 1.825 & 5.121 & 6.504 & 7.845 & 5.072 & 6.465 & 7.783 & 1.418 & 1.029 & 0.869 \\
& SSN        & 386 & 358 & 349 & 3.493 & 3.368 & 3.451 & 0.728 & 0.827 & 0.822 & 1.241 & 1.581 & 1.637 & 2.997 & 4.586 & 5.718 & 2.878 & 4.496 & 5.604 & 1.462 & 1.162 & 0.986 \\
& Station    & 380 & 355 & 340 & 3.799 & 3.999 & 4.106 & 0.731 & 0.853 & 0.855 & 1.219 & 1.744 & 1.810 & 3.897 & 5.858 & 7.211 & 3.799 & 5.800 & 7.104 & 1.461 & 1.018 & 0.857 \\
& Street     & 380 & 371 & 387 & 4.107 & 4.147 & 4.070 & 0.747 & 0.850 & 0.855 & 1.220 & 1.701 & 1.779 & 4.566 & 6.094 & 7.422 & 4.512 & 6.063 & 7.338 & 1.456 & 1.030 & 0.873 \\
\bottomrule
\end{tabular}
}
\caption{Detailed noise-type-wise analysis on the noisy VCTK dataset for AP-BWE and
UDSS-BWE.} 
\label{tab:appendix_noise_wise}
\end{table*}

\subsection{Details on Computational Complexity}
\label{append:analysisoncomputationcomplexity}

Table \ref{tab:computation} presents the detailed analysis of computational complexity.

\textbf{Model Size (Parameters):} AP-BWE has 72.07M parameters, which is more than 4$\times$ that of UDSS-BWE (18.5M). This indicates that AP-BWE requires substantially more memory for model weights and intermediate activations. In contrast, UDSS-BWE is more memory-efficient, making it easier to deploy on resource-limited devices.

\textbf{Computational Cost (MACs \& FLOPs):} AP-BWE requires 14.2B MACs and 28.5B FLOPs per inference, whereas UDSS-BWE requires only 5.3B MACs and 10.7B FLOPs. This corresponds to approximately a 62.7\% reduction in MACs and a 62.5\% reduction in FLOPs, making UDSS-BWE substantially more suitable for resource-constrained settings.

\textbf{Real-Time Factor (RTF, GPU):} RTF is defined as the ratio between generation time and total signal duration. AP-BWE runs at 0.0023--0.0025$\times$ RTF, while UDSS-BWE achieves 0.0035--0.0033$\times$ RTF. Hence, both models are highly efficient on GPU. UDSS-BWE shows a comparable RTF to AP-BWE, demonstrating that it is suitable for real-time downstream generation tasks.


\textbf{Feature Extractor Complexity (Training Only):} Table \ref{tab:feature_extractor_complexity} summarizes the computational cost of the feature extractors used in the decision-science and uncertainty discriminators. Among them, CVaR\_FE is the most computationally intensive, requiring 49.696M MACs and 99.392M FLOPs, with a latency of 0.659 ms. ED\_FE and DRO\_FE have moderate complexity, each requiring about 14M MACs and less than 0.5 ms latency. In contrast, CC\_FE and MCU\_FE are extremely lightweight in terms of arithmetic operations, with only about 0.14M MACs each, although their measured latency is slightly higher due to STFT-based feature formation and auxiliary processing overhead. Overall, the additional feature extraction modules remain lightweight and practical for training, while contributing richer decision-aware supervision.

\textbf{Training vs. Inference Impact:} These computations are invoked only within the discriminators during training, where they provide robustness, uncertainty awareness, and constraint-guided supervision; therefore, they introduce no deployment-time overhead. During inference, UDSS-BWE executes only the generator forward pass and does not perform CCD/MCUD STFT feature extraction, CVaR pooling, evidential uncertainty estimation, or entropic-risk aggregation. As a result, deployment RTF is dominated by the generator alone and is therefore comparable to, or lower than, that of the baseline.

\vspace{-0.4em}
\begin{table}[ht]
\centering
\scriptsize
\setlength{\tabcolsep}{4pt}
\renewcommand{\arraystretch}{0.9}
\begin{tabular}{lccc}
\toprule
Feature Extractor & MACs (M) & FLOPs (M) & Latency (ms) \\
\midrule
CVaR\_FE              & 49.696 & 99.392 & 0.659 \\
CC\_FE    & 0.141  & 0.282  & 0.823 \\
MCU\_FE  & 0.139  & 0.278  & 0.889 \\
ED\_FE   & 14.376 & 28.752 & 0.441 \\
DRO\_FE           & 14.248 & 28.496 & 0.348 \\
\bottomrule
\end{tabular}
\vspace{-0.4em}
\caption{Computational complexity analysis of the feature extractors used in the decision-science and uncertainty discriminators only.}
\label{tab:feature_extractor_complexity}
\vspace{-1.0em}
\end{table}

\subsection{Study for Different Frequency Ranges}
\label{subsec:Analysis for Different Frequency Ranges}

{\color{black}Table~\ref{tab:SoTA_different_fr_range}} examines performance across multiple frequency ranges and reveals a consistent trend: \textit{performance improves as the gap between the narrowband input and the target band decreases}. Accordingly, the widest reconstruction setting (2--48,kHz) is the most challenging and yields the weakest results, whereas the narrowest reconstruction setting (24--48,kHz) is the least challenging and achieves the strongest performance. The table further shows that all seven evaluation metrics improve from 2--48,kHz to 24--48,kHz as the reconstruction gap becomes smaller.

\begin{table}[t]
\centering
\scriptsize
\setlength{\tabcolsep}{3pt}
\renewcommand{\arraystretch}{0.85}
\begin{tabular}{lccccccc}
\toprule
Freq. range & LSD & STOI & PESQ & SDR & SNR & N-MOS & WER \\
\midrule
2-16 kHz  & 1.08 & 0.86 & 1.5 & 7.61 & 7.59 & 4.49 & 0.60 \\
2-48 kHz  & 1.07 & 0.84 & 1.42 & 7.24 & 7.25 & 4.04 & 0.73 \\
4-48 kHz  & 1.28 & 0.89 & 1.37 & 9.23 & 9.32 & 2.57 & 0.35 \\
8-48 kHz  & 0.94 & 0.99 & 3.37 & 15.32 & 15.24 & 4.52 & 0.05 \\
12-48 kHz & 0.87 & 0.99 & 4.23 & 16.81 & 16.69 & 4.54 & 0.03 \\
24-48 kHz & 0.67 & 0.99 & 4.47 & 22.23 & 22.21 & 4.52 & 0.01 \\
\bottomrule
\end{tabular}
\caption{Performance over different frequency ranges.}
\label{tab:SoTA_different_fr_range}
\end{table}

\vspace{-0.25em}
\subsection{ITU-Inspired Evidence of UDSS-BWE Improvement over AP-BWE}
\label{subsec:itu_inspired_udss_ap_evidence}
\vspace{-0.32em}
This subsection compares AP-BWE and UDSS-BWE using three ITU-inspired sub-band diagnostic metrics: the BS.1387-inspired \cite{bs1387_2023} spectral-flux error $D_{\mathrm{SF},i}$, the BS.1770-inspired \cite{itu_bs1770_2023} short-term loudness standard-deviation error $D_{\sigma L,i}$, and the P.862/PESQ-inspired \cite{itu_p862_2001} missing-disturbance error $D_{\mathrm{miss},i}$. These metrics are selected because they capture complementary HF reconstruction cues: frame-to-frame spectral continuity, short-term loudness stability, and missing perceptual components. All three are errors or loss measures; therefore, lower values indicate better alignment with the 48 kHz reference.

Let $y[n]$ and $\hat{y}[n]$ denote the 48 kHz reference and reconstructed signals, respectively. For sub-band $\mathcal{B}_i$, let $Y_i(k,m)$ and $\hat{Y}_i(k,m)$ be the corresponding short-time spectral magnitudes at frequency bin $k$ and frame $m$. The normalized spectral shape vector is
$
\mathbf{s}_i(m)=
\frac{\left[|Y_i(k,m)|\right]_{k\in\Omega_i}}
{\left\|\left[|Y_i(k,m)|\right]_{k\in\Omega_i}\right\|_2+\epsilon},
$
$
\hat{\mathbf{s}}_i(m)=
\frac{\left[|\hat{Y}_i(k,m)|\right]_{k\in\Omega_i}}
{\left\|\left[|\hat{Y}_i(k,m)|\right]_{k\in\Omega_i}\right\|_2+\epsilon}.
$
The BS.1387-inspired spectral-flux error is then
$D_{\mathrm{SF},i}=\frac{1}{M-1}\sum_{m=2}^{M}\left\|\mathbf{s}_i(m) \mathbf{s}_i(m-1)\right\|_2-$ $\frac{1}{M-1}\sum_{m=2}^{M} \left\|\hat{\mathbf{s}}_i(m)-\hat{\mathbf{s}}_i(m-1)\right\|_2.$ This metric measures whether the reconstructed HF spectrum changes over time like the reference. For the BS.1770-inspired short-term loudness statistic, let $L_i(m)$ and $\hat{L}_i(m)$ denote K-weighted short-term loudness values in band $\mathcal{B}_i$. The loudness-variation error is
$D_{\sigma L,i} = \mathrm{std}_{m}\!\left(L_i(m)\right) -
\mathrm{std}_{m}\!\left(\hat{L}_i(m)\right).$
This metric evaluates whether the reconstructed signal preserves the reference dynamics of short-term loudness. Finally, for the P.862/PESQ-inspired missing-disturbance metric, define the log-spectral difference
$\Delta_i(k,m)=20\log_{10}\!\left(|\hat{Y}_i(k,m)|+\epsilon\right)-20\log_{10}\!\left(|Y_i(k,m)|+\epsilon\right).$ The missing disturbance is $D_{\mathrm{miss},i}=\frac{1}{|\Omega_i|M}\sum_{m=1}^{M}\sum_{k\in\Omega_i}
\max\!\left(-\Delta_i(k,m),0\right).$
This metric penalizes reference spectral components that are under reconstructed by the model.

\begin{table}[t]
\centering
\scriptsize
\setlength{\tabcolsep}{1.1pt}

\begin{tabular}{clcccccc}
\toprule
Row & Band &
\multicolumn{2}{c}{$D_{\mathrm{SF},i}$} &
\multicolumn{2}{c}{$D_{\sigma L,i}$} &
\multicolumn{2}{c}{$D_{\mathrm{miss},i}$} \\
\cmidrule(lr){3-4}\cmidrule(lr){5-6}\cmidrule(lr){7-8}
& & AP & UDSS
& AP & UDSS
& AP & UDSS \\
\midrule
\textcircled{\scriptsize 1} & 0--4
& \textbf{$1.04{\times}10^{-4}$} & $1.23{\times}10^{-4}$
& 0.0521 & \textbf{0.0514}
& \textbf{0.227} & 0.302 \\
\textcircled{\scriptsize 2} & 4--8 
& \textbf{$1.57{\times}10^{-4}$} & $1.93{\times}10^{-4}$
& 0.375 & \textbf{0.171}
& 0.704 & \textbf{0.561} \\
\textcircled{\scriptsize 3} & 8--12 
& 0.00114 & \textbf{0.000906}
& 0.831 & \textbf{0.774}
& \textbf{3.979} & 4.233 \\
\textcircled{\scriptsize 4} & 12--16 
& 0.00124 & \textbf{0.000866}
& \textbf{0.689} & 1.018
& 3.214 & \textbf{2.961} \\
\textcircled{\scriptsize 5} & 16--20 
& 0.00113 & \textbf{0.000795}
& \textbf{0.775} & 0.793
& 3.616 & \textbf{2.560} \\
\textcircled{\scriptsize 6} & 20--24
& 0.00133 & \textbf{0.000817}
& 0.942 & \textbf{0.711}
& 3.342 & \textbf{2.979} \\
\bottomrule
\end{tabular}
\caption{Sub-band AP-BWE vs. UDSS-BWE comparison for ITU-inspired metrics}
\vspace{-1em}
\label{tab:ap_udss_v18_selected_itu_metrics}
\end{table}

Table~\ref{tab:ap_udss_v18_selected_itu_metrics} reports all six 4 kHz sub-bands, rows \textcircled{\scriptsize 1}--\textcircled{\scriptsize 6}. UDSS-BWE achieves lower error in 12 of the 18 comparisons, and in 9 of the 12 comparisons within the generated high-frequency region above 8 kHz, rows \textcircled{\scriptsize 3}--\textcircled{\scriptsize 6}. Thus UDSS-BWE provides stronger temporal spectral-continuity cues. This is supported by $D_{\mathrm{SF},i}$, where we can claim that UDSS-BWE improves every generated band: 8--12, 12--16, 16--20, and 20--24 kHz, rows \textcircled{\scriptsize 3}--\textcircled{\scriptsize 6}. The intuition is that HF speech is not only defined by average magnitude, but also by how spectral details evolve from frame to frame. Lower spectral-flux error indicates that UDSS-BWE reconstructs HF transitions with more reference-like temporal smoothness. Simultaneously, UDSS-BWE provides more faithful short-term loudness-dynamics cues. For $D_{\sigma L,i}$, UDSS-BWE improves four of six bands, including the 4--8 kHz transition band, row \textcircled{\scriptsize 2}, the 8--12 kHz generated band, row \textcircled{\scriptsize 3}, and the uppermost 20--24 kHz band, row \textcircled{\scriptsize 6}. The reduction from 0.375 to 0.171 in 4--8 kHz, row \textcircled{\scriptsize 2}, and from 0.942 to 0.711 in 20--24 kHz, row \textcircled{\scriptsize 6}, suggests that UDSS-BWE better matches the dynamic loudness variation of the reference. This is important because perceptual HF quality depends on whether fricatives, bursts, and sibilant components rise and decay with realistic strength, not merely whether energy is present. Lastly, UDSS-BWE provides stronger missing-component recovery cues. For $D_{\mathrm{miss},i}$, UDSS-BWE improves four of six bands and three of the four generated HF bands, rows \textcircled{\scriptsize 4}--\textcircled{\scriptsize 6}. In particular, the error decreases from 3.214 to 2.961 in 12--16 kHz, row \textcircled{\scriptsize 4}, from 3.616 to 2.560 in 16--20 kHz, row \textcircled{\scriptsize 5}, and from 3.342 to 2.979 in 20--24 kHz, row \textcircled{\scriptsize 6}. Since $D_{\mathrm{miss},i}$ penalizes reference components that are absent or under-represented in the reconstruction, these gains indicate that UDSS-BWE recovers more perceptually relevant HF contents. Overall, three selected ITU-inspired metrics support the conclusion that UDSS-BWE improves reconstructed performance by providing more coherent HF temporal-spectral cues, more accurate short-term loudness dynamics, and fewer missing high-frequency components. These improvements are especially concentrated in the generated region above the 8 kHz input Nyquist limit, rows \textcircled{\scriptsize 3}--\textcircled{\scriptsize 6}, where bandwidth extension is most challenging.

\vspace{-0.85em}
\subsection{\textbf{Statistical Analysis of MOS Scores}}
\label{subsec:nisqa_stats_udss_ap}
\vspace{-0.15em}

We compute MOS statistics on a 1,000-utterance matched subset of AP-BWE and UDSS-BWE outputs. The subset was sampled without replacement using a fixed random seed from the matched evaluation files. An availability-constrained approximately balanced speaker allocation was used: speaker p362 contributed all 123 available utterances, speaker p364 contributed 147 utterances, and speakers p360, p361, p363, p374, and p376 each contributed 146 utterances.

To perform the statistical analysis of the MOS scores, we calculate the mean
$\bar{x}_m=\frac{1}{N}\sum_{i=1}^{N}x_{i,m}$, standard deviation
$SD_m=\sqrt{\frac{1}{N-1}\sum_{i=1}^{N}(x_{i,m}-\bar{x}_m)^2}$, standard error of the mean
$SEM_m=\frac{SD_m}{\sqrt{N}}$, and 95\% confidence interval
$95\%\ \mathrm{CI}=\bar{x}_m\pm t_{0.975,N-1}SEM_m$, where $N$ is the number of utterances and $m$ denotes the method. For paired testing, the MOS difference for matched item $i$ is
$d_i=x_i^{\mathrm{UDSS}}-x_i^{\mathrm{AP}},\ i=1,\ldots,N.$
After removing zero differences, the absolute differences are ranked in ascending order:
$R_i=\operatorname{rank}(|d_i|).$
The positive and negative rank sums are
$W^+=\sum_{i:d_i>0}R_i,\qquad W^-=\sum_{i:d_i<0}R_i,$
and the two-sided Wilcoxon signed-rank statistic is
$T=\min(W^+,W^-).$
The null and alternative hypotheses are
$H_0:\operatorname{median}(d_i)=0,\qquad
H_1:\operatorname{median}(d_i)\neq 0.$

The statistical analysis results and paired Wilcoxon signed-rank test are tabulated in Table~\ref{tab:nisqa_mos_stats_udss_1000} and Table~\ref{tab:wilcoxon_nisqa_udss_1000}, respectively. On the 1,000-utterance subset, UDSS-BWE increases mean MOS over AP-BWE by 0.038 points. The utterance-level Wilcoxon test indicates a statistically significant paired difference $(p=1.94\times10^{-35})$, with UDSS-BWE obtaining higher MOS on 718 of 1,000 matched utterances. The speaker-level comparison also shows a significant paired difference $(p=0.0313)$, with UDSS-BWE improving 6 of the 7 speaker-level averages. These results indicate that UDSS-BWE provides a consistent MOS improvement over AP-BWE.

\vspace{-0.5em}
\begin{table}[ht!]
\centering
\resizebox{\linewidth}{!}{
\setlength{\tabcolsep}{4pt}
\renewcommand{\arraystretch}{1}
\begin{tabular}{llccccc}
\hline
Agg. & Method & $N$ & MOS $\uparrow$ & SD & SEM & 95\% CI \\
\hline
Utt. & AP-BWE   & 1000 & 4.458 & 0.472 & 0.015 & [4.429, 4.488] \\
Utt. & UDSS-BWE & 1000 & 4.497 & 0.486 & 0.015 & [4.466, 4.527] \\
Spk. & AP-BWE   & 7    & 4.448 & 0.196 & 0.074 & [4.267, 4.630] \\
Spk. & UDSS-BWE & 7    & 4.487 & 0.196 & 0.074 & [4.305, 4.668] \\
\hline
\end{tabular}
}
\caption{MOS descriptive statistics. Utt. = utterance-level; Spk. = speaker-level.}
\vspace{-0.5em}
\label{tab:nisqa_mos_stats_udss_1000}
\end{table}

\begin{table}[ht!]
\centering
\resizebox{\linewidth}{!}{
\setlength{\tabcolsep}{4pt}
\renewcommand{\arraystretch}{1}
\begin{tabular}{lccccccc}
\hline
Agg. & $N$ & $\overline{d}$ & Median$(d)$ & $W^+$ & $W^-$ & $T$ & $p$ \\
\hline
Utt. & 1000 & 0.038 & 0.038 & 363749 & 136751 & 136751 & $1.94{\times}10^{-35}$ \\
Spk. & 7    & 0.038 & 0.039 & 27     & 1      & 1      & 0.0313 \\
\hline
\end{tabular}
}
\caption{Paired Wilcoxon signed-rank test for AP-BWE and UDSS-BWE MOS scores.}
\label{tab:wilcoxon_nisqa_udss_1000}
\end{table}

\subsection{PCM and Telephony-Codec Results}
\label{subsec:telephony_codec_results}

\paragraph{Experimental details:}
We evaluate the channel-sensitivity protocol in
Table~\ref{tab:pcm_telephony_sensitivity} using the UDSS-BWE's $(4\rightarrow16)~\mathrm{kHz}$ VCTK setting and the same 50-epoch training configuration as the main experiments. We compare five training experiment settings. 
\textbf{R0} is the reference model trained with sinc resampling only; \textbf{R1} is trained with a uniform mixture of sinc resampling, PCM-8, and PCM-24;
\textbf{R2} is trained with a uniform mixture
of sinc resampling, G.711 $\mu$-law, and G.711 A-law;
\textbf{R3} extends the mixture to sinc
resampling, PCM-8, both G.711 laws, and AMR-NB at 12.2~kb/s; and \textbf{R4} applies the R3 mixture together
with 300--3400~Hz telephone filtering and 3\% packet loss with repeat-previous-packet concealment. The current reported results contain one trained model for each experiment (seed 1234). Each model is evaluated on
2,937 test utterances per condition, and Tables~\ref{tab:pcm_codec_results_seed1234}
and~\ref{tab:telephony_loss_results_seed1234} report utterance-level averages. The ideal condition uses sinc resampling and writes the enhanced signal in a
24-bit PCM container; PCM-24 and PCM-8 additionally quantize the narrowband
conditioning signal; and both G.711 conditions use an actual 8-kHz codec
round trip. The packet-loss conditions apply G.711 $\mu$-law coding, a
300--3400~Hz telephone passband, independent loss of 20-ms packets, and
repeat-previous-packet concealment. Codec, quantization, filtering, and packet
loss are applied only to the narrowband model input, while the clean wideband
target remains unchanged. 

\begin{table*}[t]
\centering
\scriptsize
\setlength{\tabcolsep}{3.2pt}
\renewcommand{\arraystretch}{1.02}
\caption{PCM and codec results on the VCTK test set for seed 1234
($n=2{,}937$ utterances per condition). 
Within each test condition, the best value is shown in bold.
Lower LSD is better; all other metrics are higher-is-better.}
\label{tab:pcm_codec_results_seed1234}
\begin{tabularx}{\textwidth}{@{}
>{\centering\arraybackslash}p{0.035\textwidth}
>{\raggedright\arraybackslash}p{0.090\textwidth}
>{\centering\arraybackslash}p{0.040\textwidth}
*{6}{>{\centering\arraybackslash}X}@{}}
\toprule
Row & Test condition & Exp. & LSD & STOI & PESQ & SI-SDR & SI-SNR & N-MOS \\
\midrule
\textcircled{\scriptsize 1} & \multirow{5}{*}{Ideal}
& R0 & \textbf{0.9471} & 0.9445 & \textbf{2.4250} & \textbf{13.0330} & \textbf{12.9754} & 4.2779 \\
\textcircled{\scriptsize 2} & & R1 & 0.9533 & 0.9412 & 2.3440 & 12.4628 & 12.3991 & 4.4166 \\
\textcircled{\scriptsize 3} & & R2 & 0.9514 & \textbf{0.9458} & 2.3629 & 12.4504 & 12.3844 & 4.3511 \\
\textcircled{\scriptsize 4} & & R3 & 0.9586 & 0.9397 & 2.2754 & 12.4433 & 12.3627 & \textbf{4.4190} \\
\textcircled{\scriptsize 5} & & R4 & 1.2455 & 0.8781 & 1.5612 & $-19.7806$ & $-20.4072$ & 2.6345 \\
\midrule
\textcircled{\scriptsize 6} & \multirow{5}{*}{PCM-24}
& R0 & \textbf{0.9471} & 0.9446 & \textbf{2.4249} & \textbf{13.0326} & \textbf{12.9751} & 4.2783 \\
\textcircled{\scriptsize 7} & & R1 & 0.9533 & 0.9412 & 2.3439 & 12.4628 & 12.3990 & \textbf{4.4173} \\
\textcircled{\scriptsize 8} & & R2 & 0.9514 & \textbf{0.9458} & 2.3626 & 12.4500 & 12.3840 & 4.3505 \\
\textcircled{\scriptsize 9} & & R3 & 0.9585 & 0.9397 & 2.2755 & 12.4433 & 12.3627 & 4.4177 \\
\textcircled{\scriptsize 10} & & R4 & 1.2455 & 0.8781 & 1.5611 & $-19.7725$ & $-20.4068$ & 2.6327 \\
\midrule
\textcircled{\scriptsize 11} & \multirow{5}{*}{PCM-8}
& R0 & 1.0630 & 0.8844 & 1.4756 & \textbf{12.4510} & \textbf{12.3745} & 1.8301 \\
\textcircled{\scriptsize 12} & & R1 & \textbf{1.0031} & \textbf{0.9131} & \textbf{1.9602} & 12.3922 & 12.3414 & \textbf{3.7645} \\
\textcircled{\scriptsize 13} & & R2 & 1.0891 & 0.8842 & 1.4589 & 11.9486 & 11.8720 & 1.9492 \\
\textcircled{\scriptsize 14} & & R3 & 1.0072 & 0.9127 & 1.8807 & 12.3418 & 12.2702 & 3.5892 \\
\textcircled{\scriptsize 15} & & R4 & 1.1835 & 0.8422 & 1.4701 & $-18.7199$ & $-19.3331$ & 2.0266 \\
\midrule
\textcircled{\scriptsize 16} & \multirow{5}{*}{G.711-$\mu$}
& R0 & \textbf{0.9475} & 0.9409 & \textbf{2.4210} & \textbf{12.9862} & \textbf{12.9319} & 4.2045 \\
\textcircled{\scriptsize 17} & & R1 & 0.9535 & 0.9379 & 2.3454 & 12.4326 & 12.3719 & 4.3588 \\
\textcircled{\scriptsize 18} & & R2 & 0.9525 & \textbf{0.9433} & 2.3644 & 12.4062 & 12.3434 & 4.3218 \\
\textcircled{\scriptsize 19} & & R3 & 0.9601 & 0.9373 & 2.2733 & 12.4128 & 12.3345 & \textbf{4.3974} \\
\textcircled{\scriptsize 20} & & R4 & 1.2455 & 0.8758 & 1.5673 & $-19.5694$ & $-20.1538$ & 2.6283 \\
\midrule
\textcircled{\scriptsize 21} & \multirow{5}{*}{G.711-A}
& R0 & \textbf{0.9492} & 0.9406 & \textbf{2.4235} & \textbf{13.0082} & \textbf{12.9544} & 4.1879 \\
\textcircled{\scriptsize 22} & & R1 & 0.9552 & 0.9377 & 2.3492 & 12.4545 & 12.3951 & 4.3377 \\
\textcircled{\scriptsize 23} & & R2 & 0.9533 & \textbf{0.9433} & 2.3674 & 12.4231 & 12.3608 & 4.3134 \\
\textcircled{\scriptsize 24} & & R3 & 0.9606 & 0.9373 & 2.2748 & 12.4290 & 12.3516 & \textbf{4.3847} \\
\textcircled{\scriptsize 25} & & R4 & 1.2481 & 0.8751 & 1.5708 & $-19.4626$ & $-20.1289$ & 2.6392 \\
\bottomrule
\end{tabularx}
\end{table*}

\begin{table*}[t]
\centering
\scriptsize
\setlength{\tabcolsep}{3.2pt}
\renewcommand{\arraystretch}{1.02}
\caption{Robustness to filtered G.711 $\mu$-law channels with packet loss on
the VCTK test set for seed 1234 ($n=2{,}937$ utterances per condition).
Within each loss rate, the best value is shown in bold.}
\label{tab:telephony_loss_results_seed1234}
\begin{tabularx}{\textwidth}{@{}
>{\centering\arraybackslash}p{0.035\textwidth}
>{\raggedright\arraybackslash}p{0.090\textwidth}
>{\centering\arraybackslash}p{0.040\textwidth}
*{6}{>{\centering\arraybackslash}X}@{}}
\toprule
Row & Test condition & Exp. & LSD$\downarrow$ & STOI$\uparrow$ & PESQ$\uparrow$ & SI-SDR$\uparrow$ & SI-SNR$\uparrow$ & N-MOS$\uparrow$ \\
\midrule
\textcircled{\scriptsize 1} & \multirow{5}{*}{Loss-3}
& R0 & 1.1206 & 0.8921 & 1.7298 & $-8.7184$ & $-8.7466$ & 2.9929 \\
\textcircled{\scriptsize 2} & & R1 & 1.1186 & 0.8864 & 1.6759 & $-8.6251$ & $-8.6830$ & \textbf{3.5332} \\
\textcircled{\scriptsize 3} & & R2 & \textbf{1.1012} & 0.8962 & 1.7421 & $-8.8512$ & $-8.8997$ & 3.3267 \\
\textcircled{\scriptsize 4} & & R3 & 1.1426 & 0.8868 & 1.5979 & $-8.6734$ & $-8.7331$ & 3.4352 \\
\textcircled{\scriptsize 5} & & R4 & 1.1346 & \textbf{0.9104} & \textbf{1.8614} & \textbf{7.8475} & \textbf{7.7938} & 2.9225 \\
\midrule
\textcircled{\scriptsize 6} & \multirow{5}{*}{Loss-5}
& R0 & 1.1209 & 0.8824 & 1.6314 & $-8.9576$ & $-8.9889$ & 2.7699 \\
\textcircled{\scriptsize 7} & & R1 & 1.1194 & 0.8772 & 1.5914 & $-8.8360$ & $-8.8940$ & \textbf{3.3018} \\
\textcircled{\scriptsize 8} & & R2 & \textbf{1.1016} & 0.8864 & 1.6394 & $-9.0888$ & $-9.1402$ & 3.0710 \\
\textcircled{\scriptsize 9} & & R3 & 1.1432 & 0.8774 & 1.5233 & $-8.9188$ & $-8.9773$ & 3.1619 \\
\textcircled{\scriptsize 10} & & R4 & 1.1408 & \textbf{0.9036} & \textbf{1.7872} & \textbf{6.9913} & \textbf{6.9318} & 2.7730 \\
\bottomrule
\end{tabularx}
\end{table*}
\paragraph{Codec and quantization robustness:}
For \textbf{Table~\ref{tab:pcm_codec_results_seed1234},
Rows~\textcircled{\scriptsize 1}--\textcircled{\scriptsize 10}},
the ideal and PCM-24 conditions are effectively identical. This is expected
because a 24-bit output container changes numerical representation but does not recover or remove information in the narrowband conditioning signal. G.711 coding is also a comparatively mild mismatch for
\textbf{Table~\ref{tab:pcm_codec_results_seed1234},
Rows~\textcircled{\scriptsize 1}--\textcircled{\scriptsize 4},
\textcircled{\scriptsize 16}--\textcircled{\scriptsize 19}, and
\textcircled{\scriptsize 21}--\textcircled{\scriptsize 24}}
(R0--R3): their LSD, STOI, and PESQ values remain close to the ideal-condition
values under both $\mu$-law and A-law. This suggests that the generator can
still infer high-frequency structure when companding (non-linear compression $\rightarrow$ quantization $\rightarrow$ expension) preserves the relevant
low-frequency phonetic cues. However, there are no experimental settings in which a single one dominates all metrics.
\textbf{R0 in Table~\ref{tab:pcm_codec_results_seed1234},
Rows~\textcircled{\scriptsize 16} and~\textcircled{\scriptsize 21}} retains the best PESQ, LSD,
SI-SDR, and SI-SNR under both G.711 laws, whereas
\textbf{R3 in Table~\ref{tab:pcm_codec_results_seed1234},
Rows~\textcircled{\scriptsize 19} and~\textcircled{\scriptsize 24}} produces the highest N-MOS
(4.3974 for $\mu$-law and 4.3847 for
A-law). Thus, codec augmentation can improve estimated naturalness without
necessarily improving sample-aligned spectral or waveform fidelity. PCM-8 quantization is substantially more destructive. Relative to
\textbf{Table~\ref{tab:pcm_codec_results_seed1234},
Row~\textcircled{\scriptsize 11} (R0)} under PCM-8, the matched PCM-mix
\textbf{Table~\ref{tab:pcm_codec_results_seed1234},
Row~\textcircled{\scriptsize 12} (R1)} reduces LSD by 0.0599 (5.6\%), increases
STOI by 0.0287 and PESQ by 0.4846 (32.8\%), and raises N-MOS from 1.8301 to
3.7645. Its SI-SDR changes by only $-0.0588$~dB. The large perceptual gain with
nearly unchanged SI-SDR indicates that waveform-level distortion metrics can
understate audible quantization artifacts. \textbf{Table~\ref{tab:pcm_codec_results_seed1234},
Row~\textcircled{\scriptsize 14} (R3)} exhibits a similar, though slightly weaker, benefit, while G.711-only
\textbf{Table~\ref{tab:pcm_codec_results_seed1234},
Row~\textcircled{\scriptsize 13} (R2)} does not protect
against PCM-8 degradation. The improvement is therefore attributable to
exposure to quantized inputs rather than to generic codec augmentation.
\paragraph{Packet-loss robustness and intuition:} Increasing the packet-loss rate from 3\% to 5\% consistently reduces STOI, PESQ, SI-SDR, SI-SNR, and N-MOS for all five experimental settings, as expected when repeat-packet concealment creates longer local discontinuities. Among \textbf{Table~\ref{tab:telephony_loss_results_seed1234},
Rows~\textcircled{\scriptsize 1}--\textcircled{\scriptsize 4} and
\textcircled{\scriptsize 6}--\textcircled{\scriptsize 9}}
(R0--R3), \textbf{R2 in Table~\ref{tab:telephony_loss_results_seed1234},
Rows~\textcircled{\scriptsize 3} and~\textcircled{\scriptsize 8}} gives the strongest STOI and PESQ at both loss rates, while \textbf{R1 in Table~\ref{tab:telephony_loss_results_seed1234},
Rows~\textcircled{\scriptsize 2} and~\textcircled{\scriptsize 7}} gives the highest N-MOS. The full-telephony
\textbf{R4 in Table~\ref{tab:telephony_loss_results_seed1234},
Rows~\textcircled{\scriptsize 5} and~\textcircled{\scriptsize 10}} yields the strongest matched-channel STOI and PESQ: compared with the best non-R4 result in
\textbf{Table~\ref{tab:telephony_loss_results_seed1234},
Rows~\textcircled{\scriptsize 3} and~\textcircled{\scriptsize 8}},
STOI improves by 0.0142
and PESQ by 0.1193 at 3\% loss, and by 0.0172 and 0.1478 at 5\% loss. R4 also
changes SI-SDR from the approximately $-9$~dB range of
\textbf{Table~\ref{tab:telephony_loss_results_seed1234},
Rows~\textcircled{\scriptsize 1}--\textcircled{\scriptsize 4} and
\textcircled{\scriptsize 6}--\textcircled{\scriptsize 9}} to
7.8475 and 6.9913~dB. However, these gains do not transfer to N-MOS;
\textbf{R1 in Table~\ref{tab:telephony_loss_results_seed1234},
Rows~\textcircled{\scriptsize 2} and~\textcircled{\scriptsize 7}} remain higher by 0.6107 and 0.5288 points at the two loss rates. This divergence suggests that
matched telephony augmentation teaches R4 to preserve waveform alignment and
intelligibility under packet erasures, but may encourage conservative or
channel-colored outputs that are judged less natural by NISQA.
\paragraph{Diagnostic Reservation:}
\textbf{R4 in Table~\ref{tab:pcm_codec_results_seed1234},
Rows~\textcircled{\scriptsize 5}, \textcircled{\scriptsize 10},
\textcircled{\scriptsize 15}, \textcircled{\scriptsize 20}, and
\textcircled{\scriptsize 25}, and Table~\ref{tab:telephony_loss_results_seed1234},
Rows~\textcircled{\scriptsize 5} and~\textcircled{\scriptsize 10}} specializes strongly to the degraded channel and performs poorly on ideal,
PCM, and loss-free G.711 inputs. In particular, its SI-SDR changes from
$-19.7806$~dB for ideal resampling in \textbf{Table~\ref{tab:pcm_codec_results_seed1234},
Row~\textcircled{\scriptsize 5}} to 7.8475~dB for the 3\%-loss condition in
\textbf{Table~\ref{tab:telephony_loss_results_seed1234},
Row~\textcircled{\scriptsize 5}}.
Although this may reflect severe distribution specialization, the magnitude
and sign reversal are sufficiently unusual that the final multi-seed analysis
should verify identical waveform lengths, gain normalization, target pairing,
and compensation for codec or packetization delay across all R4 evaluations.
If the result persists after these checks, it provides evidence that always-on
telephone filtering and packet-loss augmentation overfits the matched channel;
otherwise, it may reveal an evaluation-alignment mismatch. Overall, the current
results favor condition-matched augmentation, but they also show that robustness
should be selected using both intrusive metrics and perceptual quality rather
than any single score.
\subsection{Parameter Breakdown for Discriminators}
\label{subsec:Parameter Breakdown for Discriminators}

A layerwise parameter breakdown for each discriminator and total for all discriminators in UDSS-BWE are shown in Table \ref{tab:riskcvar_breakdown}- 

\clearpage
\onecolumn

\begin{table}
\begin{tabular}{@{}
p{0.19\textwidth}
p{0.10\textwidth}
p{0.20\textwidth}
p{0.10\textwidth}
p{0.08\textwidth}
p{0.08\textwidth}
p{0.08\textwidth}
r@{}}
\hline
\textbf{Discriminator} & \textbf{Stage} & \textbf{Layer Type} & \textbf{In→Out} & \textbf{Kernel} & \textbf{Stride} & \textbf{Padding} & \textbf{Params} \\
\hline
\multirow{17}{*}{CVaRD}
  & Block 1 & Depthwise Conv1d                 & 1→1     & 7   & 2 & 3 & 8        \\
  &         & Pointwise Conv1d                 & 1→32    & 1   & 1 & 0 & 64       \\
  &         & BatchNorm1d + LReLU(0.2)         & 32→32   & –   & – & – & 64       \\
\cline{2-8}
  & Block 2 & Depthwise Conv1d                 & 32→32   & 7   & 2 & 3 & 256      \\
  &         & Pointwise Conv1d                 & 32→64   & 1   & 1 & 0 & 2\,112   \\
  &         & BatchNorm1d + LReLU(0.2)         & 64→64   & –   & – & – & 128      \\
\cline{2-8}
  & Block 3 & Depthwise Conv1d                 & 64→64   & 7   & 2 & 3 & 512      \\
  &         & Pointwise Conv1d                 & 64→128  & 1   & 1 & 0 & 8\,320   \\
  &         & BatchNorm1d + LReLU(0.2)         & 128→128 & –   & – & – & 256      \\
\cline{2-8}
  & Block 4 & Depthwise Conv1d                 & 128→128 & 7   & 2 & 3 & 1\,024   \\
  &         & Pointwise Conv1d                 & 128→256 & 1   & 1 & 0 & 33\,024  \\
  &         & BatchNorm1d + LReLU(0.2)         & 256→256 & –   & – & – & 512      \\
\cline{2-8}
  & Block 5 & Depthwise Conv1d                 & 256→256 & 7   & 2 & 3 & 2\,048   \\
  &         & Pointwise Conv1d                 & 256→256 & 1   & 1 & 0 & 65\,792  \\
  &         & BatchNorm1d + LReLU(0.2)         & 256→256 & –   & – & – & 512      \\
\cline{2-8}
  & FC      & Linear                           & 256→128 & –   & – & – & 32\,896   \\
  &         & Linear                           & 128→1   & –   & – & – & 129       \\
\cline{2-8}
\multicolumn{7}{r}{\textbf{CVaRD total}} & 147\,657 \\
\hline
\end{tabular}
\caption{Layer-wise parameter breakdown for the CVaRD.}
\label{tab:riskcvar_breakdown}
\end{table}

\clearpage
\onecolumn
\begin{table}
\begin{tabular}{@{}
p{0.19\textwidth}
p{0.10\textwidth}
p{0.20\textwidth}
p{0.10\textwidth}
p{0.08\textwidth}
p{0.08\textwidth}
p{0.08\textwidth}
r@{}}
\hline
\textbf{Discriminator} & \textbf{Stage} & \textbf{Layer Type} & \textbf{In→Out} & \textbf{Kernel} & \textbf{Stride} & \textbf{Padding} & \textbf{Params} \\
\hline
\multirow{12}{*}{CCD}
  & Block 1 & Depthwise Conv1d                 & 1→1     & 9   & 2 & 4 & 10       \\
  &         & Pointwise Conv1d                 & 1→32    & 1   & 1 & 0 & 64       \\
  &         & BatchNorm1d + LReLU(0.2)         & 32→32   & –   & – & – & 64       \\
\cline{2-8}
  & Block 2 & Depthwise Conv1d                 & 32→32   & 7   & 2 & 3 & 256      \\
  &         & Pointwise Conv1d                 & 32→64   & 1   & 1 & 0 & 2\,112   \\
  &         & BatchNorm1d + LReLU(0.2)         & 64→64   & –   & – & – & 128      \\
\cline{2-8}
  & Block 3 & Depthwise Conv1d                 & 64→64   & 5   & 2 & 2 & 384      \\
  &         & Pointwise Conv1d                 & 64→64   & 1   & 1 & 0 & 4\,160   \\
  &         & BatchNorm1d + LReLU(0.2)         & 64→64   & –   & – & – & 128      \\
\cline{2-8}
  & Block 4 & Depthwise Conv1d                 & 64→64   & 3   & 1 & 1 & 256      \\
  &         & Pointwise Conv1d                 & 64→1    & 1   & 1 & 0 & 65       \\
  &         & BatchNorm1d                      & 1→1     & –   & – & – & 2        \\
\cline{2-8}
\multicolumn{7}{r}{\textbf{CCD total}} & 7\,629 \\
\hline
\multirow{11}{*}{MCUD}
  & Weights & Learnable logits\_w (K=3)        & 3→3     & –   & – & – & 3        \\
\cline{2-8}
  & Block 1 & Depthwise Conv1d                 & 1→1     & 9   & 2 & 4 & 10       \\
  &         & Pointwise Conv1d                 & 1→32    & 1   & 1 & 0 & 64       \\
  &         & BatchNorm1d + LReLU(0.2)         & 32→32   & –   & – & – & 64       \\
\cline{2-8}
  & Block 2 & Depthwise Conv1d                 & 32→32   & 7   & 2 & 3 & 256      \\
  &         & Pointwise Conv1d                 & 32→64   & 1   & 1 & 0 & 2\,112   \\
  &         & BatchNorm1d + LReLU(0.2)         & 64→64   & –   & – & – & 128      \\
\cline{2-8}
  & Block 3 & Depthwise Conv1d                 & 64→64   & 5   & 2 & 2 & 384      \\
  &         & Pointwise Conv1d                 & 64→64   & 1   & 1 & 0 & 4\,160   \\
  &         & BatchNorm1d + LReLU(0.2)         & 64→64   & –   & – & – & 128      \\
\cline{2-8}
  & Final   & Linear                           & 64→1    & –   & – & – & 65       \\
\cline{2-8}
\multicolumn{7}{r}{\textbf{MCUD total}} & 7\,374 \\
\hline
\end{tabular}
\caption{Layer-wise parameter breakdown for the CCD and MCUD.}
\label{tab:chance_utility_breakdown}
\end{table}

\clearpage
\onecolumn
\begin{table}[p]
\begin{tabular}{@{}
p{0.19\textwidth}
p{0.10\textwidth}
p{0.20\textwidth}
p{0.10\textwidth}
p{0.08\textwidth}
p{0.08\textwidth}
p{0.08\textwidth}
r@{}}
\hline
\textbf{Discriminator} & \textbf{Stage} & \textbf{Layer Type} & \textbf{In→Out} & \textbf{Kernel} & \textbf{Stride} & \textbf{Padding} & \textbf{Params} \\
\hline
\multirow{9}{*}{EDD}
  & Block 1 & Depthwise Conv1d                 & 1→1     & 9   & 2 & 4 & 10       \\
  &         & Pointwise Conv1d                 & 1→32    & 1   & 1 & 0 & 64       \\
\cline{2-8}
  & Block 2 & Depthwise Conv1d                 & 32→32   & 7   & 2 & 3 & 256      \\
  &         & Pointwise Conv1d                 & 32→64   & 1   & 1 & 0 & 2\,112   \\
\cline{2-8}
  & Block 3 & Depthwise Conv1d                 & 64→64   & 5   & 2 & 2 & 384      \\
  &         & Pointwise Conv1d                 & 64→64   & 1   & 1 & 0 & 4\,160   \\
\cline{2-8}
  & Block 4 & Depthwise Conv1d                 & 64→64   & 3   & 2 & 1 & 256      \\
  &         & Pointwise Conv1d                 & 64→128  & 1   & 1 & 0 & 8\,320   \\
\cline{2-8}
  & Final   & Linear                           & 128→2   & –   & – & – & 258      \\
\cline{2-8}
\multicolumn{7}{r}{\textbf{EDD total}} & 15\,820 \\
\hline
\multirow{9}{*}{DROD}
  & Block 1 & Depthwise Conv1d                 & 1→1     & 9   & 2 & 4 & 10       \\
  &         & Pointwise Conv1d                 & 1→32    & 1   & 1 & 0 & 64       \\
\cline{2-8}
  & Block 2 & Depthwise Conv1d                 & 32→32   & 7   & 2 & 3 & 256      \\
  &         & Pointwise Conv1d                 & 32→64   & 1   & 1 & 0 & 2\,112   \\
\cline{2-8}
  & Block 3 & Depthwise Conv1d                 & 64→64   & 5   & 2 & 2 & 384      \\
  &         & Pointwise Conv1d                 & 64→64   & 1   & 1 & 0 & 4\,160   \\
\cline{2-8}
  & Block 4 & Depthwise Conv1d                 & 64→64   & 3   & 2 & 1 & 256      \\
  &         & Pointwise Conv1d                 & 64→128  & 1   & 1 & 0 & 8\,320   \\
\cline{2-8}
  & Final   & Conv1d                           & 128→1   & 1   & 1 & 0 & 129      \\
\cline{2-8}
\multicolumn{7}{r}{\textbf{DROD total}} & 15\,691 \\
\hline
\end{tabular}
\caption{Layer-wise parameter breakdown for the EDD and DROD.}
\label{tab:evidential_dro_breakdown}
\end{table}

\begin{table}[p]
\begin{tabular}{@{}
p{0.19\textwidth}
p{0.10\textwidth}
p{0.20\textwidth}
p{0.10\textwidth}
p{0.08\textwidth}
p{0.08\textwidth}
p{0.08\textwidth}
r@{}}
\hline
\textbf{Discriminator} & \textbf{Stage} & \textbf{Layer Type} & \textbf{In→Out} & \textbf{Kernel} & \textbf{Stride} & \textbf{Padding} & \textbf{Params} \\
\hline
\multirow{6}{*}{MRAD (per res)}
  & Conv 1     & Conv2d WeightNorm              & 1→64    & 7×5 & 2×2 & 3×2 & 2\,368   \\
  & Conv 2     & Conv2d WeightNorm              & 64→64   & 5×3 & 2×1 & 2×1 & 61\,568  \\
  & Conv 3     & Conv2d WeightNorm              & 64→64   & 5×3 & 2×2 & 2×1 & 61\,568  \\
  & Conv 4     & Conv2d WeightNorm              & 64→64   & 3×3 & 2×1 & 1×1 & 36\,992  \\
  & Conv 5     & Conv2d WeightNorm              & 64→64   & 3×3 & 2×2 & 1×1 & 36\,992  \\
  & Conv\_post & Conv2d WeightNorm              & 64→1    & 3×3 & 1×1 & 1×1 & 578      \\
\cline{2-8}
\multicolumn{7}{r}{\textbf{MRAD total (per res)}} & 200\,066 \\
\multicolumn{7}{r}{\textbf{MRAD total (all 3 res)}} & 600\,198 \\
\hline
\end{tabular}
\caption{Layer-wise parameter breakdown for MRAD.}
\label{tab:mrad_breakdown}
\end{table}

\clearpage
\onecolumn
\begin{table}[p]
\begin{tabular}{@{}
p{0.19\textwidth}
p{0.10\textwidth}
p{0.20\textwidth}
p{0.10\textwidth}
p{0.08\textwidth}
p{0.08\textwidth}
p{0.08\textwidth}
r@{}}
\hline
\textbf{Discriminator} & \textbf{Stage} & \textbf{Layer Type} & \textbf{In→Out} & \textbf{Kernel} & \textbf{Stride} & \textbf{Padding} & \textbf{Params} \\
\hline
\multirow{6}{*}{MRPD (per res)}
  & Conv 1     & Conv2d WeightNorm              & 1→64    & 7×5 & 2×2 & 3×2 & 2\,368   \\
  & Conv 2     & Conv2d WeightNorm              & 64→64   & 5×3 & 2×1 & 2×1 & 61\,568  \\
  & Conv 3     & Conv2d WeightNorm              & 64→64   & 5×3 & 2×2 & 2×1 & 61\,568  \\
  & Conv 4     & Conv2d WeightNorm              & 64→64   & 3×3 & 2×1 & 1×1 & 36\,992  \\
  & Conv 5     & Conv2d WeightNorm              & 64→64   & 3×3 & 2×2 & 1×1 & 36\,992  \\
  & Conv\_post & Conv2d WeightNorm              & 64→1    & 3×3 & 1×1 & 1×1 & 578      \\
\cline{2-8}
\multicolumn{7}{r}{\textbf{MRPD total (per res)}} & 200\,066 \\
\multicolumn{7}{r}{\textbf{MRPD total (all 3 res)}} & 600\,198 \\
\hline
\end{tabular}
\caption{Layer-wise parameter breakdown for MRPD.}
\label{tab:mrpd_breakdown}
\end{table}

\begin{table}[p]
\begin{tabular}{@{}l r@{}}
\hline
\textbf{Discriminator} & \textbf{Total Params} \\
\hline
CVaRD & 147\,657 \\
CCD & 7\,629 \\
MCUD & 7\,374 \\
EDD & 15\,820 \\
DROD & 15\,691 \\
MRAD  & 600\,198 \\
MRPD  & 600\,198 \\
\hline
\textbf{Grand total} & \textbf{1\,394\,567} \\
\hline
\end{tabular}
\caption{Per-discriminator totals and grand total for all instantiated discriminators.}
\label{tab:disc_param_summary}
\end{table}

\clearpage
\twocolumn

\subsection{Parameter Breakdown for Generators}
\label{subsec:Parameter Breakdown for Generators}

A layer-wise parameter breakdown for the UDSS-BWE generator, including LatticeBlock1D parameters alongside pre-processing, ConformerNeXt blocks, and post-processing are shown in Table  \ref{table:generator_breakdown_swin1d}.

\begin{table*}[ht!]
\centering
\resizebox{\textwidth}{!}{
\begin{tabular}{l l c c c c c r}
\hline
\textbf{Stage / Component} & \textbf{Layer Type} & \textbf{In$\rightarrow$Out} & \textbf{Kernel} & \textbf{Stride} & \textbf{Padding} & \textbf{Heads} & \textbf{Params} \\
\hline
\multicolumn{8}{c}{\textbf{Pre-processing (NB magnitude/phase feature lift)}} \\
\hline
Pre-mag convolution        & Conv1d      & 513$\rightarrow$512 & 7 & 1 & 3 & -- & 1\,839\,104 \\
Pre-pha convolution        & Conv1d      & 513$\rightarrow$512 & 7 & 1 & 3 & -- & 1\,839\,104 \\
Pre-mag LayerNorm          & LayerNorm   & 512$\rightarrow$512 & -- & -- & -- & -- & 1\,024 \\
Pre-pha LayerNorm          & LayerNorm   & 512$\rightarrow$512 & -- & -- & -- & -- & 1\,024 \\
\hline

\multicolumn{8}{c}{\textbf{Swin1DBlock (per block breakdown; dim=512, heads=8, window=8, mlp\_ratio=4)}} \\
\hline
\multicolumn{8}{c}{\textbf{(S)W-MSA + Norm}} \\
Norm$_1$                   & LayerNorm   & 512$\rightarrow$512 & -- & -- & -- & -- & 1\,024 \\
QKV projection             & Linear      & 512$\rightarrow$1\,536 & -- & -- & -- & 8 & 787\,968 \\
Relative position bias     & Learned bias table & (2W$-$1)$\times$H & -- & -- & -- & 8 & 120 \\
Output projection          & Linear      & 512$\rightarrow$512 & -- & -- & -- & 8 & 262\,656 \\
LayerScale $\gamma_1$      & Learned scale & 512$\rightarrow$512 & -- & -- & -- & -- & 512 \\
\hline
\multicolumn{8}{c}{\textbf{FFN + Norm}} \\
Norm$_2$                   & LayerNorm   & 512$\rightarrow$512 & -- & -- & -- & -- & 1\,024 \\
FFN Linear$_1$ + GELU + Dropout(0.1) & Linear & 512$\rightarrow$2\,048 & -- & -- & -- & -- & 1\,050\,624 \\
FFN Linear$_2$ + Dropout(0.1)        & Linear & 2\,048$\rightarrow$512 & -- & -- & -- & -- & 1\,049\,088 \\
LayerScale $\gamma_2$      & Learned scale & 512$\rightarrow$512 & -- & -- & -- & -- & 512 \\
\hline
\multicolumn{7}{r}{\textbf{Total per Swin1DBlock}} & 3\,153\,528 \\
\multicolumn{7}{r}{\textbf{Total per Swin1DStack (2 Swin1DBlocks; shift=[0, W/2])}} & 6\,307\,056 \\
\hline

\multicolumn{8}{c}{\textbf{LatticeBlock1D (per block)}} \\
\hline
Two-stream lattice coupling & Shared Swin1DStack branch (used twice) & 512$\rightarrow$512 & -- & -- & -- & 8 & 6\,307\,056 \\
Lattice scalars $(a_1,a_2,b_1,b_2)$ & 4 learned scalars & -- & -- & -- & -- & -- & 4 \\
\hline
\multicolumn{7}{r}{\textbf{Total per LatticeBlock1D}} & 6\,307\,060 \\
\multicolumn{7}{r}{\textbf{Total LatticeBlock1D (2 blocks)}} & 12\,614\,120 \\
\hline

\multicolumn{8}{c}{\textbf{Post-processing (WB heads)}} \\
\hline
Post-mag LayerNorm          & LayerNorm   & 512$\rightarrow$512 & -- & -- & -- & -- & 1\,024 \\
Post-mag FFN head           & Linear      & 512$\rightarrow$513 & -- & -- & -- & -- & 263\,169 \\
Post-pha LayerNorm          & LayerNorm   & 512$\rightarrow$512 & -- & -- & -- & -- & 1\,024 \\
Post-pha head (real)        & Linear      & 512$\rightarrow$513 & -- & -- & -- & -- & 263\,169 \\
Post-pha head (imag)        & Linear      & 512$\rightarrow$513 & -- & -- & -- & -- & 263\,169 \\
\hline
\multicolumn{7}{r}{\textbf{Total generator parameters}} & 17\,085\,931 \\
\hline
\end{tabular}}
\caption{Layer-wise parameter breakdown for the Swin-T(iny)-Lattice APNet-BWE generator derived from the provided code: pre Conv1d+LN for magnitude/phase, two LatticeBlock1D stages (each using a shared Swin1DStack called twice but counted once in parameters) and post linear heads for WB magnitude and phase (real/imag $\rightarrow$ atan2).}
\label{table:generator_breakdown_swin1d}
\end{table*}

\subsection{Hyperparameters and Configuration}
\label{subsec:hyperparameters}

\begin{table}[t]
\raggedright
\begingroup
\scriptsize
\setlength{\tabcolsep}{2.5pt}
\renewcommand{\arraystretch}{1.12}
\setlength{\emergencystretch}{1em}
\sloppy
\begin{tabularx}{\columnwidth}{@{}P{0.36\columnwidth} P{0.22\columnwidth} X@{}}
\toprule
\textbf{Hyperparameter} & \textbf{Value} & \textbf{Use Case \& Rationale} \\
\midrule
Max epochs
& 50
& Upper bound on training iterations to ensure convergence while controlling compute budget. \\
Number of GPUs
& auto
& Hardware-aware scaling: training automatically adapts to available accelerators and enables multi-GPU training when present. \\
Batch size (per GPU)
& auto
& Per-device minibatch is adjusted to maintain stable memory footprint and consistent throughput across different GPU counts. \\
Random seed
& 1234
& Ensures deterministic initialization and repeatable experimental results for fair ablations. \\
Distributed backend
& \texttt{nccl}
& High-performance multi-GPU communication backend optimized for NVIDIA devices. \\
Initialization URL
& \texttt{tcp://127.0.0.1:\allowbreak 54321}
& Single-node rendezvous endpoint for distributed process coordination. \\
World size
& 1
& Single-node baseline configuration (scales naturally to multi-process when needed). \\
DataLoader workers
& 0
& Baseline I/O setting; can be increased to overlap host-side loading with GPU execution. \\
\bottomrule
\end{tabularx}
\endgroup
\caption{Training-system configuration hyperparameters.}
\label{tab:train_system_hparams_app}
\end{table}

\begin{table}[t]
\raggedright
\begingroup
\scriptsize
\setlength{\tabcolsep}{2.5pt}
\renewcommand{\arraystretch}{1.12}
\setlength{\emergencystretch}{1em}
\sloppy
\begin{tabularx}{\columnwidth}{@{}P{0.36\columnwidth} P{0.22\columnwidth} X@{}}
\toprule
\textbf{Hyperparameter} & \textbf{Value} & \textbf{Use Case \& Rationale} \\
\midrule
Optimizer (G / D)
& AdamW / AdamW
& Decoupled weight decay improves optimization stability and generalization under adversarial training dynamics. \\
Generator learning rate
& $2\times10^{-4}$
& Conservative step size to promote stable convergence of the generator under multi-loss supervision. \\
Discriminator learning rate
& $3.5\times10^{-4}$
& Slightly higher step size to keep discriminators responsive and maintain informative gradients for the generator. \\
Adam $(\beta_1,\beta_2)$
& (0.8, 0.99)
& Momentum and second-moment settings tuned for GAN non-stationarity, balancing fast adaptation and variance control. \\
Learning-rate schedule
& Exponential
& Smooth annealing reduces oscillations late in training and encourages fine-grained refinement near convergence. \\
Learning-rate decay factor
& 0.999
& Gentle per-epoch decay to avoid premature stagnation while still improving late-stage stability. \\
Gradient clipping
& 10.0
& Prevents gradient explosion and improves robustness under adversarial and multi-objective losses. \\
Adversarial warmup (steps)
& 5000
& Gradually introduces adversarial objectives to stabilize early training and avoid discriminator overpowering. \\
\bottomrule
\end{tabularx}
\endgroup
\caption{Optimization hyperparameters.}
\label{tab:optim_hparams_app}
\end{table}

\begin{table}[t]
\raggedright
\begingroup
\scriptsize
\setlength{\tabcolsep}{2.5pt}
\renewcommand{\arraystretch}{1.12}
\setlength{\emergencystretch}{1em}
\sloppy
\begin{tabularx}{\columnwidth}{@{}P{0.36\columnwidth} P{0.22\columnwidth} X@{}}
\toprule
\textbf{Hyperparameter} & \textbf{Value} & \textbf{Use Case \& Rationale} \\
\midrule
Mixed precision
& \makecell[l]{enabled\\(FP16)}
& Improves throughput and reduces memory; numerically sensitive spectral transforms are kept in full precision for stability. \\
TF32
& enabled
& Accelerates matrix operations on compatible GPUs with negligible impact on model quality in practice. \\
cuDNN benchmarking
& enabled
& Selects efficient kernels to maximize training speed for fixed input shapes. \\
Model compilation
& disabled
& Prioritizes reproducibility and broad compatibility across environments over potential speed gains. \\
Fused optimizers
& disabled
& Keeps the training stack portable and consistent across different CUDA/toolchain versions. \\
\bottomrule
\end{tabularx}
\endgroup
\caption{Numerical and runtime settings.}
\label{tab:runtime_hparams_app}
\end{table}

\begin{table}[ht!]
\raggedright
\begingroup
\scriptsize
\setlength{\tabcolsep}{2.5pt}
\renewcommand{\arraystretch}{1.12}
\setlength{\emergencystretch}{1em}
\sloppy
\begin{tabularx}{\columnwidth}{@{}P{0.36\columnwidth} P{0.22\columnwidth} X@{}}
\toprule
\textbf{Hyperparameter} & \textbf{Value} & \textbf{Use Case \& Rationale} \\
\midrule
Magnitude reconstruction weight
& 45
& Emphasizes accurate log-magnitude recovery to preserve spectral envelope and formant structure. \\
Phase reconstruction weight
& 100
& Prioritizes phase consistency (including instantaneous and derivative terms) to reduce temporal smearing and artifacts. \\
Complex-spectrum reconstruction weight
& 90
& Encourages coherent complex STFT prediction, improving perceptual fidelity beyond magnitude-only supervision. \\
STFT-consistency weight
& 90
& Enforces analysis--synthesis consistency to suppress hallucinated components and stabilize waveform reconstruction. \\
\bottomrule
\end{tabularx}
\endgroup
\caption{Generator objective weights.}
\label{tab:gen_loss_weights_app}
\end{table}

\begin{table}[ht!]
\raggedright
\begingroup
\scriptsize
\setlength{\tabcolsep}{2.5pt}
\renewcommand{\arraystretch}{1.12}
\setlength{\emergencystretch}{1em}
\sloppy
\begin{tabularx}{\columnwidth}{@{}P{0.36\columnwidth} P{0.22\columnwidth} X@{}}
\toprule
\textbf{Hyperparameter} & \textbf{Value} & \textbf{Use Case \& Rationale} \\
\midrule
Segment size (samples)
& 8000
& Fixed-length training segments provide a consistent receptive field and enable efficient batching. \\
High-rate sampling rate (Hz)
& 16000
& Target wideband sampling rate defining the output bandwidth for reconstruction. \\
Low-rate sampling rate (Hz)
& 4000
& Input narrowband sampling rate defining the degraded bandwidth used as conditioning. \\
Subsampling ratio (HR/LR)
& 4
& Specifies the bandwidth expansion factor from narrowband input to wideband target. \\
STFT FFT size
& 1024
& Trade-off between frequency resolution and computational efficiency for time--frequency modeling. \\
STFT hop size
& 80
& Dense frame sampling improves temporal tracking while maintaining manageable compute. \\
STFT window size
& 320
& Balances time--frequency localization to support both transient detail and harmonic structure. \\
STFT centering
& enabled
& Symmetric framing reduces boundary artifacts and yields more stable time--frequency estimates. \\
STFT windowing (MRAD/MRPD)
& rectangular
& Uses an unbiased reference window to capture multi-resolution spectral statistics consistently. \\
STFT windowing (chance/utility)
& Hann
& Smooth tapering reduces spectral leakage for energy- and criterion-based measurements. \\
\bottomrule
\end{tabularx}
\endgroup
\caption{Signal processing and data hyperparameters.}
\label{tab:signal_hparams_app}
\end{table}

\begin{table}[ht!]
\raggedright
\begingroup
\scriptsize
\setlength{\tabcolsep}{2.5pt}
\renewcommand{\arraystretch}{1.12}
\setlength{\emergencystretch}{1em}
\sloppy
\begin{tabularx}{\columnwidth}{@{}P{0.36\columnwidth} P{0.22\columnwidth} X@{}}
\toprule
\textbf{Hyperparameter} & \textbf{Value} & \textbf{Use Case \& Rationale} \\
\midrule
Feature dimension
& 512
& Hidden width controlling representational capacity for joint magnitude/phase modeling. \\
Pre-projection kernel size
& 7
& Enlarged temporal context for initial feature extraction before attention-based processing. \\
Pre-projection input channels
& 513
& One-sided STFT bin count corresponding to the chosen FFT size. \\
Layer normalization epsilon
& $10^{-6}$
& Stabilizes normalization in low-variance regimes and improves training robustness. \\
Number of lattice coupling blocks
& 2
& Two-stage cross-stream coupling to progressively refine magnitude and phase representations. \\
Coupling scalars
& \makecell[l]{$a_1,a_2$\\$b_1,b_2$ (init 1.0)}
& Learnable interaction strengths enabling adaptive information flow between magnitude and phase streams. \\
Initialization (Conv/Linear)
& trunc-normal (0.02)
& Standardized initialization improves convergence speed and reduces sensitivity to random seeds. \\
\bottomrule
\end{tabularx}
\endgroup
\caption{Generator core architectural hyperparameters.}
\label{tab:gen_core_hparams_app}
\end{table}

\begin{table}[ht!]
\raggedright
\begingroup
\scriptsize
\setlength{\tabcolsep}{2.5pt}
\renewcommand{\arraystretch}{1.12}
\setlength{\emergencystretch}{1em}
\sloppy
\begin{tabularx}{\columnwidth}{@{}P{0.36\columnwidth} P{0.22\columnwidth} X@{}}
\toprule
\textbf{Hyperparameter} & \textbf{Value} & \textbf{Use Case \& Rationale} \\
\midrule
\multicolumn{3}{@{}l@{}}{\textbf{Swin-1D branch (per lattice block)}} \\
Swin stack depth
& 2 blocks
& Alternating non-shifted and shifted window attention improves cross-window context while preserving locality. \\
Heads / window / shift
& 8 / 8 / 4
& Multi-head attention over short temporal windows; shift enables boundary-crossing interactions with minimal cost. \\
MLP ratio
& 4.0
& Expands channel capacity in the feed-forward sublayer for improved nonlinear modeling. \\
Dropout / attention dropout
& 0.1 / 0.0
& Regularization via projection/MLP dropout while keeping attention deterministic for stability. \\
LayerScale initialization
& $10^{-6}$
& Residual scaling to stabilize deep residual learning and mitigate early training instability. \\
Relative position bias init (std)
& 0.02
& Provides a mild inductive bias for temporal ordering within attention windows. \\
\midrule
\multicolumn{3}{@{}l@{}}{\textbf{Output heads}} \\
Magnitude head
& residual linear
& Predicts a correction in the log-magnitude domain to preserve input structure while extending bandwidth. \\
Phase head
& \makecell[l]{two linear\\(real/imag) + atan2}
& Produces a stable phase estimate via a normalized angle parameterization. \\
Complex reconstruction
& \makecell[l]{$(e^m\cos\theta,$\\$e^m\sin\theta)$}
& Converts predicted log-magnitude and phase into a complex spectrum for waveform synthesis. \\
\bottomrule
\end{tabularx}
\endgroup
\caption{Swin-1D branch and output-head hyperparameters.}
\label{tab:gen_swin_output_hparams_app}
\end{table}

\begin{table}[ht!]
\raggedright
\begingroup
\scriptsize
\setlength{\tabcolsep}{2.5pt}
\renewcommand{\arraystretch}{1.12}
\setlength{\emergencystretch}{1em}
\sloppy
\begin{tabularx}{\columnwidth}{@{}P{0.36\columnwidth} P{0.22\columnwidth} X@{}}
\toprule
\textbf{Hyperparameter} & \textbf{Value} & \textbf{Use Case \& Rationale} \\
\midrule
Period set
& \makecell[l]{2, 3, 5,\\7, 11}
& Captures periodic artifacts across multiple temporal granularities common in neural vocoders and bandwidth expansion. \\
Kernel / stride
& 5 / 3
& Controls local pattern sensitivity and progressive temporal downsampling. \\
Channels
& \makecell[l]{32$\rightarrow$128$\rightarrow$512\\$\rightarrow$1024$\rightarrow$1024}
& Increasing capacity enables hierarchical discrimination from local textures to global structure. \\
LeakyReLU slope
& 0.1
& Mild negative slope improves gradient flow and avoids dead activations. \\
Normalization
& weight norm
& Stabilizes discriminator training by constraining effective weight magnitudes. \\
\bottomrule
\end{tabularx}
\endgroup
\caption{Multi-Period Discriminator (MPD) hyperparameters.}
\label{tab:mpd_hparams_app}
\end{table}

\begin{table}[ht!]
\raggedright
\begingroup
\scriptsize
\setlength{\tabcolsep}{2.5pt}
\renewcommand{\arraystretch}{1.12}
\setlength{\emergencystretch}{1em}
\sloppy
\begin{tabularx}{\columnwidth}{@{}P{0.36\columnwidth} P{0.22\columnwidth} X@{}}
\toprule
\textbf{Hyperparameter} & \textbf{Value} & \textbf{Use Case \& Rationale} \\
\midrule
MRAD resolutions
& \makecell[l]{(512,128,512)\\(1024,256,1024)\\(2048,512,2048)}
& Multi-scale amplitude evaluation to detect artifacts at short, mid, and long analysis windows. \\
MRPD resolutions
& \makecell[l]{(512,128,512)\\(1024,256,1024)\\(2048,512,2048)}
& Multi-scale phase evaluation to penalize phase inconsistency across resolutions. \\
Base channels
& 64
& Provides sufficient capacity for time--frequency feature extraction without excessive compute. \\
STFT window
& rectangular
& Uses a consistent reference spectrogram definition for multi-resolution comparisons. \\
2D kernels
& \makecell[l]{(7,5), (5,3), (5,3),\\(3,3), (3,3)}
& Controls receptive fields over time--frequency neighborhoods for artifact detection. \\
2D strides
& \makecell[l]{(2,2), (2,1), (2,2),\\(2,1), (2,2)}
& Balances downsampling across frequency and time to maintain discriminative detail. \\
\bottomrule
\end{tabularx}
\endgroup
\caption{Multi-resolution spectral discriminator hyperparameters (MRAD/MRPD).}
\label{tab:mrad_mrpd_hparams_app}
\end{table}

\begin{table}[ht!]
\raggedright
\begingroup
\scriptsize
\setlength{\tabcolsep}{2.5pt}
\renewcommand{\arraystretch}{1.12}
\setlength{\emergencystretch}{1em}
\sloppy
\begin{tabularx}{\columnwidth}{@{}P{0.36\columnwidth} P{0.22\columnwidth} X@{}}
\toprule
\textbf{Hyperparameter} & \textbf{Value} & \textbf{Use Case \& Rationale} \\
\midrule
\multicolumn{3}{@{}l@{}}{\textbf{CVaRD}} \\
Base channels
& 32
& Lightweight separable-convolution backbone to detect waveform artifacts with low overhead. \\
CVaR tail fraction / mode
& 0.2 / abs
& Emphasizes extreme deviations (two-sided) to improve sensitivity to rare but perceptually salient artifacts. \\
LeakyReLU slope
& 0.2
& Stronger negative slope improves gradient propagation in compact discriminators. \\
\midrule
\multicolumn{3}{@{}l@{}}{\textbf{CCD + Primal--Dual Control}} \\
HF analysis STFT $(n\_fft,hop,win)$
& (1024,80,320)
& Matches the main analysis resolution to ensure consistent spectral measurements during constraint enforcement. \\
HF band fraction
& 0.25
& Defines the high-frequency band used to quantify disproportionate HF energy (a proxy for harshness/artifacts). \\
Barrier sharpness / threshold
& 8.0 / 0.40
& Smooth barrier penalizes HF-energy exceedance; parameters control sensitivity and activation point. \\
Target barrier
& 0.05
& Sets the desired operating point for constraint satisfaction during training. \\
Augmented Lagrangian penalty $\rho$
& 0.5
& Stabilizes constraint optimization by penalizing large violations quadratically. \\
Dual step size
& 0.01
& Controls the update rate of the dual variable for stable primal--dual dynamics. \\
Dual variable init / max
& 0.0 / 10.0
& Conservative initialization with an upper bound prevents instability from excessively strong constraint pressure. \\
\bottomrule
\end{tabularx}
\endgroup
\caption{Tail-risk and chance-constraint discriminator hyperparameters.}
\label{tab:disc_risk_constraint_hparams_app}
\end{table}

\begin{table}[ht!]
\raggedright
\begingroup
\scriptsize
\setlength{\tabcolsep}{2.5pt}
\renewcommand{\arraystretch}{1.12}
\setlength{\emergencystretch}{1em}
\sloppy
\begin{tabularx}{\columnwidth}{@{}P{0.36\columnwidth} P{0.22\columnwidth} X@{}}
\toprule
\textbf{Hyperparameter} & \textbf{Value} & \textbf{Use Case \& Rationale} \\
\midrule
\multicolumn{3}{@{}l@{}}{\textbf{MCUD}} \\
Criteria ($K$)
& 3
& Aggregates complementary spectral descriptors to approximate perceptual utility beyond adversarial realism alone. \\
Rolloff threshold
& 0.85
& Captures spectral energy distribution by measuring the frequency containing 85\% of cumulative energy. \\
Utility weights
& softmax
& Learnable convex combination enables data-driven prioritization among competing criteria. \\
Sample rate
& 16000
& Defines Nyquist normalization for frequency-domain criteria such as centroid/rolloff. \\
\midrule
\multicolumn{3}{@{}l@{}}{\textbf{EDD}} \\
Base channels
& 32
& Compact backbone to estimate evidence for real vs.\ fake while remaining stable under GAN training. \\
Classes
& 2
& Binary evidential classification aligned with adversarial supervision. \\
Evidence activation
& softplus
& Ensures non-negative evidence and well-defined Dirichlet concentrations. \\
KL regularization weight
& 0.001
& Prevents overconfident predictions by regularizing towards a high-uncertainty prior. \\
Uncertainty weight (G)
& 0.1
& Encourages the generator to reduce epistemic uncertainty by producing more distributionally plausible outputs. \\
\bottomrule
\end{tabularx}
\endgroup
\caption{Utility-based and evidential discriminator hyperparameters.}
\label{tab:disc_utility_evidential_hparams_app}
\end{table}

\begin{table}[ht!]
\raggedright
\begingroup
\scriptsize
\setlength{\tabcolsep}{2.5pt}
\renewcommand{\arraystretch}{1.12}
\setlength{\emergencystretch}{1em}
\sloppy
\begin{tabularx}{\columnwidth}{@{}P{0.36\columnwidth} P{0.22\columnwidth} X@{}}
\toprule
\textbf{Hyperparameter} & \textbf{Value} & \textbf{Use Case \& Rationale} \\
\midrule
\multicolumn{3}{@{}l@{}}{\textbf{DROD}} \\
Base channels
& 32
& Efficient per-segment scoring with separable convolutions for robust aggregation. \\
Entropic risk parameter $\lambda$
& 1.0
& Tunes robustness: higher values emphasize worst-case segments, improving resilience to hard examples. \\
Aggregation
& \makecell[l]{entropic risk\\(log-sum-exp)}
& Smooth worst-case pooling that prioritizes difficult regions without introducing non-differentiable maxima. \\
\midrule
\multicolumn{3}{@{}l@{}}{\textbf{Loss reweighting}} \\
Evidential / DRO weights (D)
& 1.0 / 1.0
& Balances auxiliary discriminators against spectral and waveform discriminators for stable multi-critic training. \\
Evidential / DRO weights (G)
& 1.0 / 1.0
& Controls the influence of uncertainty-aware and robustness-aware objectives on generator updates. \\
\bottomrule
\end{tabularx}
\endgroup
\caption{Entropic-risk discriminator and auxiliary loss-weighting hyperparameters.}
\label{tab:disc_dro_weights_hparams_app}
\end{table}

We use an Anaconda virtual environment with Python 3.9.21, PyTorch 2.0.0+cu118, Torchaudio 0.15.0+cu118, Torchvision 0.15.0+cu118, and CUDA Toolkit 11.8.0. For distributed training and potential scalability, we use NCCL for multi-GPU training and TCP to initialize communication between processes. The training and model hyperparameters for the UDSS-BWE setup, with use cases and rationale are provided in Table \ref{tab:train_system_hparams_app}-\ref{tab:disc_dro_weights_hparams_app}.

\subsection{Hyperparameter Sensitivity}
\label{subsec:sensitivity_codec}

The reference hyperparameters were selected through preliminary development experiments rather than an exhaustive grid or Bayesian search. A comprehensive joint search would be computationally prohibitive because the proposed system contains several interacting discriminators, auxiliary objectives, and optimization controls. We therefore adopt a controlled intuitive few values for sensitivity test. 

All sensitivity experiments use the paper's $(4\rightarrow16)~kHz$ VCTK
training and evaluation split, a 50-epoch training, and the same
implementation used for the main experiments. Unless otherwise stated, the reference value is the middle value shown in Table~\ref{tab:hyperparameter_sensitivity}. The values were chosen to produce meaningful changes in each experiment without decreasing the performance. They should therefore be interpreted as a local robustness analysis around the reported configuration rather than as an attempt to identify a globally optimal parameter set. The reference value used
in the main experiments is shown in bold. The final column states the
anticipated qualitative effect of moving from the low to the high setting; these descriptions are hypotheses to be evaluated rather than empirical conclusions.

\begin{table*}[t]
\centering
\scriptsize
\setlength{\tabcolsep}{2.8pt}
\renewcommand{\arraystretch}{1.08}
\caption{Hyperparameter sensitivity design.}
\label{tab:hyperparameter_sensitivity}
\begin{tabularx}{\textwidth}{@{}
p{0.10\textwidth}
p{0.19\textwidth}
p{0.22\textwidth}
>{\raggedright\arraybackslash}X@{}}
\toprule
Module & Hyperparameter & Low / reference / high & Anticipated sensitivity \\
\midrule
CVaRD
  & Tail fraction \(\alpha\)
  & \(0.10,\ \mathbf{0.20},\ 0.40\)
  & A smaller value concentrates pooling on a more extreme temporal tail,
    increasing sensitivity to localized artifacts but also increasing
    gradient variance. A larger value approaches average pooling and may
    improve stability at the cost of weaker tail-risk awareness. \\

CVaRD
  & Tail-selection mode
  & right,\ \(\mathbf{absolute}\),\ left
  & Absolute selection responds to large activations of either sign and is
    expected to be the least dependent on feature-map polarity. Right- and
    left-tail modes may reveal whether the learned activation sign carries
    consistent risk information, but they may be less robust across
    layers and random seeds. \\

\midrule

CCD
  & High-frequency-band fraction
    \(r_{\mathrm{HF}}\)
  & \(0.15,\ \mathbf{0.25},\ 0.35\)
  & A smaller fraction restricts the constraint to the highest-frequency
    bins. A larger fraction also constrains the upper mid-band and is
    therefore expected to suppress a broader range of spectral
    amplification, potentially reducing both artifacts and useful
    high-frequency reconstruction. \\

CCD
  & Barrier sharpness \(\kappa\)
  & \(4,\ \mathbf{8},\ 16\)
  & A small \(\kappa\) produces a smooth penalty and gradual gradients.
    Increasing \(\kappa\) makes the transition around the threshold more
    selective, but an excessively sharp barrier may create abrupt
    gradients and greater seed-to-seed variability. \\

CCD
  & Activation threshold \(\tau_{0}\)
  & \(0.30,\ \mathbf{0.40},\ 0.50\)
  & A lower threshold activates the high-frequency constraint more often
    and is expected to reduce over-amplification, although it may
    over-regularize legitimate reconstructed harmonics. A higher threshold is more permissive and may improve spectral detail while allowing more high-frequency artifacts. \\

CCD
  & Target mean barrier
  & \(0.025,\ \mathbf{0.05},\ 0.10\)
  & A lower target defines a stricter feasible operating point and should
    increase constraint pressure. A higher target tolerates more
    high-frequency excess and may improve reconstruction freedom at the
    expense of weaker artifact control. \\

CCD
  & Augmented-Lagrangian coefficient
    \(\rho_{\mathrm{cc}}\)
  & \(0.25,\ \mathbf{0.50},\ 1.00\)
  & Increasing \(\rho_{\mathrm{cc}}\) strengthens the quadratic penalty for residual constraint violations. Moderate increases may improve compliance, whereas an overly large value may allow the constraint term to dominate the reconstruction and adversarial objectives. \\

CCD
  & Dual update step
    \(\eta_{\mathrm{cc}}\)
  & \(0.005,\ \mathbf{0.010},\ 0.020\)
  & A small step updates the multiplier conservatively and may respond too slowly to persistent violations. A large step reacts more rapidly but can produce oscillatory multiplier dynamics and unstable generator gradients. \\

CCD
  & Dual initialization / cap
  & \(0/5,\ \mathbf{0/10},\ 0/20\)
  & The zero initialization avoids imposing strong constraint pressure
    before meaningful generator outputs are available. A lower cap offers greater protection against constraint domination, whereas a higher cap permits stronger correction of sustained violations but increases the risk of over-regularization. \\

\midrule

EDD
  & Evidential KL weight
  & \(10^{-4},\ \mathbf{10^{-3}},\ 10^{-2}\)
  & A small KL weight permits concentrated evidence and may produce
    overconfident discriminator predictions. A large weight pulls the
    Dirichlet distribution more strongly toward the high-uncertainty prior, improving regularization but potentially weakening discrimination. \\

EDD
  & Generator uncertainty weight
  & \(0.05,\ \mathbf{0.10},\ 0.20\)
  & Increasing this weight places greater pressure on the generator to
    produce outputs for which the evidential discriminator has low
    epistemic uncertainty. Moderate weighting may improve reliability,
    while excessive weighting may cause the generator to optimize
    discriminator confidence rather than perceptual fidelity. \\

\midrule

DROD
  & Entropic-risk parameter \(\lambda\)
  & \(0.50,\ \mathbf{1.00},\ 2.00\)
  & As \(\lambda\) decreases, entropic aggregation approaches an average
    segment loss. Larger values increasingly emphasize difficult temporal
    segments and may improve worst-case robustness, but they can also
    amplify outlier gradients and reduce training stability. \\

\midrule

Optimization
  & Adversarial warm-up
  & \(2{,}500,\ \mathbf{5{,}000},\ 10{,}000\) steps
  & A shorter warm-up introduces adversarial feedback earlier and may
    accelerate perceptual sharpening, but it increases the risk of early
    instability. A longer warm-up prioritizes reconstruction before
    adversarial learning and is expected to be more stable but may delay
    high-frequency refinement. \\

Optimization
  & Generator/discriminator learning rates
  & \(\begin{array}{c}
       1.0{\times}10^{-4}/2.0{\times}10^{-4}\\
       \mathbf{2.0{\times}10^{-4}/3.5{\times}10^{-4}}\\
       3.0{\times}10^{-4}/5.0{\times}10^{-4}
     \end{array}\)
  & Lower rates should improve numerical stability but may underfit within
    the fixed 50-epoch budget. Higher rates may accelerate convergence but
    are expected to increase adversarial oscillation, particularly because
    several discriminators are updated jointly. \\

Optimization
  & AdamW \((\beta_{1},\beta_{2})\)
  & \(\begin{array}{c}
       (0.70,0.98)\\
       \mathbf{(0.80,0.99)}\\
       (0.90,0.999)
     \end{array}\)
  & Lower momentum values respond more rapidly to changing adversarial
    gradients but produce noisier updates. Higher values smooth the
    optimization trajectory, although excessive smoothing may make the
    discriminator and generator respond too slowly to each other. \\

Optimization
  & Exponential learning-rate decay
  & \(0.995,\ \mathbf{0.999},\ 0.9995\)
  & A smaller decay factor reduces the learning rate more rapidly and may stabilize late training, but it can prematurely restrict adaptation. A value closer to one maintains learning capacity longer and may yield better final refinement if adversarial training remains stable. \\

\bottomrule
\end{tabularx}
\end{table*}

Among the decision-science controls, the CCD threshold
$(\tau_{0})$, barrier sharpness $(\kappa)$, and dual update step
$(\eta_{\mathrm{cc}})$ are have the most immediate effect because
they directly determine when and how strongly high-frequency constraint gradients are applied. The CVaR tail fraction $(\alpha)$ and entropic-risk parameter $(\lambda)$ control a common trade-off between average performance and sensitivity to difficult temporal regions. The evidential weights affect calibration and adversarial stability more strongly than conventional signal-reconstruction metrics.


\paragraph{PCM and telephony-channel sensitivity.}
The original experiment constructs the narrowband conditioning signal using
sinc resampling. We retain this procedure as the reference experiment R0 and vary
only the degradation applied to the narrowband model input. The native source waveform and clean wideband target are not quantized or codec-processed. R1 samples uniformly from sinc resampling, 8-bit linear PCM, and 24-bit linear PCM. R2 samples uniformly from sinc resampling, G.711 $(\mu)$-law, and G.711
A-law. R3 extends the mixture to sinc resampling, PCM-8, both G.711 laws, and AMR-NB at 12.2~kb/s. R4 uses the R3 mixture together with a
300--3400~Hz telephone passband and independent packet losses applied to 20-ms packets with probability 0.03. A lost packet is concealed by repeating the previously received packet. G.711 degradation is implemented as an actual 8-kHz encode/decode round trip rather than as an approximate companding transform applied at 16~kHz. AMR-NB is used only when the installed FFmpeg build exposes the \texttt{libopencore\_amrnb} encoder. AMR-dependent training or evaluation conditions are reported as unavailable when this encoder is absent; additive noise is not substituted as an AMR or telephony surrogate.

\begin{table*}[t]
\centering
\scriptsize
\setlength{\tabcolsep}{2.5pt}
\renewcommand{\arraystretch}{1.04}
\caption{PCM and telephony-channel sensitivity matrix. Operations in a
training mixture are sampled uniformly. 
}
\label{tab:pcm_telephony_sensitivity}

\begin{tabularx}{\textwidth}{@{}
>{\raggedright\arraybackslash}p{0.045\textwidth}
>{\raggedright\arraybackslash}p{0.075\textwidth}
>{\raggedright\arraybackslash}X
>{\centering\arraybackslash}p{0.085\textwidth}
>{\centering\arraybackslash}p{0.095\textwidth}
>{\centering\arraybackslash}p{0.09\textwidth}
>{\centering\arraybackslash}p{0.10\textwidth}@{}}
\toprule
Phase & ID & Narrowband channel operation
& AMR rate & Passband (Hz) & Packet loss & Concealment \\
\midrule

\multicolumn{7}{@{}l}{\textit{Training regimes
  (50 epochs; seeds 1234, 2345, and 3456)}} \\

Train & R0
  & sinc resampling
  & -- & -- & -- & -- \\

Train & R1
  & uniform\{resampling, PCM-8, PCM-24\}
  & -- & -- & -- & -- \\

Train & R2
  & uniform\{resampling, G.711 \(\mu\)-law, G.711 A-law\}
  & -- & -- & -- & -- \\

Train & R3
  & uniform\{resampling, PCM-8, G.711 \(\mu\)-law,
    G.711 A-law, AMR-NB\}
  & 12.2 kb/s & -- & -- & -- \\

Train & R4
  & R3 mixture with telephone filtering and packet loss
  & 12.2 kb/s & 300--3400 & 3\% (20 ms) & repeat previous \\

\midrule

\multicolumn{7}{@{}l}{\textit{Evaluation stressors applied to every
  available trained regime}} \\

Test & Ideal
  & sinc resampling
  & -- & -- & -- & -- \\

Test & PCM-24
  & 24-bit linear PCM quantization
  & -- & -- & -- & -- \\

Test & PCM-8
  & 8-bit linear PCM quantization
  & -- & -- & -- & -- \\

Test & G.711-\(\mu\)
  & 8-kHz G.711 \(\mu\)-law round trip
  & -- & -- & -- & -- \\

Test & G.711-A
  & 8-kHz G.711 A-law round trip
  & -- & -- & -- & -- \\

Test & AMR-12.2
  & 8-kHz AMR-NB round trip
  & 12.2 kb/s & -- & -- & -- \\

Test & AMR-4.75
  & 8-kHz AMR-NB round trip with telephone filtering
  & 4.75 kb/s & 300--3400 & -- & -- \\

Test & Loss-3
  & G.711 \(\mu\)-law with telephone filtering
  & -- & 300--3400 & 3\% (20 ms) & repeat previous \\

Test & Loss-5
  & G.711 \(\mu\)-law with telephone filtering
  & -- & 300--3400 & 5\% (20 ms) & repeat previous \\

\bottomrule
\end{tabularx}
\end{table*}

Codec, filtering, and packet-loss operations are applied only to the
narrowband conditioning waveform presented to the model. The clean
wideband target retains its native waveform precision. Similarly, exporting an enhanced waveform in a 24-bit PCM container controls the output representation but does not restore information removed by earlier quantization, resampling, or codec processing.

Every trained channel is evaluated using the ideal sinc-resampling
condition, PCM-24, PCM-8, G.711 $(\mu)$-law, G.711 A-law, AMR-NB at
12.2 and 4.75~kb/s, and filtered G.711 $(\mu)$-law with 3\% and 5\%
packet loss, subject to codec availability. The evaluation matrix therefore measures both matched-channel performance and cross-channel generalization.
In particular, R0 measures sensitivity to an unseen deployment channel,
whereas R2--R4 measure whether channel augmentation improves robustness without materially degrading performance under ideal resampling. For each condition, PESQ, STOI, LSD, SI-SDR, SI-SNR, and SNR are reported as
the mean and standard deviation across the three independently trained seeds. WER is additionally reported when the paper's local ASR evaluator is enabled.
Utterance-level 95\% confidence intervals are estimated using 1,000 bootstrap resamples. To characterize training reliability, the sensitivity analysis also reports failed or numerically unstable runs and, where applicable, the mean CCD barrier value and final dual multiplier. These diagnostic quantities help distinguish genuine performance trade-offs from hyperparameter settings that merely destabilize adversarial optimization.


\subsection{Subjective Evaluation Details}
\label{subsec:subj_eval_details}

\begin{figure*}[ht!]
  \centering
  \includegraphics[width=0.98\textwidth]{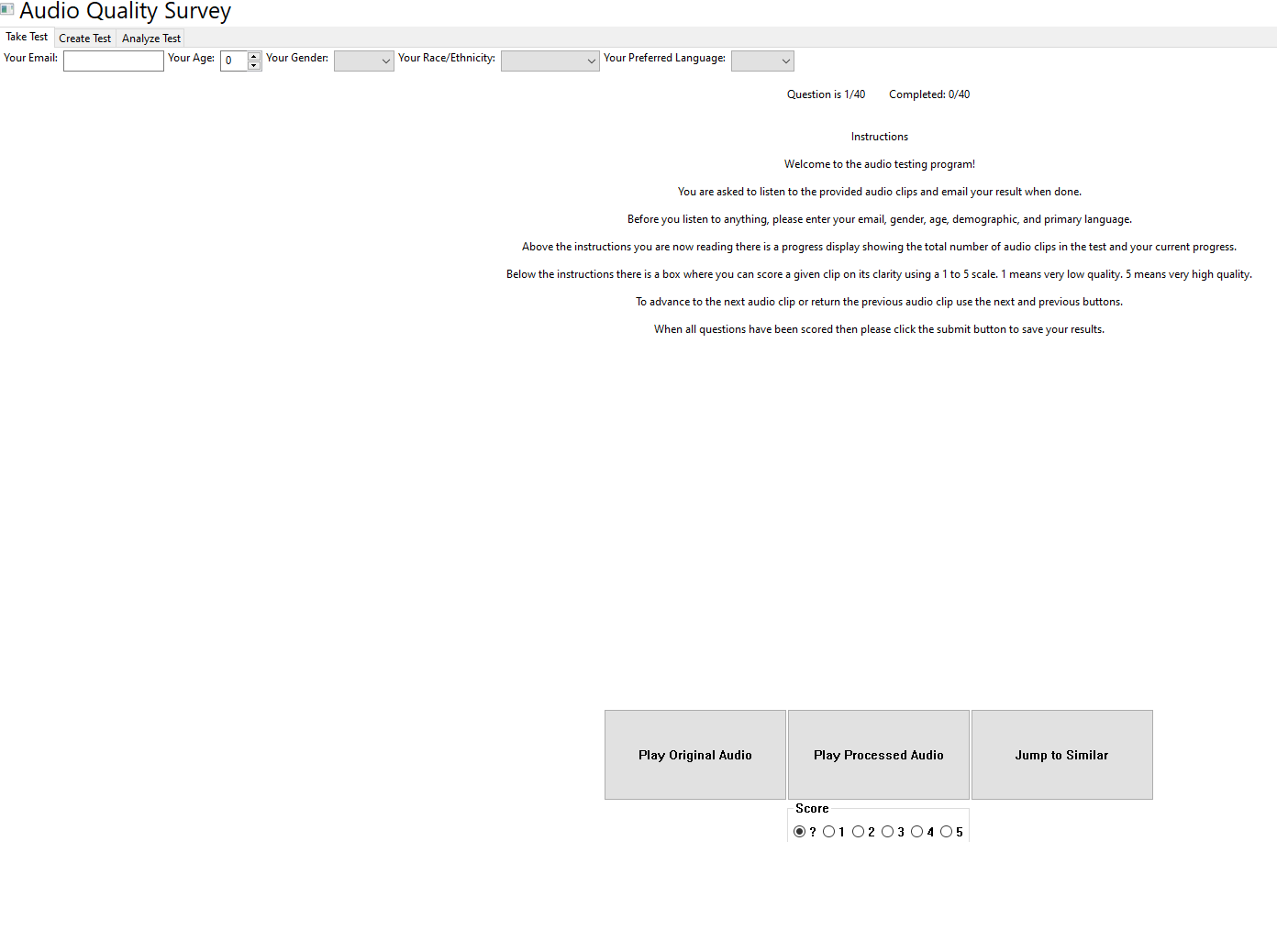}
  \caption{The software interface used in Subjective Tests.}
  \label{fig:subj_eval_interface_ss}
\end{figure*}

To evaluate the performance of the proposed UDSS-BWE, a formal pair-preference listening test was conducted. A total of ten English-speaking undergraduate students, who self-reported normal-hearing (NH) volunteers — comprising four males and six females with an average age of 24—participated in the study. The participants voluntarily join to rate the audios without any compensation. At first they are trained on how to assign scores based on perceived perceptual quality of audios. They are also briefed about the purpose of the experiments, potential risks, and about the outcome of this paper.
All participants were native English speakers and used soundproof headsets to ensure consistent and distraction-free listening conditions.
Each participant evaluated 30 sets of speech samples. Every set included three randomly presented versions of the same utterance: (i) the unprocessed (noisy) signal, (ii) the baseline-enhanced signal using APBWE processing, and (iii) the speech processed by the proposed network. For reference, a clean version of the speech signal was also available, though it was not part of the evaluation. Participants rated the perceptual quality of each sample on a 5-point Mean Opinion Score (MOS) scale, where 1 indicates the lowest and 5 the highest quality. Additionally, they were asked to select the most preferred version from the three presented options.
The test used 5–7-second speech clips selected from the VCTK dataset, which were processed under three different frequency band conditions: 2–16 kHz, 12–48 kHz, and 24–48 kHz. Individual results were analyzed separately for both male and female participants across all band configurations. In the Figure \ref{fig:subj_eval_interface_ss}, we have presented the software interface that we use to conduct subjective evaluation.

\begin{figure}[ht!]
  \centering
  \includegraphics[width=0.48\textwidth]{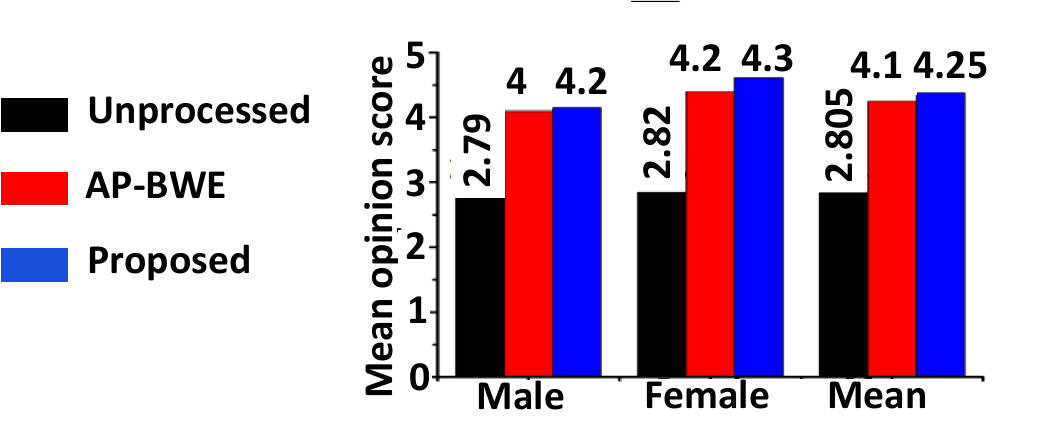}
  \caption{Subjective test score.}
  \label{fig:comparison}
\end{figure}

The findings in Figure \ref{fig:comparison} clearly show that the speech enhanced by the proposed network was consistently and significantly preferred over both the unprocessed and baseline-processed versions. In particular, the proposed UDSS-BWE achieved a $\sim$51\% improvement in user preference compared to unprocessed speech and a 3.6\% improvement over the AP-BWE baseline. These results highlight the network’s robust ability to enhance perceptual speech quality.

\end{document}